\documentclass{CSML}

\def\dOi{11(4:3)2015}
\lmcsheading%
{\dOi}
{1--45}
{}
{}
{Mar.~\phantom06, 2014}
{Oct.~26, 2015}
{}

\ACMCCS{[{\bf Theory of computation}]: Logic---Equational logic and
  rewriting\,/\,Type theory\,/\,Automated reasoning\,/\,Logic and verification}

\usepackage[breaklinks=true]{hyperref}
\usepackage[anythingbreaks]{breakurl}
\usepackage{amsmath}
\usepackage{amssymb}
\usepackage{t1enc}
\usepackage{MnSymbol}
\usepackage{stmaryrd}

\newcommand\comment[1]{}

\newcommand\hsp[1][3ex]{\hspace*{#1}}
\newcommand\vsp[1][1mm]{\vspace*{#1}}

\newcommand\vide{\emptyset}
\newcommand\eg{{\em e.g.} }
\newcommand\ie{{\em i.e.} }

\newcommand\furl[1]{\footnote{\url{http://#1}}}

\newcommand\dom{\mr{dom}}

\newcommand\FV{\mr{FV}}
\newcommand\Pos{\mr{Pos}}
\newcommand\lex{\mr{lex}}
\newcommand\mul{\mr{mul}}
\renewcommand\prod{\mr{prod}}
\newcommand\stat{\mr{stat}}

\renewcommand\a{\rightarrow}
\newcommand\A{\Rightarrow}

\renewcommand\to{\mapsto}

\newcommand\ab{\a_\b}

\renewcommand\ae{\a_\eta}

\newcommand\I[1]{\llbracket{#1}\rrbracket}

\newcommand\ex{\exists}
\newcommand\all{\forall}

\newcommand\et{\wedge}

\newcommand\sle{\subseteq}

\newcommand\tle{\unlhd}
\newcommand\tge{\unrhd}
\newcommand\tlt{\lhd}
\newcommand\tgt{\rhd}

\newcommand\qge{\sqsupseteq}

\newcommand\qgt{\sqsupset}

\newcommand\al{\alpha}
\renewcommand\b{\beta}

\newcommand\vep{\varepsilon}

\renewcommand\t{\theta}

\renewcommand\l{\lambda}

\newcommand\s{\sigma}
\renewcommand\S{\Sigma}

\newcommand\mc{\mathcal}

\newcommand\mr{\mathrm}
\newcommand\mb{\mathbb}

\newcommand\mk{\mathfrak}
\newcommand\ms{\mathsf}

\newcommand\bN{\mb{N}}

\newcommand\cF{\mc{F}}

\newcommand\cL{\mc{L}}
\newcommand\cM{\mc{M}}
\newcommand\cN{\mc{N}}

\newcommand\cP{\mc{P}}

\newcommand\cS{\mc{S}}
\newcommand\cT{\mc{T}}

\newcommand\cX{\mc{X}}

\newcommand\ka{\mk{a}}
\newcommand\kb{\mk{b}}
\newcommand\kc{\mk{c}}

\renewcommand\sa{\ms{a}}
\renewcommand\sb{\ms{b}}
\renewcommand\sc{\ms{c}}
\newcommand\sd{\ms{d}}
\newcommand\se{\ms{e}}
\renewcommand\sf{\ms{f}}
\newcommand\sg{\ms{g}}
\newcommand\sh{\ms{h}}

\newcommand\sj{\ms{j}}
\newcommand\sk{\ms{k}}
\renewcommand\sl{\ms{l}}

\renewcommand\ss{\ms{s}}

\newcommand\sA{\ms{A}}
\newcommand\sB{\ms{B}}
\newcommand\sC{\ms{C}}

\newcommand\sL{\ms{L}}

\newcommand\sN{\ms{N}}
\newcommand\sO{\ms{O}}

\newcommand\vt{{\vec{t}}}
\newcommand\vu{{\vec{u}}}
\newcommand\vv{{\vec{v}}}

\newcommand\vx{{\vec{x}}}

\newcommand\vB{{\vec{B}}}

\newcommand\vS{{\vec{S}}}
\newcommand\vT{{\vec{T}}}
\newcommand\vU{{\vec{U}}}
\newcommand\vV{{\vec{V}}}
\newcommand\vW{{\vec{W}}}

\newenvironment{rul}
  {$\begin{array}{rcl}}
  {\end{array}$}

\newenvironment{rew}[1][~~\a~~]
  {$\begin{array}{r@{#1}l}}
  {\end{array}$}
\newenvironment{rewc}[1][~~\a~~]
  {\begin{center}\begin{rew}[#1]}
  {\end{rew}\end{center}}

\comment{
\newcounter{counter}

{\theorembodyfont{\rmfamily} 
  
  \newtheorem{lem}[counter]{Lemma}
  \newtheorem{thm}[counter]{Theorem}
  \newtheorem{cor}[counter]{Corollary}
  
  \newtheorem{prop}[counter]{Property}

}}

\newenvironment{prf}{\proof}{}

\newcommand\af{{\al(\sf)}}

\newcommand\tf{\tau(\sf)}

\newcommand\tx{\tau(x)}
\newcommand\ty{\tau(y)}
\newcommand\tz{\tau(z)}

\newcommand\mset[1]{\{\!\!|#1|\!\!\}}

\newcommand\Acc{\mr{Acc}}
\newcommand\rk{\mr{rk}}
\newcommand\SN{\mr{SN}}
\newcommand\CC{\mr{CC}}
\newcommand\Sort{\mr{Sort}}
\newcommand\horpo{\mr{horpo}}
\newcommand\chorpo{\mr{chorpo}}

\newcommand\SPos{\mr{SPos}}
\newcommand\RPos{\mr{RPos}}
\newcommand\LPos{\mr{LPos}}
\newcommand\NPos{\mr{NPos}}
\newcommand\CPos{\mr{CPos}}

\newcommand\gtt{>}
\newcommand\get\ge
\newcommand\lttwf\lessdot
\newcommand\letwf\leqdot
\newcommand\gttwf\gtrdot
\newcommand\getwf\geqdot
\newcommand\gtf{>_\cF}
\newcommand\gef{\ge_\cF}
\newcommand\eqf{\simeq_\cF}
\newcommand\gts\tgt
\newcommand\ges\tge
\newcommand\lts\tlt
\newcommand\gta{\gts_a}
\newcommand\gea{\ges_a}

\newcommand\gtb{\gts^s_b}
\newcommand\geb{\ges^s_b}
\newcommand\gti[1][]{\tgt_@^{#1}}
\newcommand\gto\qgt
\newcommand\geo\qge
\newcommand\gtp[1][]{\succ^{#1}}
\newcommand\gep[1][]{\succeq^{#1}}
\newcommand\gtapp[1][]{\succ^{#1}_@}
\newcommand\gtpt[1][]{\succ^{#1}_\tau}
\newcommand\gept[1][]{\succeq^{#1}_\tau}

\newcommand{\hide}[1]{}

\newcommand\target[1]{\hypertarget{#1}{}}
\newcommand\link[1]{\hyperlink{#1}{(#1)}}
\renewcommand\url[1]{\href{#1}{\tt #1}}

\newcommand\ind[1][]

\begin{document}

\title[The Computability Path Ordering]{The Computability Path Ordering}

\author[F.~Blanqui]{Fr\'ed\'eric Blanqui\rsuper a}
\address{{\lsuper a}INRIA, Deducteam, France}
\email{frederic.blanqui@inria.fr}

\author[J.-P.~Jouannaud]{Jean-Pierre Jouannaud\rsuper b}
\address{{\lsuper b}\'Ecole Polytechnique, LIX, and Universit\'e Paris-Sud, France}
\email{jeanpierre.jouannaud@gmail.com}

\author[A.~Rubio]{Albert Rubio\rsuper c}
\address{{\lsuper c}Technical University of Catalonia, Spain}
\email{albert@cs.upc.edu}

\thanks{{\lsuper c}Fr\'ed\'eric Blanqui thanks the Institute of Software of the
  Chinese Academy of Sciences for hosting him from June 2012 to August
  2013. This research was supported by the Spanish MINECO under grant
  TIN2013-45732-C4-3-P}

\begin{abstract}
  This paper aims at carrying out termination proofs for simply typed
  higher-order calculi automatically by using ordering comparisons. To
  this end, we introduce the computability path ordering (CPO), a
  recursive relation on terms obtained by lifting a precedence on
  function symbols. A first version, core CPO, is essentially obtained
  from the higher-order recursive path ordering (HORPO) by eliminating
  type checks from some recursive calls and by incorporating the
  treatment of bound variables as in the so-called computability closure. The
  well-foundedness proof shows that core CPO captures the essence of
  computability arguments {\em \`a la} Tait and Girard, therefore
  explaining its name. We further show that no further type check can be
  eliminated from its recursive calls without loosing
  well-foundedness, but one for which we found no counter-example
  yet. Two extensions of core CPO are then introduced which allow one
  to consider: the first, higher-order inductive types; the second, a
  precedence in which some function symbols are smaller than
  application and abstraction.
\end{abstract}

\maketitle

\section{Introduction}
\label{sec-intro}

This paper addresses the problem of automating termination proofs for
typed higher-order calculi by reducing them to ordering
comparisons between lefthand and righthand sides of rules.

It also addresses another, more fundamental problem of mathematical
importance.  Consider the set of terms generated by a denumerable set
of variables, application, abstraction and some set of function
symbols with arities, our version of the pure $\lambda$-calculus. We
shall use a (possibly infinite) set $R$ of pairs of $\lambda$-terms
called rewrite rules used as our computing device.  Given a term as
input, whether our computing device will eventually terminate and
return an answer is \emph{in general} undecidable, even if $R$ is a
singleton set~\cite{dauchet88mfcs}.  It may even be undecidable for
specific rewrite systems, such as the well-known $\beta$-reduction
rule (formally defined here as the infinite set of its instances). A
major question is the following: can we approximate the set of
$\beta$-terminating terms by some meaningful subset?  An important
partial answer was given by Turing: the set of simply typed
$\lambda$-terms, where the word \emph{simply} refers to a specific
typing discipline introduced by Church in the
$\lambda$-calculus~\cite{church40jsl}, is terminating when a specific
strategy is employed~\cite{turing42unpub}. The complete answer, the
fact that the very same set is indeed terminating under any strategy,
is due to Sanchis~\cite{sanchis67ndj}. Tait and Girard gave later
proofs~\cite{tait72lc,girard72phd} which have been the basis of many
further generalizations, by considering more rules ($\eta$-reduction,
recursors, general schema), and more terms characterized by more
elaborate type disciplines (polymorphic, dependent, inductive type
systems). When considering $\b$-reduction alone, the obtained
approximations of the set of terminating $\l$-terms are quite
satisfactory. But proving the corresponding statement that
computations terminate when given a typed $\l$-term as input, requires
using an extremely powerful technique called
\emph{reducibility}\footnote{In fact, Tait speaks of
  ``convertibility'' in \cite{tait67jsl}, ``realizability'' in
  \cite{tait72lc}; and Girard of ``r\'eductibilit\'e'' and
  ''reducibility'' in
  \cite{girard71sls,girard72phd,girard88book}. Following G\"odel,
  ``computability'' is used by Troelstra in \cite{troelstra73chapter},
  p. 100.}, introduced by Tait for simply typed terms, and further
developed by Girard for the richer type disciplines.  Given a set of
terms and a set of rewrite rules $R$, a reducibility predicate is
defined by axioms that it should satisfy, mainly closure under term
constructions, closure under rewriting with $R$, and containment in
the set of terminating terms. Girard exhibited a particular predicate
for $\beta$-reduction which can be easily adapted for other sets of
rules, but there are sets of rules for which some typable terms
originate an infinite computation. We therefore turn to a new
undecidable question: which sets of rules admit a computability
predicate?

The question we answer in this paper is whether this set $S$ (of sets
$R$ of rules) admits some non-trivial decidable subset: our
approximation of $S$ is the set of sets $R$ of rules such that pairs
in $R$ are ordered by (some instance of) the computability path
ordering CPO.

The work itself takes its roots in early attempts by Breazu-Tannen and
Gallier \cite{tannen89icalp} and independently Okada
\cite{okada89issac} to consider mixed typed $\lambda$-calculi with
algebraic rewriting. Both works used Girard's
computability predicates method to show that the strong normalization
property of algebraic rewriting was preserved in the union.  These
results grew into a whole new area, by extending the type discipline
on the one hand, and the kind of rules that could be taken care of on
the other hand. The type discipline was extended independently by
Barbanera and Dougherty in order to cover the whole calculus of
constructions \cite{barbanera90ctrs,dougherty92ic}, while the rule
format was extended as described next.

Higher-order rewrite rules satisfying the {\em general schema}, a
generalization of G\"odel's primitive recursion rules for higher
types, were introduced by Jouannaud and Okada in the case of a
polymorphic type discipline \cite{jouannaud91lics,jouannaud97tcs}. The
latter work was then extended first by Barbanera and Fern\'andez
\cite{barbanera93tlca,barbanera93icalp} and finally by Barbanera,
Fern\'andez and Geuvers to cover the whole calculus of constructions
\cite{barbanera94lics}. Recursors for {\em basic} inductive types,
which constructors admit arguments of a non-functional type only,
could be taken care of by the general schema, but arbitrary strictly
positive inductive types could not, prompting for an extension of the
schema, which was reformulated for that purpose by Blanqui, Jouannaud
and Okada \cite{blanqui02tcs}. This new formulation was based on the
notion of {\em computability closure} of a term $\sf(\vt)$, defined as a
set of terms containing $\vt$ and closed under computability preserving
operations in the sense of Tait and Girard. Membership to the general
schema was then defined for an arbitrary rewrite rule as membership of
its righthand side to the computability closure of its lefthand
side.  This elegant, flexible and powerful definition of the general
schema was finally extended by Blanqui in a series of papers, until it
covered the entire calculus of inductive constructions including
strong elimination
rules~\cite{blanqui05mscs,blanqui05fi}, rewriting modulo
some equational theories and rewriting with higher-order
pattern-matching \cite{blanqui15tcs}.

Introduced by Jouannaud and Rubio, HORPO was the next step, the very
first order on simply typed $\l$-terms defined by induction on the
term structure, as does Dershowitz recursive path ordering for
first-order terms~\cite{dershowitz82tcs}. Comparing two terms with
HORPO starts by comparing their types in a given well-founded ordering
on types before to proceed recursively on the structure of the
compared terms, in a way which depends on a comparison of the roots of
both terms in a given well-founded order on the algebraic signature
called the precedence~\cite{jouannaud99lics}. HORPO was extended to
the calculus of constructions by Walukiewicz \cite{walukiewicz03jfp},
and to use semantic interpretations of terms instead of a precedence
on function symbols by Borralleras and Rubio \cite{borralleras01lpar}.
An axiomatic presentation of the rules underlying HORPO can be found
in \cite{goubault01csl}. A more recent work in the same direction is
\cite{dershowitz12tcs}. A more general version of HORPO appears in
\cite{jouannaud07jacm}, which uses the computability closure to
strengthen its expressivity. Blanqui proved that the first version of
HORPO is contained in an order defined as a fixpoint of the
computability closure definition~\cite{blanqui06tr}. Indeed, HORPO and
the computability closure share many similar constructs, raising
expectations for a simpler and yet more expressive definition, instead
of a pair of mutually inductive definitions for the computability
closure and the ordering itself. On the positive side, the
computability closure makes little use of type comparisons, hence may
succeed when HORPO fails for type reason. Unfortunately, its fixpoint is not a
syntax-oriented definition, hence has a more limited practical usage.

Originally formulated in~\cite{blanqui06lpar-horpo}, the question of
finding a syntax oriented recursive definition of HORPO that would
inherit the advantages of the computability closure paved the way to
CPO, the computability path ordering. The first definition was given
in~\cite{blanqui07lpar}, later improved as CPO
in~\cite{blanqui08csl}. A major improvement of CPO is that
type comparisons are no more systematic, but occur in very specific
cases. This does not only speed up computations, but also boosts
the ordering capabilities in an essential way. Further, bound
variables are handled explicitly by CPO, allowing for arbitrary
abstractions in the righthand sides together with a more uniform definition.

In this paper, we present an in-depth study of an improved version of
CPO for a simple extension of Church's simple type discipline
\cite{church40jsl}, before we extend it to inductive types along the
lines suggested in~\cite{blanqui08csl} following a technique dating
back to Mendler \cite{mendler87phd,mendler91apal} and extended to
rewriting by Blanqui \cite{blanqui05mscs}. In particular,
we first show that many improvements of CPO cannot be well-founded: type
comparisons are necessary when recursive calls deconstruct the
lefthand side, but are not otherwise. While this all came out of the
well-foundedness proof, it indeed shows a strong relationship between
the recursive structure of CPO and the computability predicates method
of Tait and Girard that is used to carry out the proof, which explains
the name CPO. We then address the treatment of inductive types which
remained {\em ad hoc} so far, thanks to the use of accessibility, a
relationship introduced by Blanqui which generalizes the notion of
inductive type~\cite{blanqui05fi}. We finally introduce
another novelty: small symbols. In all previous definitions, function
symbols were bigger in the precedence than application and
abstraction. Such symbols are now called \emph{big}, while
\emph{small} symbols behave differently, being possibly smaller than
both. Small symbols were suggested by Li to carry out a generalization of CPO to dependent
types~\cite{jouannaud15tlca}.

In the recent years, the success of HORPO has prompted interest in the
generalization to higher-order computations of various other methods
used for first-order computations, most notably Art and Giesl's
dependency pairs \cite{arts00tcs,giesl06jar,hirokawa07ic} yielding for
instance 
\cite{sakai05ieice,kusakari09ieice,kop10wst,suzuki11pro,kop11rta,kop12phd},
and interpretation methods
\cite{manna70hicss,lankford79tr,zantema95fi,contejean05jar} yielding
for instance 
\cite{vandepol96phd,hamana07ppdp,roux11phd,fuhs12rta}.

The paper is organized as follows. First, we define the sets of types
and terms that we consider (simply typed $\l$-terms with function
symbols of fixed arity), and the class of orderings on types that can
be used in CPO. We then give a first definition of our ordering (core
CPO), and show that it can hardly be improved while keeping the same
recursive structure and well-foundedness. We then show how to prove
its well-foundedness by extending Tait and Girard's technique of
computability predicates. In the following sections, we consider two
extensions of core CPO. In the first one, core CPO is extended by
using accessible subterms which allows to handle strictly inductive
types. In the second, application or abstraction are allowed to be
bigger than a function call. Concluding remarks are given in
Section~\ref{sec-conclu}.

We recommend surveys~\cite{dershowitz90chapter,terese03book}
for rewriting and~\cite{barendregt92chapter} for typed $\lambda$-calculus.

\section{Types and admissible type orderings}
\label{sec-typ}

CPO is a relation on well typed terms but, instead of allowing the
comparison of terms of the same type only, it allows the type to
decrease in some well-founded ordering. However, not any type ordering
is admissible.

In this section, we first recall the definition of (simple) {\em
  types} and some basic functions on types. Then, we define what are
the (strict) orderings on types that can be used in CPO, study some of
their properties and give an example based on a well-founded
precedence on type constants.

\begin{defi}[Types]
Let $\cS$ be a set of {\em sorts}. The set $\cT$ of {\em types}, the
{\em arity} $\al(\_)$ and the {\em order} $o(\_)$ of a type are
inductively defined as follows:
\begin{itemize}
\item a sort $\sA\in\cS$ is a type of arity $\al(\sA)=0$ and order
  $o(\sA)=0$.
\item if $T$ and $U$ are types, then $T\a U$ is a type of arity
  $\al(T\a U)=1+\al(U)$ and order $o(T\a U)=\max\{1+o(T),o(U)\}$.
\end{itemize}
\end{defi}

\noindent We use capital letters for types and a different font for sorts (\eg
$T$ and $\sA$), and $\vT$ for a (possibly empty) sequence of types
$T_1,\ldots,T_n$, of \emph{length} $|\vT|=n$.

As usual, $\a$ associates to the right so that $\sA\a\sA\a\sA$ and
$\sA\a(\sA\a\sA)$ are the same.

Given a relation $R$, let $R^+$ (resp. $R^*$) denote the transitive
(resp. transitive and reflexive) closure of $R$.

\begin{defi}[Admissible type orderings]
\label{def-typ-ord}
\renewcommand\ind[1]{{\bf(#1)}\index{(#1)|inddef[def-typ-ord]}} Let
$\gts_l$ and $\gts_r$ be the relations on types such that ${T\a
  U}\,\gts_l\,{T}$ and ${T\a U}\,\gts_r\,{U}$ respectively, and $\gts$
be the transitive closure of their union. A (strict) ordering
$>$  on types is {\em admissible} if:
\begin{itemize}
\item\target{typ-right-subterm} ${\gts_r}\sle{>}$ \hfill\ind{typ-right-subterm}
\item\target{typ-sn}${\gttwf}={({>}\cup{\gts_l})^+}$ is well-founded\hfill\ind{typ-sn}
\item\target{typ-arrow} if $T\a U>V$, then $U\ge V$ or $V=T\a U'$ with
  $U>U'$ \hfill\ind{typ-arrow}
\end{itemize}
where $\ge$ is the reflexive closure of $>$. We say that a type $T$ is
compatible (resp. strictly compatible) with a sort $\sA$, written
$\Sort_{\letwf\sA}(T)$ (resp. $\Sort_{\lttwf\sA}(T)$) if $\sB\letwf\sA$
(resp. $\sB\lttwf\sA$) for every sort $\sB$ occurring in $T$.
\end{defi}

Admissible type orderings originate from~\cite{jouannaud07jacm}. Note
that a sort can be bigger than an arrow type. If $\sA$ is a sort
occurring in $T$, then $T\getwf\sA$. Finally, note that the relation
$\gttwf$ is a {\em simplification ordering} \cite{dershowitz82tcs}.

We now give an example of admissible ordering based on a well-founded
precedence on sorts. For a concrete use case, see Example
\ref{ex-cont} below.

\begin{lem}
\label{lem-typ-ord}
Given a well-founded ordering $>_\cS$ on sorts, let $>$ be the
smallest ordering $>$ on types containing $>_\cS$ and $\gts_r$ and
such that, for all $U,V,V'$, it holds that $V>V'$ implies $U\a V>U\a
V'$. Then, $>$ is admissible.
\end{lem}

\begin{prf}\hfill
\begin{itemize}
\item(typ-sn) $\gttwf$ is included in the RPO extending $>_\cS$
  \cite{dershowitz79focs}, hence is well-founded.
\item(typ-right-subterm) By definition.
\item(typ-arrow) Let $T\a U > V$. The proof is by induction on the
  definition of $>$.
\begin{enumerate}
\item $>$ is ${>_\cS}$. Impossible since $T\a U$ is not a sort.
\item $>$ is $\gts_r$. Then $U=V$, hence $U\ge V$.
\item $V = T\a W$ and $U>W$. Immediate.
\item $T\a U > W > V$. By induction hypothesis applied to $T\a U>W$,
  there are two cases:
\begin{itemize}
\item $U\ge W$. Then, by transitivity, $U>V$.
\item $W=T\a U'$ for some $U'<U$. By induction hypothesis on \mbox{$T\a U'>V$,}
  there are two cases:
\begin{itemize}
\item $U'\ge V$. By transitivity, $U>V$.
\item $V=T\a V'$ for some $V'<U'$. By transitivity, $U> V'$ and we are done.
\qed
\end{itemize}
\end{itemize}
\end{enumerate}
\end{itemize}
\end{prf}\medskip

\noindent In the following, we prove some properties of admissible type orderings:

\begin{lem}
\label{lem-typ-arrow}
Let $>$ be an admissible type ordering. If $T\a U> T'\a U'$, then
$U>U'$.
\end{lem}

\begin{prf}
By \link{typ-arrow}\index{(typ-arrow)|indlem[lem-typ-arrow]}, either $T=T'$
and $U>U'$ or $U\ge T'\a U'$,
in which case 
we conclude by \link{typ-right-subterm}\index{(typ-right-subterm)|indlem[lem-typ-arrow]} and transitivity. 
\qed
\end{prf}

\begin{lem}
\label{lem-typ-occ-sort}
Let $>$ be an admissible type ordering. If $\sA>U$, then $\Sort_{\lttwf\sA}(U)$.
\end{lem}

\begin{prf}
  Let $\sB$ be a sort occurring in $U$. Then, $U\getwf\sB$. Hence, by
  transitivity, $\sA\gttwf\sB$.\qed
\end{prf}

\begin{lem}
\label{lem-typ-le-sort}
Let $>$ be an admissible type ordering. If $T>U$ and
$\Sort_{\letwf\sA}(T)$, then $\Sort_{\letwf\sA}(U)$.
\end{lem}

\begin{prf}
Let $\sC$ be a sort occurring in $U$. We proceed by induction on $T$.
\begin{itemize}
\item $T=\sB$. Since $\Sort_{\letwf\sA}(T)$, $\sB\letwf\sA$. By Lemma
  \ref{lem-typ-occ-sort}, $\Sort_{\lttwf\sB}(U)$ and
  $C\lttwf\sB$. Therefore, by transitivity, $C\letwf\sA$.
\item $T=S\a T'$. Then, $\Sort_{\letwf\sA}(S)$ and $\Sort_{\letwf\sA}(T')$. By
  \link{typ-arrow}\index{(typ-arrow)|indlem[lem-typ-occ]}, there are two
  cases:
\begin{itemize}
\item $T'\ge U$. Then, by induction hypothesis,
  $\Sort_{\letwf\sA}(U)$.
\item $U=S\a U'$ and $T'>U'$. By induction hypothesis,
  $\Sort_{\letwf\sA}(U')$. Hence $\Sort_{\letwf\sA}(U)$.\qed
\end{itemize}
\end{itemize}
\end{prf}

\section{Terms}

In this section, we define the set of terms on which CPO operates. We
consider simply typed $\l$-terms
\cite{church40jsl,barendregt92chapter} with function symbols of fixed
arity, that is, a function symbol of arity $n$ always comes with $n$
arguments.  We assume that every variable or function symbol comes
equipped with a fixed type and that $\alpha$-equivalence replaces a
variable by another variable of the same type.

\begin{defi}[Terms]
  Let $\cX$ be an infinite set of {\em variables}, each variable $x$
  being equipped with a type $\tau(x)\in\cT$ so that there is an
  infinite number of variables of each type. Let also $\cF$ be a
  (finite or infinite) set of {\em function symbols} disjoint from
  $\cX$, each function symbol $\sf$ being equipped with a type
  $\tf\in\cT$ and an arity $\af\le\al(\tf)$. The {\em declaration}
  $\sf^n:T$ indicates the arity $n$ and type $T$ of $\sf$. The set
  $\cL$ of terms is defined inductively as follows:
\begin{itemize}
\item
a variable $x$ is a term of type $\tx$;
\item
if $\sf^n : T_1\a\ldots\a T_n \a U$ and
$t_1,\ldots,t_n$ are terms of type $T_1,\ldots,T_n$ respectively,
then $\sf(t_1,\ldots,t_n)$ is a term of type $U$;
\item
if $t$ and $u$ are terms of types $U\a V$ and $U$ respectively, then
$tu$ is a term of type $V$;
\item
if $x$ is a variable and $t$ is a term of type $T$, then $\l xt$ is a
term of type $\tx\a T$.
\end{itemize}

\noindent
We denote by $\tau(t)$ the type of a term $t$, and write $t:T$ when
$\tau(t)=T$.
\end{defi}

We usually write $\sf:T$ for the declaration $\sf^0:T$, omitting the
arity $n=0$, and $\sf$ for $\sf()$. Note that a term $\sf(\vt)$ may
have a functional type, hence can be applied. Application associates
to the left so that $tuv$ is the same as $(tu)v$.

We use the letters $x,y,z,\ldots$ for variables, $\sf,\sg,\ldots$ for
function symbols, and $a,b,\ldots$, $s,t,u,v,\ldots$, $t',u',\ldots$
for terms.

We denote by $\FV(t)$ the set of free variables in $t$, by $\tlt$ the
strict subterm relationship on terms, and by $\tle$ its reflexive closure.
The height of a term $t$, written $|t|$, is the height of its tree
representation: $|x|=0$, $|\sf|=0$ if $\af=0$,
$|\sf(\vt)|=1+\max\{|t_i|\mid 1\le i\le\af\}$ if $\af>0$,
$|tu|=1+\max\{|t|,|u|\}$ and $|\l xt|=1+|t|$.

\begin{defi}[Substitution]\hfill
\begin{itemize}
\item A substitution is a function $\s:\cX\a\cL$ such that
  $\dom(\s)=\{x\in\cX\mid\s(x)\neq x\}$ is finite and, for every $x$,
  $\tau(\s(x))=\tau(x)$. As usual, the application of a substitution
  $\s$ to a term $t$, written $t\s$, is defined so as to avoid
  free-variable captures when renaming some bound variables of $t$ by
  new variables of the same type \cite{curry58book}.
\item A substitution $\s$ is {\em away from} a finite set of variables
  $X$ if $(\dom(\s)\cup\FV(\s))\cap X=\vide$, where
  $\FV(\s)=\bigcup\,\{\FV(\s(x))\mid x\in\dom(\s)\}$.
\item A relation $>$ on terms (or sequences of terms) is stable by
  substitution away from $X$ if $a\s>b\s$ whenever $a>b$ and $\s$ is
  away from $X$. A relation is stable by substitution if it is stable
  by substitution away from $\vide$.
\end{itemize}
\end{defi}

\noindent We will use the letters $\s,\t,\ldots$ for substitutions, and denote
the substitution mapping the variables $\vx$ of its domain to the
terms in $\vt$ (hence $|\vx|=|\vt|$) by $(_\vx^\vt)$.

Note that stability by substitution reduces to the standard
definition: $>$ is stable by substitution if $a\s>b\s$ whenever $a>b$
(because any substitution is away from $\vide$).

The equivalence relation identifying terms up to type-preserving
renaming of their bound variables is called $\al$-equivalence and
written $=_\al$ as usual \cite{curry58book}.

Given a relation $R$, let $\SN_T(R)$ be the set of terms of type $T$
from which there is no infinite sequence of $R$-steps, and
$\SN(R)=\bigcup\{\SN_T(R)\mid T\in\cT\}$.

\section{Relations}

One ingredient of CPO is a well-founded quasi-ordering on function
symbols and, for each equivalence class generated by the corresponding
equivalence relation, a {\em status}
$\stat\in\{\mul\}\cup\{\lex(n)\mid n>2\}$ prescribing how to compare
the arguments of two equivalent symbols, by either its multiset
\cite{dershowitz79cacm} or lexicographic extension. We hereafter
recall the necessary definitions and state some simple but important
properties of these operations. The product extension is introduced
here for technical reasons.

Given a relation $\gtp$ on terms, let:
\begin{itemize}
\item $\vt~\gtp_\prod\,\vu$ if $|\vt|=|\vu|$ and there is
  $j\in\{1,\ldots,|\vu|\}$ s.t. $t_j\gtp u_j$ and, for all
  $i\neq j$, $t_i=u_i$.

\item $\vt\gtp_\mul\vu$ if $\mset\vt\,(\gtp[1]_\cM)^+\,\mset\vu$
  where $\mset\vt$ is the multiset made of the elements in $\vt$ and
  $M+\mset{x}\,\gtp[1]_\cM\,M+\mset{y_1,\ldots,y_n}$ ($n\ge 0$) if,
  for all $i$, $x\gtp y_i$ ($+$ being the multiset union);

\item $\vt~\gtp_{\lex(n)}\,\vu$ if there is $j\in\{1,\ldots,n\}$
  such that $t_j\gtp u_j$ and, for all $i<j$, $t_i=u_i$.
\end{itemize}

Note that both $\gtp_\mul$ and $\gtp_{\lex(n)}$ may compare all the
arguments whatever their types are (from left to right for
$\gtp_{\lex(n)}$). In \cite{blanqui15tcs}, the first author describes a
more general version of these statuses that take types into account
and allow reordering and filtering of the arguments
\cite{arts00tcs}. We could also consider statuses combining both
lexicographic and multiset comparisons~\cite{fernandez94adt}.

In the following, we will omit $n$ in $\gtp_{\lex(n)}$ and simply
write $\gtp_\lex$.

Here are the properties of statuses we will rely on:

\begin{prop}
Given a relation $\gtp$ on terms:
\begin{itemize}
\item\label{lem-stat-wf} $\gtp_\stat$ preserves termination: if $\gtp$
  is well-founded, then $\gtp_\stat$ is well-founded.
\item\label{lem-stat-prod} $\gtp_\stat$ contains $\gtp_\prod$.
\item\label{lem-stat-subs} $\gtp_\stat$ preserves stability: if $\gtp$ is
  stable by substitution away from $X$, then so is $\gtp_\stat$.
\end{itemize}
\end{prop}

\section{Computability path ordering}
\label{sec-core}

\newcommand\bigsub{($\cF_b$$\gts$)}
\newcommand\bigeq{($\cF_b$=)}
\newcommand\biggt{($\cF_b$>)}
\newcommand\bigapp{($\cF_b$@)}
\newcommand\biglam{($\cF_b$$\l$)}
\newcommand\bigvar{($\cF_b$$\cX$)}

\newcommand\appsub{(@$\gts$)}
\newcommand\appeq{(@=)}
\newcommand\applam{(@$\l$)}
\newcommand\appvar{(@$\cX$)}
\newcommand\appbeta{(@$\b$)}

\newcommand\lamsub{($\l$$\gts$)}
\newcommand\lameq{($\l$=)}
\newcommand\lamneq{($\l$$\neq$)}
\newcommand\lamvar{($\l\cX$)}
\newcommand\lameta{($\l\eta$)}

\newcommand\bigeqmul{($\cF_b$=$\mul$)}
\newcommand\bigeqlex{($\cF_b$=$\lex$)}

In this section, we give the core definition of the computability path
ordering (CPO) before to explore its limits by means of examples
and compare it with its father definition, HORPO.

\subsection{Definition of core CPO}

We assume given:
\begin{itemize}
\item an admissible ordering on types $>$;
\item a quasi-ordering $\gef$ on $\cF$, called \emph{precedence},
  whose equivalence ${\gef\cap\gef^{-1}}$ is written $\eqf$ and strict
  part ${\gef\setminus\gef^{-1}}$ is written $\gtf$ and assumed
  well-founded;
\item for every $\sf\in\cF$, a status $\stat(\sf)\in\{\mul\}\cup\{\lex(n)\mid n\ge 2\}$
  such that symbols equivalent in $\eqf$ have the same status.
\end{itemize}

\begin{defi}[Core computability path relation]
\label{def-cpr-core}
The core computability path relation is the relation $\gtpt[\vide]$
($\gtpt$ for short) where:
\begin{itemize}
\item the set $\cF_b$ of \emph{big} symbols is identical to $\cF$,\footnote{In core CPO, all symbols are big. {\em Small} symbols will show up in Section \ref{sec-small}.}
\item for any given finite set $X$ of variables, $\gtp[X]$ is
  inductively defined in Figure~\ref{fig-cpr-core},
\item $t\,\gtp[X]_\tau\,u$ if $t\,\gtp[X]\,u$ and $\tau(t)\get\tau(u)$,
\item $\gep[X]$ (resp. $\gep[X]_\tau$) is the reflexive closure of
  $\gtp[X]$ (resp. $\gtpt[X]$).
\end{itemize}
\end{defi}

\begin{figure}[ht]
\caption{Core CPO\label{fig-cpr-core}}
\fbox{\begin{tabular}{rl}
\bigsub & $\sf(\vt)\gtp[X] v$ if $\sf\in\cF_b$ and $(\ex i)~t_i\gept v$
\\\bigeq & $\sf(\vt)\gtp[X] \sg(\vu)$
if $\sf\in\cF_b$, $\sf\eqf\sg$, $(\all i)~\sf(\vt)\gtp[X] u_i$
and $\vt\,(\gtpt)_{\stat(\sf)}\,\vu$
\\\biggt & $\sf(\vt)\gtp[X] \sg(\vu)$
if $\sf\in\cF_b$, $\sf\gtf\sg$ and $(\all i)~\sf(\vt)\gtp[X] u_i$
\\\bigapp & $\sf(\vt)\gtp[X] uv$
if $\sf\in\cF_b$, $\sf(\vt)\gtp[X] u$ and $\sf(\vt)\gtp[X] v$
\\\biglam & $\sf(\vt)\gtp[X] \l yv$
if $\sf\in\cF_b$, $\sf(\vt)\gtp[X\cup\{z\}] v_y^z$,
$\ty=\tz$ and $z$ fresh
\\\bigvar & $\sf(\vt)\gtp[X] y$ if $\sf\in\cF_b$ and $y\in X$
\medskip

\\\appsub & $tu\gtp[X] v$ if $t\gep[X] v$ or $u\gept[X] v$
\\\appeq & $tu\gtp[X] t'u'$
if $t=t'$ and $u\gtp[X] u'$, or $tu\gtapp[X] t'$ and $tu\gtapp[X] u'$
\\& where $tu\gtapp[X] v$ if $t\gtpt[X] v$ or $u\gept[X] v$ or $tu\gtpt[X] v$
\\\applam & $tu\gtp[X] \l yv$ if $tu \gtp[X] v_y^z$ and $z$ fresh
\\\appvar & $tu\gtp[X] y$ if $y\in X$
\\\appbeta & $(\l xt)u\gtp[X] v$ if $t_x^u\gep[X] v$
\medskip

\\\lamsub & $\l xt\gtp[X] v$
if $t_x^z\gept[X] v$, $\tx=\tz$ and $z$ fresh
\\\lameq & $\l xt\gtp[X] \l yv$
if $t_x^z\gtp[X] v_y^z$, $\tx=\ty=\tz$ and $z$ fresh
\\\lamneq & $\l xt\gtp[X] \l yv$
if $\l xt\gtp[X] v_y^z$, $\tx\neq\ty$, $\ty=\tz$ and $z$ fresh
\\\lamvar & $\l xt\gtp[X] y$ if $y\in X$
\\\lameta & $\l x(tx)\gtp[X] v$ if $t\gep[X] v$ and $x\notin\FV(t)$
\end{tabular}}
\end{figure}

The parameter $X$ serves as a meta-level binder to keep track of the variables
that were previously bound in the righthand side but have become free
when destructuring a righthand side abstraction. We shall say that a
variable $x$ is {\em fresh} with respect to a comparison $u\gtp[X]v$ if
$x\not\in\FV(u)\cup X\cup\FV(v)$.

Note that the parameter $X$ is carried along computations without
change, except in rule \biglam. Hence, any comparison $u\gtp[X]v$
generated from an initial comparison $s\gtp[\emptyset]t$ implies
$X\cap\FV(u)=\emptyset$.

Explicit variable renamings and the associated freshness conditions
are used to make the relation invariant by $\al$-equivalence, the
smallest congruence generated by the equation $\l xt=\l yt_x^y$ if
$\tau(x)=\tau(y)$ and $y\notin\FV(\l xt)$ \cite{curry58book}, and by
appropriate renaming of the variables in $X$, as we shall prove
later.

Note the seemingly complex behaviour of application in rule (@=),
which allows to search the lefthand side for appropriate arguments
bigger than those of the righthand side. This enhancement of CPO
intends to mimic the corresponding rule of HORPO without flattening
lefthand sides.

Having function symbols equipped with an arity is more general than
having uncurried function symbols (\ie of null arity) only: any
uncurried system can be dealt with as it is. However, in this case,
the ($\cF_b$\_) rules are very limited: \bigsub\ is not applicable,
\bigeq\ and \biggt\ reduce to the precedence itself. Moreover,
applications of the form $\sf\,\vt$ with $|\vt|>0$ can only be
compared by using the (@\_) rules which are more constrained than the
corresponding ($\cF_b$\_) rules, especially \applam\ and \appeq.
Considering function symbols with
non-null arities provides more structure to the terms, and this
structure can be used for proving termination~\cite{hirokawa08lpar}.

Lemma~\ref{lem-fv-core} below will show that $\FV(v)\subseteq
\FV(u)\cup X$ whenever $u\gtp[X] v$. Hence, an alternative formulation
of rules \applam\ and \lamneq\ could therefore be given by replacing the
condition ``$z$ fresh'' by $y\notin\FV(v)$.

Another, perhaps surprising fact is that the definition of core CPO can
be simplified by replacing $\gtp[X]$ by $\gtp$ everywhere but in
\biglam. This is true at the start since we are interested
in $\gtpt$. This is then an invariant of the computation, for two
reasons: $X$ is never increased, except in \biglam; $X$ is reset to
the empty set by \bigsub\ and \bigeq, which are the only rules which
may move from a ($\cF_b$) comparison to a (@) or ($\l$) comparison.
We could therefore simplify our definition by removing the superfluous
$X$ subscripts. This will however no more be true of the extension of
core CPO to inductive types, and we prefer to have a uniform
definition over the various sections. Further, the present definition
will allow us to study a relaxation of \applam\ in the next section.

\smallskip
Surprisingly, core CPO is powerful enough already to prove termination of
examples that usually require techniques like the ones developed in
Section~\ref{sec-ind}.

\begin{exa}
\label{ex-cont}
Consider the breadth-first search of labeled trees using continuations
\cite{hofmann95note}, using the sorts $\sL$ for lists of labels and
$\sC$ for continuations,  the abbreviation $\neg T=T\a\sL$,
and the symbols $\sd:\sC$ and
$\sc^1:\neg\neg\sC\a\sC$ for
building continuations. Let now $\se:\neg\sC$ defined by the rule:

\begin{rewc}
\se\,\sc(x) & x\,\se\\[3mm]
\end{rewc}

Its termination can be checked by core CPO by taking $\sC\ge\sL$ and
$\sc\gtf\se$. Indeed, $\se\,\sc(x)\gtpt x\,\se$ holds by \appsub\ since
$\sc(x)\gtpt x\,\se$ for $\tau(\sc(x))=\sC\ge\tau(x\,\se)=\sL$ and, by
\bigapp, $\sc(x)\gtp x$ by \bigsub, and $\sc(x)\gtp\se$ by \biggt.
\end{exa}

\subsection{Tightness of core CPO}
\label{sec-limit}

In this section, we show that almost all possible relaxations (by
replacing $\gtpt$ by $\gtp$, and $\gtp$ by $\gtp[X]$) of the above
definition lead to non-termination by providing appropriate examples
that are also meant to help understanding how CPO works.  To this end,
we will consider three different systems, using $o:*$ to
declare a sort $o$:

\begin{enumerate}
\renewcommand\labelenumi{S\arabic{enumi}:}
\item\label{sa}
$o:*;~
\sa:o,\; \sf^1:o\a o,\; \sg:o\a o\a o ;~
\sa \gtf \sf \gtf \sg$.

\item\label{sb}
$o,o':*;~
\sa\!:\!o,\; \sf^1\!:\!o\a o,\; \sj^1\!\!:\!(o'\a o\a o)\a o ;~
\sa \gtf \sf \gtf \sj$.

\item\label{sc}
$o:*;~
\sa:o,\; \sh^2:o\a o\a o,\; \sk^2:(o\a o)\a o\a o;~ 
\sa \gtf \sh \eqf \sk;~
\stat(\sh)\!=\!\stat(\sk)\!=\!\mul$.
\end{enumerate}

\noindent
For each rule, we now consider all its natural relaxations.

\begin{itemize}
\item\bigsub\ $\sf(\vt)\gtp[X] v$ if $(\ex i)t_i\gept v$

\begin{itemize}
\item Replace $\gept$ by $\gept[X]$. Then, in S\ref{sa}, we have:
\begin{enumerate}
\item $\sf(\sa)\gtpt (\l x\sf(x))\sa$ since
  $\tau(\sf(\sa))=o\ge\tau((\l x\sf(x))\sa)=o$ and, by \bigapp:
\begin{enumerate}
\item $\sf(\sa)\gtp\l x\sf(x)$ since, by \biglam,
  $x\notin\FV(\sf(\sa))$ and
\begin{enumerate}
\item $\sf(\sa)\gtp[\{x\}]\sf(x)$ since, by relaxing \bigsub:
\begin{enumerate}
\item[] $\sa\gept[\{x\}]\sf(x)$ since $\tau(\sa)=o\ge\tau(\sf(x))=o$
  and, by \biggt, $\sa\gtp[\{x\}]x$ by \bigvar,
\end{enumerate}
\end{enumerate}
\item $\sf(\sa)\gtp\sa$ by \bigsub.
\end{enumerate}

\item $(\l x\sf(x))\sa\gtpt\sf(\sa)$ since $\tau((\l x\sf(x))\sa)=o
  \ge\tau(\sf(\sa))=o$ and by \appbeta.
\end{enumerate}

\vsp[2mm]
\item Replace $\gept$ by $\gep$. Then, in S\ref{sa}, we have:
\begin{enumerate}
\item $\sf(\sa)\gtpt (\l x\sf(x))\sa$ since
  $\tau(\sf(\sa))=o\ge\tau((\l x\sf(x))\sa)=o$ and, by \bigapp:
\begin{enumerate}
\item $\sf(\sa)\gtp\l x\sf(x)$ since, by relaxing \bigsub:
\begin{enumerate}
\item $\sa\gep\l x\sf(x)$ since, by \biglam, $x\notin\FV(\sa)$ and
\begin{enumerate}
\item[] $\sa\gtp[\{x\}]\sf(x)$ and, by \biggt, $\sa\gtp[\{x\}]x$ by \bigvar
\end{enumerate}
\end{enumerate}
\item $\sf(\sa)\gtp\sa$ by \bigsub.
\end{enumerate}

\item $(\l x\sf(x))\sa\gtpt\sf(\sa)$ since $\tau((\l x\sf(x))\sa)=o
  \ge\tau(\sf(\sa))=o$ and by \appbeta.
\end{enumerate}
\end{itemize}


\vsp[2mm]
\item\bigeq\ $\sf(\vt)\gtp[X] \sg(\vu)$
if $\sf\eqf\sg$, $\sf(\vt)\gtp[X] \vu$ and $\vt\,(\gtpt)_{\stat(\sf)}\,\vu$

\begin{itemize}
\item Replace $\gtpt$ by $\gtpt[X]$. Then, in S\ref{sa}, we have:
\begin{enumerate}
\item $\sf(\sa)\gtpt (\l x\sf(x))\sa$ since
  $\tau(\sf(\sa))=o\ge\tau((\l x\sf(x))\sa)=o$ and, by \bigapp:
\begin{enumerate}
\item $\sf(\sa)\gtp\l x\sf(x)$ since, by \biglam,
  $x\notin\FV(\sf(\sa))$ and:
\begin{enumerate}
\item $\sf(\sa)\gtp[\{x\}]\sf(x)$ since, by relaxing \bigeq:
\begin{enumerate}
\item[] $\sf(\sa)\gtp[\{x\}] x$ by \bigvar,
\item[] $\sa\gtpt[\{x\}] x$ since $\tau(\sa)=o\ge\tau(x)=o$ and by
  \bigvar,
\end{enumerate}
\end{enumerate}
\item $\sf(\sa)\gtp\sa$ by \bigsub,
\end{enumerate}

\item $(\l x\sf(x))\sa\gtpt\sf(\sa)$ since $\tau((\l x\sf(x))\sa)=o
  \ge\tau(\sf(\sa))=o$ and by \appbeta.
\end{enumerate}

\vsp[2mm]
\item Replace $\gtpt$ by $\gtp$. We found no counter-example for this
  case, but this is due to the condition $\sf(\vt)\gtp[X]\vu$. If we
  consider \bigeqmul\ and \bigeqlex\ instead, then simple
  counter-examples like the following one in S\ref{sc} come up.

\begin{enumerate}
\item $\sh(\sa,\sa)\gtpt \sk(\l x\sh(x,x))\sa$ since
  $\tau(\sh(\sa,\sa))=o\ge\tau(\sk(\l x\sh(x,x))\sa)=o$ and, by relaxing
  \bigeqmul, $\{\sa , \sa\}\,(\gtp)_\mul\,\{ \l x\sh(x,x) , \sa \}$,
  since $\sa \gtp \l x\sh(x,x) $, by case \biglam\ because $\sa
  \gtp[\{x\}] x$ by case \bigvar.

\item $\sk(\l x\sh(x,x))\sa \gtpt (\l x\sh(x,x))\sa$ since
  $\tau(\sk(\l x\sh(x,x))\sa)=o\ge\tau((\l x\sh(x,x))\sa)=o$ and by
  case \bigapp, since
\begin{enumerate}
\item $\sk(\l x\sh(x,x))\sa \gtp \l x\sh(x,x)$ by \bigsub.
\item $\sk(\l x\sh(x,x))\sa \gtp \sa$ by \bigsub.
\end{enumerate}

\item $(\l x\sh(x,x))\sa\gtpt \sh(\sa,\sa)$ since 
$\tau((\l x\sh(x,x))\sa)=o\ge\tau(\sh(\sa,\sa))=o$ and by \appbeta.

\end{enumerate}

Note that this counter-example can be also applied on case
\bigeqlex\ if we take $\stat(\sh)=\stat(\sk)=\lex$. Unfortunately it does
not work on \bigeq\ since we cannot prove $\sh(\sa,\sa)\gtp \l x\sh(x,x)$.

\end{itemize}

\vsp[2mm]
\item\appsub\ $tu\gtp[X] v$ if $t\gep[X] v$ or $u\gept[X] v$

\begin{itemize}
\item Replace $\gept[X]$ by $\gep[X]$. Then, in S\ref{sa}, we have:
\begin{enumerate}
\item $\sf(\sa)\gtpt \sg\sa\sa$ since
  $\tau(\sf(\sa))=o\ge\tau(\sg\sa\sa)=o$ and, by \bigapp:
\begin{enumerate}
\item $\sf(\sa)\gtp \sg\sa$ since, by \bigapp:
\begin{enumerate}
\item $\sf(\sa)\gtp \sg$ by \biggt,
\item\label{app-lam-2} $\sf(\sa)\gtp \sa$ by \bigsub,
\end{enumerate}
\item $\sf(\sa)\gtp \sa$ by \ref{app-lam-2},
\end{enumerate}

\item $\sg\sa\sa\gtpt (\l x\sf(x))\sa$ since
  $\tau(\sg\sa\sa)=o\ge\tau((\l x\sf(x))\sa)=o$ and, by \appeq:
\begin{enumerate}
\item $\sg\sa\gtpt \l x\sf(x)$ since $\tau(\sg\sa)=o\a o\ge
  \tau(\l x\sf(x))=o\a o$ and, by relaxing \appsub:
\begin{enumerate}
\item $\sa\gtp \l x\sf(x)$ since, by \biglam:
\begin{enumerate}
\item[] $\sa\gtp[\{x\}] \sf(x)$ since, by \biggt, $\sa\gtp[\{x\}] x$ by
  \bigvar,
\end{enumerate}
\end{enumerate}
\end{enumerate}

\item $(\l x\sf(x))\sa\gtpt \sf(\sa)$ since $\tau((\l x\sf(x))\sa)=o
  \ge\tau(\sf(\sa))=o$ and by \appbeta.
\end{enumerate}
\end{itemize}

\vsp[2mm]
\item\appeq\ $tu\gtp[X] t'u'$ if if $t=t'$ and $u\gtp[X] u'$, or
  $tu\gtapp[X] t'$ and $tu\gtapp[X] u'$,\\where $tu\gtapp[X] v$ if
  $t\gtpt[X] v$ or $u\gept[X] v$ or $tu\gtpt[X] v$
  $tu\,(\gtpt)_\mul\,t'u'$.

\begin{itemize}
\item Replace $tu\gtpt[X] t'$ by $tu\gtp[X] t'$. Then, taking $t:o\a
  o$, we get $tu\gtpt tu$ since, by relaxing \appeq, we have $tu\gtp
  t$ by \appsub.

\item Replace $tu\gtpt[X] t'$ by $tu\gtp[X] t'$. Then, taking $t:(o\a
  o)\a o$, we get $tu\gtpt tu$ since, by relaxing \appeq, we have
  $tu\gtp u$ by \appsub.

\item We found no counter-example yet for the other cases.
\end{itemize}
 
\vsp[2mm]
\item\applam\ $tu\gtp[X] \l yv$ if $tu\gtp[X] v_y^z$, $\ty=\tz$ and $z$ fresh

\begin{itemize}
\item Replace $\gtp[X]$ by $\gtp[X\cup\{z\}]$. Then, in S\ref{sa}, we
  have:
\begin{enumerate}
\item $\sf(\sa)\gtpt \sg\sa\sa$ since
  $\tau(\sf(\sa))=o\ge\tau(\sg\sa\sa)=o$ and, by \bigapp:
\begin{enumerate}
\item $\sf(\sa)\gtp \sg\sa$ since, by \bigapp:
\begin{enumerate}
\item $\sf(\sa)\gtp \sg$ by \biggt,
\item\label{app-lam-1} $\sf(\sa)\gtp \sa$ by \bigsub,
\end{enumerate}
\item $\sf(\sa)\gtp \sa$ by \ref{app-lam-1},
\end{enumerate}

\item $\sg\sa\sa\gtpt (\l x\sf(x))\sa$ since
  $\tau(\sg\sa\sa)=o\ge\tau((\l x\sf(x))\sa)=o$ and, by \appeq:
\begin{enumerate}
\item $\sg\sa\gtpt \l x\sf(x)$ since $\tau(\sg\sa)=o\a o\ge \tau(\l
  x\sf(x))=o\a o$ and, by relaxing \applam:
\begin{enumerate}
\item $\sg\sa\gtp[\{x\}] \sf(x)$ since, by \appsub:
\begin{enumerate}
\item[] $\sa\gtp[\{x\}] \sf(x)$ since, by \biggt, $\sa\gtp[\{x\}] x$ by
  \bigvar,
\end{enumerate}
\end{enumerate}
\end{enumerate}

\item $(\l x\sf(x))\sa\gtpt \sf(\sa)$ since $\tau((\l x\sf(x))\sa)=o
  \ge\tau(\sf(\sa))=o$ and by \appbeta.
\end{enumerate}
\end{itemize}

\vsp[2mm]
\item\lamsub\ $\l xt\gtp[X] v$ if $t_x^y\gept[X] v$, $\tx=\ty$ and $y$ fresh

\begin{itemize}
\item Replace $\gept[X]$ by $\gep[X]$. Then, in S\ref{sb}, we have:
\begin{enumerate}
\item $\sf(\sa)\gtpt \sj(\l x\l y\sa)$ since
  $\tau(\sf(\sa))=o\ge\tau(\sj(\l x\l y\sa))=o$ and, by \biggt:
\begin{enumerate}
\item $\sf(\sa)\gtp \l x\l y\sa$ since, by \biglam\ twice:
\begin{enumerate}
\item $\sf(\sa)\gtp[\{x,y\}]\sa$ by \bigsub,
\end{enumerate}
\end{enumerate}

\item $\sj(\l x\l y\sa)\gtpt (\l z\sf(z))\sa$ since
  $\tau(\sj(\l x\l y\sa))=o\ge\tau((\l z\sf(z))\sa)=o$ and, by \bigapp:
\begin{enumerate}
\item $\sj(\l x\l y\sa)\gtp \l z\sf(z)$ since, by \bigsub:
\begin{enumerate}
\item $\l x\l y\sa\gtpt \l z\sf(z)$ since $\tau(\l x\l
  y\sa)=o'\a o\a o\ge\tau(\l z\sf(z))=o\a o$ and, by relaxed \lamsub,
  $x\notin\FV(\l z\sf(z))$ and:
\begin{enumerate}
\item[] $\l y\sa\gtp\l z\sf(z)$ since, by relaxed \lamsub,
  $y\notin\FV(\l z\sf(z))$ and
 $\sa\gtp\l z\sf(z)$ since, by \biglam,
 $\sa\gtp[\{z\}]\sf(z)$ since, by \biggt,
$\sa\gtp[\{z\}] z$ by \bigvar,
\end{enumerate}
\end{enumerate}
\item $\sj(\l x\l y\sa)\gtp \sa$ since, by \bigsub:
\begin{enumerate}
\item $\l x\l y\sa\gtpt \sa$ since $\tau(\l x\l y\sa)=o'\a o\a
  o\ge\tau(\sa)=o$ and, by \lamsub, $x\notin\FV(\sa)$ and:
\begin{enumerate}
\item[] $\l y\sa\gtpt \sa$ since $\tau(\l y\sa)=o\a
  o\ge\tau(\sa)=o$ and, by \lamsub again.
\end{enumerate}
\end{enumerate}
\end{enumerate}

\item $(\l z\sf(z))\sa\gtpt \sf(\sa)$ since $\tau((\l z\sf
  z)\sa)=o\ge\tau(\sf(\sa))=o$ and, by \appbeta.
\end{enumerate}
\end{itemize}

\vsp[2mm]
\item\lamneq\ $\l xt\gtp[X] \l yv$
if $\l xt\gtp[X] v_y^z$, $\ty\neq\tz$ and $z$ fresh

\begin{itemize}
\item Replace $\gtp[X]$ by $\gtp[X\cup\{z\}]$. Then, in S\ref{sb}, we have:
\begin{enumerate}
\item $\sf(\sa)\gtpt \sj(\l x\l y\sa)$ since
  $\tau(\sf(\sa))=o\ge\tau(\sj(\l x\l y\sa))=o$, by \biggt:
\begin{enumerate}
\item $\sf(\sa)\gtp \l x\l y\sa$ since, by \biglam\ twice:
\begin{enumerate}
\item $\sf(\sa)\gtp \sa$ by \bigsub,
\end{enumerate}
\end{enumerate}

\item $\sj(\l x\l y\sa)\gtpt (\l z\sf(z))\sa$ since, by \bigapp:
\begin{enumerate}
\item $\sj(\l x\l y\sa)\gtp \l z\sf(z)$ since, by \bigsub:
\begin{enumerate}
\item $\l x\l y\sa\gtpt \l z\sf(z)$ since $\tau(\l x\l y\sa)=o'\a o\a
  o\ge\tau(\l z\sf(z))=o\a o$ and, by relaxing \lamneq:
\begin{enumerate}
\item[] $\l x\l y\sa\gtp[\{z\}] \sf(z)$ since, by \lamsub, $\l
  y\sa\gtpt[\{z\}] \sf(z)$ since $\tau(\l y\sa)=o\a
  o\ge\tau(\sf(z))=o$ and by \lamsub\ again, $\sa\gtpt[\{z\}] \sf(z)$
  since $\tau(\sa)=o\ge\tau(\sf(z))=o$ and, by \biggt, $\sa\gtp[\{z\}]
  z$ by \bigvar,
\end{enumerate}
\end{enumerate}
\item $\sj(\l x\l y\sa)\gtp \sa$ since, by \bigsub:
\begin{enumerate}
\item $\l x\l y\sa\gtpt \sa$ since $\tau(\l x\l y\sa)=o'\a o\a
  o\ge\tau(\sa)=o$ and, by \lamsub:
\begin{enumerate}
\item[] $\l y\sa\gtpt \sa$ since $\tau(\l y\sa)=o\a o\ge\tau(\sa)=o$
  and, by \lamsub\ again.
\end{enumerate}
\end{enumerate}
\end{enumerate}

\item $(\l z\sf(z))\sa\gtpt \sf(\sa)$ by \appbeta.
\end{enumerate}

\vsp[2mm]
\item Remove the condition $\tx\neq\ty$.

Then, in S\ref{sa}, we have $\tau(\l x\sa)\ge \tau(\l x\sb)$ and
$\l x\sa\gtpt \l x\sa$ by the relaxed \lamneq\ since $\l x\sa\gtp\sa$
by \lamsub.

\end{itemize}
\end{itemize}

\subsection{Transitivity}

As HORPO, core CPO is not transitive (both include $\b$-reduction
which is not transitive). Adding transitivity as a rule yields
non-termination as shown by the following counter-example:

\begin{exa}
  In the premises of \bigapp, replace $\gtp[X]$ by
  $(\gtp[X])^+$. Then, in the system S\ref{sa} described at the
  beginning of next section, we have:
\begin{enumerate}
\item $\sf(\sa)\gtpt (\l x\sf(x))\sa$ since
  $\tau(\sf(\sa))=o\ge\tau((\l x\sf(x))\sa)=o$ and, by relaxing \bigapp:
  \begin{enumerate}
  \item $\sf(\sa)\gtp[+]\l x\sf(x)$ since
    \begin{enumerate}
    \item $\sf(\sa)\gtp \sa$ by \bigsub, and
    \item $\sa\gtp \l x\sf(x)$ by \biglam\ since $\sa\gtp[\{x\}] \sf(x)$
      by \biggt\ and then \bigvar
    \end{enumerate}
  \item $\sf(\sa)\gtp \sa$ by \bigsub.
  \end{enumerate}
\item $(\l x\sf(x))\sa\gtpt\sf(\sa)$ since $\tau((\l x\sf(x))\sa)=o
  \ge\tau(\sf(\sa))=o$ and by \appbeta.
\end{enumerate}
\end{exa}

Similar counter-examples can be built as well for \biggt\ and \bigeq,
since the key point is that, by using $(\gtp[X])^+$, we can apply case
\bigsub\ without requiring type decreasingness.

Useful implemented heuristics for under-approximating $\gtpt[+]$ are
discussed in~\cite{jouannaud07jacm}. The introduction of small symbols
in Section \ref{sec-small} will reduce the need for such heuristics,
although not completely. On the other hand, we will show soon that
core CPO is a well-founded relation on terms. So is therefore its
transitive closure.

\subsection{Comparison with HORPO}
\label{sec-comp}

In \cite{jouannaud07jacm}, the last two authors define a relation on
simply-typed {\em polymorphic} $\l$-terms, $>_\horpo$, and its
extension $>_\chorpo$ using the notion of computability closure
introduced in~\cite{blanqui02tcs}. In this section, we explain the
differences between CPO and $>_\horpo$. We will compare CPO with
$>_\chorpo$ in Section \ref{sec-chorpo}.

\begin{itemize}
\item {\bf Type discipline.} $>_\horpo$ and $>_\chorpo$ are relations
  on {\em polymorphic} $\l$-terms, where types may
  contain type variables that have to be instantiated when forming
  function calls, while CPO is a relation on simply-typed {\em
    monomorphic} $\l$-terms. In the following, we will therefore
  compare CPO with the monomorphic versions of $>_\horpo$ and
  $>_\chorpo$. Extending CPO to polymorphic types along
  the lines of~\cite{jouannaud07jacm} is routine.

\item {\bf Relation on types.} In \cite{jouannaud07jacm}, the relation
  $\ge$ on types must be a quasi-ordering satisfying the following
  conditions\footnote{Condition (2) is actually stated there as an
    equivalence, but its converse follows
    from (4).}, where ${>}={\ge\setminus\ge^{-1}}$ is its strict part and
  ${\simeq}={\ge\cap\ge^{-1}}$ its associated equivalence relation:
\begin{enumerate}
\item $>$ is well-founded;
\item $T\a U\simeq V$ implies $V=T'\a U'$ with $T\simeq T'$ and
  $U\simeq U'$;
\item $T\a U>V$ implies $U\ge V$ or $V=T'\a U'$ with $T\simeq T'$
  and $U>U'$;
\item $T\ge T'$ implies $T\a U\ge T'\a U$ and $U\a T\ge U\a T'$.
\end{enumerate}

It turns out that these conditions are inconsistent: if $T>U$ then, by
(4), $T\a V>U\a V$ and, by (3), $V\ge U\a V$, which is impossible by
(1) \cite{kop08pc}. However, the results of \cite{jouannaud07jacm} are
still true since property (4) is only used to build the simplification
ordering $\gttwf$\footnote{Written $\ge^\a_{\cT_\cS}$ in Lemma 3.15 of
  \cite{jouannaud07jacm}.} used for defining the interpretation of
types. Instead, now, we distinguish between $>$ which must contain
$\gts_r$ and satisfy (3)/(typ-arrow) ((2) is always satisfied when
$\le$ is an ordering instead of a quasi-ordering), and $\gttwf$ which
must contain ${>}\cup{\gts_l}$ and be well-founded. The monotony
property (4) is not required anymore.

In \cite{jouannaud07jacm}, $>_\horpo$ and $>_\chorpo$ are proved
well-founded not only on well-typed terms but on a larger set of terms
called candidate terms, obtained by identifying equivalent types.
Since, by (2), the arrow structure of equivalent types is invariant,
the quotient of the set of types by $\simeq$ can be obtained by simply
identifying sorts equivalent in $\simeq$, and the quasi-order becomes
then an order in the quotient structure. Since rewriting on candidate
terms coincides with rewriting in the quotient, an order on types suffices,
which removes the need for candidate terms and their technicalities.

\item {\bf Relation on terms.}
One important difference between HORPO and CPO is that, in all
sub-derivations of $>_\horpo$, types must decrease ($t>_\horpo u$ only
if $\tau(t)\ge\tau(u)$) while, in CPO, this is not the case: types
must be checked only in case the recursive call takes a subterm of the
lefthand side term (except in \appsub\ for the left argument of an application). Indeed, CPO is an
optimized version of $>_\horpo$ in this respect. 

$>_\horpo$ is defined by a set of 12 rules and each rule but (9) is
implied by a rule of CPO: (1) is implied by \bigsub, (2) by \biggt,
(3) by \bigeq\ with $\stat(\sf)=\mul$, (4) by \bigeq\ with
$\stat(\sf)=\lex$, (5) by \appsub, (6) by \lamsub, (7) by \bigapp, (8)
by \biglam\ (HORPO requires the strong condition $x\notin\FV(v)$ since
it does not manage bound variables; this is however done by the
computability closure in CHORPO), (10) by \lameq, (11) by
\appbeta\ and (12) by \lameta.

Rule (9) compares $st$ and $u_1\ldots u_n$ with $n\ge 2$ by comparing
the multisets $\mset{s,t}$ and $\mset{u_1,\ldots,u_n}$. It is implied
by \appeq. Indeed, in this case, for all $i$, either $s\gept u_i$ or
$t\gept u_i$. If there is no $i$ such that $s=u_i$ then, for all $i$,
$s\gtpt u_i$ or $t\gept u_i$, in which case one can prove that
$st\gtpt u_1\ldots u_k$ by induction on $k$. If $s=u_1$ then, for all
$i\ge 2$, $t\gtpt u_i$, in which case one can also prove that $st\gtpt
u_1\ldots u_k$ by induction on $k$. Otherwise, there is $i>1$ such
that $s=u_i$ and $t\gtpt u_1$. But, then,
$\tau(s)\gttwf\tau(t)\ge\tau(u_1)\gttwf\tau(s)$, which is not
possible by \link{typ-sn}.

\hide{
\begin{exa}
Let $o:*$, and $\sg:o\a o\a o$, $\sa:o$ and $\sb:o$, with
$\sg\gtf\sa\gtf\sb$. Then, $\sg\sa>_\horpo \sg\sb\sb$ since
$(\sg,\sa)\,(>_\horpo)_\mul\,(\sg,\sb,\sb)$, but
$\sg\sa\not{\gtpt\!\!}~\sg\sb\sb$ since we do not have
$(\sg,\sa)\,(\gtpt)_\mul\,(\sg\sb,\sb)$ for neither $\sg$ nor
$\sa$ is bigger than $\sg\sb$.
\end{exa}
}

On the other hand, the CPO rules \lamneq, \applam, \bigvar, \appvar,
\lamvar\ have no counterpart in HORPO. Therefore, HORPO is strictly
included in CPO.

\end{itemize}

\subsection{Implementation}
\label{sec-impl}

All examples given in the paper have been checked by our
implementation, which is available from the web at
\url{http://www.lsi.upc.edu/\~{}albert/cpo.zip}. In this implementation
the precedence and the status should be provided by the user. The
implemented prototype includes core CPO as well as the extended
versions of the ordering defined in Section~\ref{sec-ind}
and~\ref{sec-small}. Several more examples are also included together
with the implementation showing the power of the developed
orderings. However, like RPO, CPO cannot be compared with
transformation techniques based on, for instance, the computation of
dependency pairs~\cite{arts00tcs,giesl06jar}, but its power shows that
it should be the path ordering of choice for solving the (monotonic)
ordering comparisons which are generated by these transformation
techniques.

Given a precedence and a status for every function symbol, deciding if
a term $s$ is smaller than a term $t$ in core CPO can be made in
quadratic time (using a dynamic programming algorithm) if \appbeta\ is
not used. The proof is basically the same as for
RPO~\cite{krishnamoorthy85tcs}. Our prototype implementation written
in Prolog does not use dynamic programming. Still, some standard
optimizations over the given presentation are made, which mainly
affect case \bigeq. Let us split this case in two new cases, one for
multiset status \bigeqmul\ and one for lexicographic status \bigeqlex,
and show that even after removing all or part of the condition
$\sf(\vt)\gtp[X]\vu$, the conjunction of
both cases is equivalent to the original one.\\

\noindent
\begin{tabular}{rl}
\bigeqmul & $\sf(\vt)\gtp[X] \sg(\vu)$
if $\sf\eqf\sg$, $\stat(\sf)=\mul$ and $\vt\,(\gtpt)_\mul\,\vu$\\
\bigeqlex & $\sf(\vt)\gtp[X] \sg(\vu)$
if $\sf\eqf\sg$, $\stat(\sf)=\lex$ and:\\
&  $(\ex i)~t_i\gtpt u_i\,\et\,(\all j<i)\,t_j=u_j\,\et\,(\all j>i)\,\sf(\vt)\gtp[X]u_j$\\
\end{tabular}\medskip

\noindent In case \bigeqmul, $\vt\,(\gtpt)_\mul\,\vu$ implies that,
for every $u_j$, there is $t_i$ such that $t_i\gept u_j$, which
implies that $\sf(\vt)\gtp[X]u_j$ by \bigsub. Similarly, in case
\bigeqlex, there is $i$ such that $t_i\gtpt u_i$,
$(\all j<i)\,t_j=u_j$ and $(\all{}j>i)\,\sf(\vt)\gtp[X]u_j$, which
implies $(\all{}j)\,\sf(\vt)\gtp[X]u_j$ by \bigsub. Therefore, we have
$\sf(\vt)\gtp[X]\vu$ and hence case \bigeq\ can be applied as well.

As said, our implementation assumes that the precedence on function
symbols and the status is given. Generating the precedence and the
status automatically is a harder problem, and closely relates to the
decision problem of solving ordering constraints, which is already
NP-complete for RPO~\cite{nieuwenhuis93ipl,narendran98csl}, but
which is nowadays efficiently done in practice by encoding the problem
into SAT~\cite{codish11jar}. These kind of encodings can
be easily adapted to CPO, as done for HORPO in termination tools like
WANDA~\cite{kop12phd} and THOR~\cite{thor}.

\section{Well-foundedness of core CPO}
\label{sec-prf}

We now move to a technical analysis of the most important
properties of core CPO.

\subsection{Basic properties of core CPO}

\begin{lem}
\label{lem-well-def-core}
$\gtp[X]$ is well-defined.
\end{lem}

\begin{prf}
  $a\gtp[X] b$ is well-defined by induction on the pair $(a,b)$ with
  $({=_\al\gts}\cup{\ab}\cup{\ae},=_\al\gts)_\lex$ as well-founded
  relation, where $\lts$ is the subterm relation.\qed
\end{prf}

\begin{defi}[Monotony]
\label{def-mon}
We say that $\gtpt$ is {\em monotone} if the following properties hold:
\begin{enumerate}
\item if $\sf^{|\vT|}:\vT\a U$, $\vt:\vT$, $\vt':\vT$ and
  $\vt\,(\gtpt)_{\prod}\,\vt'$, then $\sf(\vt)\gtpt\sf(\vt')$;
\item if $t:U\a V$, $t':U\a V'$, $t\gtpt t'$, $u:U$ and $V\get V'$,
  then $tu\gtpt t'u$;
\item if $t:U\a V$, $u,u':U$ and $u\gtpt u'$, then $tu\gtpt tu'$;
\item if $t:T$, $t':T'$, $t\gtpt t'$ and $\tx\a T\get\tx\a T'$, then
  $\l xt\gtpt \l xt'$.
\end{enumerate}
\end{defi}

\begin{lem}
\label{lem-mon-core}
$\gtpt$ is monotone.
\end{lem}

\begin{prf}
\hfill
\begin{enumerate}
\item Since $\tau(\sf(\vt))=\tau(\sf(\vt'))$, it suffices to check that
  $\sf(\vt)\gtp\sf(\vt')$. By Lemma \ref{lem-stat-prod},
  $\vt\,{(\gtpt)_{\stat(\sf)}}\,\vt'$. By \bigsub, $\sf(\vt)\gtp\vt'$
  since, for each $i$, $t_i\gept t_i'$. We conclude by \bigeq.

\item Since $\tau(tu)\ge\tau(t'u)$, it suffices to check that
  $tu\gtp t'u$. This follows by by \appeq.

\item Since $\tau(tu)=\tau(tu')$, it suffices to check that
  $tu\gtp tu'$. This follows by \appeq.

\item By \lameq.\qed
\end{enumerate}
\end{prf}

\noindent
Note that Lemma \ref{lem-mon-core} holds for any relation satisfying
\bigsub, \bigeq, \appeq\ and \lameq.

\begin{lem}
\label{lem-fv-core}
If $a\gtp[X] b$, then $\FV(b)\sle\FV(a)\cup X$.
\end{lem}

\begin{prf}
By an easy induction on $a\gtp[X] b$. We detail a selection of cases:

\begin{itemize}
\item\biglam\ By the induction hypothesis,
  $\FV(v_y^z)\sle\FV(\sf(\vt))\cup X\cup\{y\}$. Now, 
$\FV(\l yv)=\FV(v_y^z)\setminus\{z\}$ since $z$ is fresh.
The result follows.
\item\lamsub\ By the induction hypothesis, $\FV(v)\sle\FV(t_x^z)\cup
  X$. Therefore, $\FV(v)\sle\FV(\l xt)\cup X$ since
  $\FV(t_x^z)\sle\FV(\l xt)\cup\{z\}$ and $z\notin\FV(v)$.
\item\lameq\ By the induction hypothesis,
  $\FV(v_y^z)\sle\FV(t_x^z)\cup X$. Now, $\FV(t_x^z)\sle\FV(\l
  xt)\cup\{z\}$ and, either $y\in\FV(v)$ and $\FV(v_y^z)=\FV(\l
  yv)\cup\{z\}$, or $\FV(v_y^z)=\FV(\l yv)$. Therefore, $\FV(\l
  yv)\sle\FV(\l xt)\cup X$ since $z\notin\FV(\l yv)$.
\item\lamneq\ By the induction hypothesis, $\FV(v_y^z)\sle\FV(\l
  xt)\cup X$. Since $z$ is fresh for $\l xt$, $X$ and $\l yv$,
  $y\notin\FV(v)$. Therefore, $\FV(\l yv)=\FV(v)-\{y\}\sle\FV(\l
  xt)\cup X$.\qed
\end{itemize}
\end{prf}

\begin{lem}
\label{lem-alpha-core}
If $a\gtp[X] b$, $a=_\al a'$ and $b=_\al b'$, then $a'\gtp[X] b'$.
\end{lem}

\begin{prf}
We prove
(i) $a\gtp[X] b$ and $a=_\al a'$ implies $a'\gtp[X] b$, and
(ii) $a\gtp[X] b$ and $b=_\al b'$ implies $a\gtp[X] b'$, separately
by induction on $a\gtp[X] b$. We only detail some cases:

\begin{itemize}
\item\biglam\ (ii) Assume that $\l yv=_\al b'$. Then, there are $y'$ and
  $v'$ such that $b'=\l y'v'$, $y\notin\FV(\l y'v')$ and $v=_\al
  {v'}_{y'}^y$. Hence, $v_y^z=_\al {{v'}_{y'}^y}_y^z=_\al {v'}_{y'}^z$
  and, by the induction hypothesis, $\sf(\vt)\gtp[X\cup\{z\}]
  {v'}_{y'}^z$. Now, $z\notin\FV(\l y'v')\cup\FV(\sf(\vt))$ since
  $\FV(\l y'v')=\FV(\l yv)$ and $z\notin\FV(\l
  yv)\cup\FV(\sf(\vt))$. Therefore, by \biglam, $\sf(\vt)\gtp[X]\l y'v'$.

\item\lamsub\ (ii) Assume that $v=_\al v'$. By the induction hypothesis,
  $t_x^z\gept[X]v'$. Now, $z$ is fresh for $\l xt$, $X$ and $v'$ since
  $\FV(v')=\FV(v)$ and $z$ is fresh for $\l xt$, $X$ and
  $v$. Therefore, by \lamsub, $\l xt\gtp[X]v'$.

  (i) Assume now that $\l xt=_\al a'$. Then, there are $x'$ and $t'$ such
  that $a'=\l x't'$, $x\notin\FV(\l x't')$ and $t=_\al
  {t'}_{x'}^x$. Hence, $t_x^z=_\al {{t'}_{x'}^x}_x^z=_\al {t'}_{x'}^z$
  and, by the induction hypothesis, ${t'}_{x'}^z\gept[X] v$. Now, $z$
  is fresh for $\l x't'$, $X$ and $v$ since $\FV(\l x't')=\FV(\l xt)$
  and $z$ is fresh for $\l xt$, $X$ and $v$. Therefore, by \lamsub,
  $\l x't'\gtp[X] v$.

\item\lameq\ (ii) Assume that $\l yv=_\al b'$. Then, there are $y'$ and
  $v'$ such that $b'=\l y'v'$, $y\notin\FV(\l y'v')$ and $v=_\al
  {v'}_{y'}^y$. Hence, $v_y^z=_\al {{v'}_{y'}^z}_y^z=_\al {v'}_{y'}^z$
  and, by the induction hypothesis, $t_x^z\gtp[X] {v'}_{y'}^z$. Now,
  $z$ is fresh for $\l xt$, $X$ and $\l y'v'$ since $\FV(\l
  y'v')=\FV(\l yv)$ and $z$ is fresh for $\l xt$, $X$ and $\l
  yv$. Therefore, by \lameq, $\l xt\gtp[X]\l y'v'$.

\item\lamneq\ (ii) Assume that $\l yv=_\al b'$. Then, there are $y'$
  and $v'$ such that $b'=\l y'v'$, $y'\notin\FV(\l yv)$ and
  $v_y^{y'}=_\al v'$. By Lemma \ref{lem-fv-core},
  $y\notin\FV(v)$. Hence, $y'\notin\FV(v')$, $v_y^z=_\al {v'}_{y'}^z$
  and, by the induction hypothesis, $\l xt\gtp[X]
  {v'}_{y'}^z$. Moreover, $z$ is fresh for $\l xt$, $X$ and $\l
  y'v'$. Therefore, by \lamneq, $\l xt\gtp[X] \l y'v'$.\qed
\end{itemize}
\end{prf}

\noindent Hence, if $t\gtp[X]u$ and $V$ is a finite set of variables, then one
can always assume without lost of generality that the bounding
variables of $t$ and $u$ do not belong to $V$.

  Invariance by variable
renaming can also be extended to $X$: 

\begin{lem}
Assume that $t\gtp[X]u$.
\begin{enumerate}
\item\label{lem-subs-core} If $\s$ is away from $X$, then
  $t\s\gtp[X]u\s$.
\item\label{lem-alpha-X-core} If $e\in X$, $e'\notin\FV(\l eu)$ and
  $\tau(e)=\tau(e')$, then $t\gtp[X-\{e\}\cup\{e'\}]u_e^{e'}$.
\end{enumerate}
\end{lem}

\begin{prf}
  Note that substitution preserves typing ($\tau(t\s)=\tau(t)$). Let
  $X_e^{e'}=X-\{e\}\cup\{e'\}$. Wlog we can assume that $e\neq
  e'$. Hence, $e'\notin\FV(u)$. We now proceed by induction on the
  deduction height of $t\gtp[X]u$. We only detail some cases:

\begin{itemize}[label=(1) \bigeq]
\item[(1) \bigeq] By induction hypothesis and Lemma \ref{lem-stat-subs},
  $(\all i)~\sf(\vt)\s\gtp[X]u_i\s$ and
  $\vt\s\,{(\gtpt)_{\stat(\sf)}}$ $\vu\s$. Therefore, by \bigeq,
  $\sf(\vt)\s\gtp[X]\sg(\vu)\s$.

\item[\biglam] Wlog we can assume $\s$ away from $\{y\}$. Hence, $(\l
  yv)\s=\l y(v\s)$. Let now $z'$ be a variable of the same type as
  $z$, fresh for $\sf(\vt)\s$, $X$, $\l y(v\s)$ and $\l zv_y^z=_\al\l
  yv$, and such that $\s$ is away from $\{z'\}$. By induction
  hypothesis (2), $\sf(\vt)>^{(X\cup\{z\})_z^{z'}}(v_y^z)_z^{z'}$. Since
  $z\notin X$, $(X\cup\{z\})_z^{z'}=X\cup\{z'\}$. Since $(_z^{z'})$ is
  away from $\{y\}$ and $z\notin\FV(v)$, $(v_y^z)_z^{z'}=v_y^{z'}$. By
  induction hypothesis, $\sf(\vt)\s>^{X\cup\{z'\}}(v_y^{z'})\s$. Since
  $\s$ is away from $\{y,z'\}$,
  $(v_y^{z'})\s=(v\s)_y^{z'}$. Therefore, by \biglam,
  $\sf(\vt)\s\gtp[X]\l y(v\s)$.

\item[\lamsub] Wlog we can assume that $\s$ is away from
  $\{x\}$. Hence, $(\l xt)\s=\l x(t\s)$. After Lemma \ref{lem-fv-core}
  and since $\s$ is away from $X$, $\FV(v)\sle\FV(\l xt)$. Hence, we
  can also assume wlog that $\dom(\s)\sle\FV(\l xt)$. Let now $z'$ be a
  variable of the same type as $z$, fresh for $\l x(t\s)$, $X$, $v\s$
  and $x$, and such that $\s$ is away from $\{z'\}$. Let
  $\s'=\s\cup\{(z,z')\}$. Since $\s'$ is away from $X$, by induction
  hypothesis, $(t_x^z)\s'\gtp[X]v\s'$. Since $\dom(\s)\sle\FV(\l xt)$
  and $\s$ is away from $\{x,z'\}$,
  $(t_x^z)\s'=(t\s)_x^{z'}$, and since $z\notin\FV(v)$, $v\s'=v\s$. 
Therefore, by \lameq, $\l
  x(t\s)\gtp[X]v\s$.
 
\item[\lameq] Wlog we can assume that $\s$ is away from
  $\{x,y\}$. Hence, $(\l xt)\s=\l x(t\s)$ and $(\l yv)\s=\l
  y(v\s)$. After Lemma \ref{lem-fv-core} and since $\s$ is away from
  $X$, $\FV(\l yv)\sle\FV(\l xt)$. Hence, we can also assume wlog that
  $\dom(\s)\sle\FV(\l xt)$. Let now $z'$ be a variable of the same
  type as $z$, fresh for $\l x(t\s)$, $X$, $\l y(v\s)$, $x$ and $y$,
  and such that $\s$ is away from $\{z'\}$. Let
  $\s'=\s\cup\{(z,z')\}$. Since $\s'$ is away from $X$, by induction
  hypothesis, $(t_x^z)\s'\gtp[X](v_y^z)\s'$. Since $\dom(s)\sle\FV(\l
  xt)$, $\dom(\s)\sle\FV(\l yv)$, and $\s$ is away from $\{x,y,z'\}$,
  $(t_x^z)\s'=(t\s)_x^{z'}$ and $(v_y^z)\s'=(v\s)_y^{z'}$. Therefore,
  by \lameq, $(\l xt)\s\gtp[X](\l yv)\s$.

\item[\lamneq] Wlog we can assume $\s$ away from $\{x,y\}$. Hence,
  $(\l xt)\s=\l x(t\s)$ and $(\l yv)\s=\l y(v\s)$. Let $z'$ fresh for
  $\l xt\s$, $X$ and $\l yv\s$. By Lemma \ref{lem-fv-core},
  $y\notin\FV(v)$, hence $v_y^z=v_y^{z'}$. By induction hypothesis,
  $(\l xt)\s\gtp[X]v_y^{z'}\s$. Therefore, by \lamneq, $\l
  x(t\s)\gtp[X]\l y(v\s)$.
\end{itemize}

\begin{itemize}[label=(1) \bigeq]
\item[(2) \biglam] Wlog we can assume $(_e^{e'})$ away from
  $\{y\}$. Hence, $(\l yv)_e^{e'}=\l y(v_e^{e'})$. Let now $z'$ be a
  variable of the same type as $z$, fresh for $\sf(\vt)$, $X_e^{e'}$,
  $\l y(v_e^{e'})$, $\l zv_y^z=_\al\l yv$ and $y$. By induction
  hypothesis, $\sf(\vt)\gtp[(X\cup\{z\})_z^{z'}](v_y^z)_z^{z'}$. Since
  $z\notin X$, $(X\cup\{z\})_z^{z'}=X\cup\{z'\}$. Since
  $z\notin\FV(\l yv)$, $(v_y^z)_z^{z'}=v_y^{z'}$. By induction hypothesis,
  $\sf(\vt)\gtp[(X\cup\{z'\})_e^{e'}](v_y^{z'})_e^{e'}$. Since
  $(_e^{e'})$ is away from $\{y,z'\}$,
  $(X\cup\{z'\})_e^{e'}=X_e^{e'}\cup\{z'\}$ and
  $(v_y^{z'})_e^{e'}=(v_e^{e'})_y^{z'}$. Therefore, by \biglam,
  $\sf(\vt)\gtp[X_e^{e'}]\l yv_e^{e'}$.

\item[\lamsub] Let $z'$ be a variable of the same type as $z$, fresh
  for $\l xt$, $X$, $X_e^{e'}$, $v_e^{e'}$ and $x$. Since $(_z^{z'})$
  is away from $X$, by induction hypothesis (1),
  $(t_x^z)_z^{z'}\gept[X]v_z^{z'}$. Since $z'\neq x$ and
  $z\notin\FV(t)$, $(t_x^z)_z^{z'}=t_x^{z'}$. Since $z\notin\FV(v)$,
  $v_z^{z'}=v$. So, by induction hypothesis,
  $t_x^{z'}\gept[X_e^{e'}]v_e^{e'}$. Therefore, by \lamsub, $\l
  xt\gtp[X_e^{e'}]v_e^{e'}$.

\item[\lameq] Wlog we can assume $(_e^{e'})$ away from $\{y\}$. Hence,
  $(\l yv)_e^{e'}=\l y(v_e^{e'})$. Let now $z'$ be a variable of the
  same type as $z$, fresh for $\l xt$, $X$, $X_e^{e'}$, $\l
  yv_e^{e'}$, $x$ and $y$. Since $(_z^{z'})$ is away from $X$, by
  induction hypothesis (1), $(t_x^z)_z^{z'}\gtp[X](v_y^z)_z^{z'}$. Since
  $z'\neq x$ and $z\notin\FV(t)$, $(t_x^z)_z^{z'}=t_x^{z'}$. Since
  $z'\neq y$ and $z\notin\FV(v)$, $(v_y^z)_z^{z'}=v_y^{z'}$. So, by
  induction hypothesis,
  $t_x^{z'}\gtp[X_e^{e'}](v_y^{z'})_e^{e'}$. Since $(_e^{e'})$ is away
  from $\{y,z'\}$, $(v_y^{z'})_e^{e'}=(v_e^{e'})_y^{z'}$. Therefore,
  by \lameq, $\l xt\gtp[X_e^{e'}]\l y(v_e^{e'})$.

\item[\lamneq] Wlog we can assume $(_e^{e'})$ away from
  $\{y\}$. Hence, $(\l yv)_e^{e'}=\l y(v_e^{e'})$. Let $z'$ fresh for
  $\l xt$, $X_e^{e'}$ and $\l y(v_e^{e'})$. By Lemma \ref{lem-fv-core},
  $y\notin\FV(v)$, hence $v_y^z=v_y^{z'}$. By the induction
  hypothesis, $\l
  xt\gtp[X_e^{e'}](v_y^{z'})_e^{e'}=(v_e^{e'})_y^{z'}$. Therefore, by
  \lamneq, $\l xt\gtp[X_e^{e'}]\l y(v_e^{e'})$.\qed
\end{itemize}
\end{prf}

\subsection{Tait and Girard's computability}

We now turn to the proof that $\gtpt$ is well-founded. This proof is
based on the meticulous analysis of the technique of computability
predicates of Tait and Girard for proving the termination of
$\b$-reduction in typed $\l$-calculi
\cite{tait67jsl,girard72phd,tait72lc,girard88book}. This technique
consists in the following three steps:

\begin{enumerate}
\item interpret each type $T$ by a set $\I{T}$ of so-called computable
  terms;
\item prove that, for every type $T$, $\I{T}$ satisfies some
  properties among which termination, \ie
  $\I{T}\sle\SN(\gtpt)$;
\item prove that every (well typed) term is computable.
\end{enumerate}

For arrow types, we will use the standard interpretation but, for
sorts, we {\em a priori} have some freedom and we will indeed use this
freedom to extend CPO to inductive types later in Section \ref{sec-ind}.

\begin{defi}[Computability]
A base type interpretation is a map $I:\cS\a\cP(\cL)$ such that, for
all sorts $\sA$, $I(\sA)$ is a set of terms of type $\sA$. A base type
interpretation naturally extends to types as follows:
\begin{itemize}
\item $\I\sA_I=I(\sA)$
\item $\I{U\a V}_I=\{t\in\cL\mid
t:U\a V\et(\all u)u\in\I{U}_I\A tu\in\I{V}_I\}$
\end{itemize}
Given a base type interpretation $I$, we say that:
\begin{itemize}
\item a term $t:T$ is {\em $I$-computable} if $t\in\I{T}_I$;
\item a substitution $\s$ is $I$-computable on a set $X$ of variables
  if, for all $x\in X$, $x\s$ is $I$-computable; it is $I$-computable
  if it is $I$-computable on $\cX$;
\item a function symbol $\sf^{|\vT|}:\vT\A U$ is $I$-computable if,
  for all $I$-computable terms $\vt:\vT$, $\sf(\vt)$ is
  $I$-computable.
\end{itemize}
\noindent
Let $\S_I$ be the set of pairs $(\sf,\vt)$ such that $\sf\in\cF$,
$\sf(\vt)$ is a term and $\vt$ are $I$-computable. Given a relation on
terms $R$, let $(\gtf,R_\stat)_\lex$ be the relation on $\S_I$ such
that $(\sf,\vt)\,(\gtf,R_\stat)_\lex\,(\sg,\vu)$ if either
$\sf\gtf\sg$, or $\sf\eqf\sg$ and $\vt\,R_{\stat(\sf)}\,\vu$.
\end{defi}

Our first lemma has a straightforward proof:

\begin{lem}
\label{lem-bints-agree}
Let $I_1$ and $I_2$ be two base type interpretations, and $T$ be a
type. Then, $\I{T}_{I_1}=\I{T}_{I_2}$ if $I_1$ and $I_2$ agree on
every sort occurring in $T$.
\end{lem}

We then consider the following properties:

\begin{defi}[Sets of neutral terms]
\label{def-valid-neutral}
\renewcommand\ind[1]{{\bf(#1)}\index{(#1)|inddef[def-valid-neutral]}}
A set $\cN$ is a set of {\em neutral} terms if it satisfies the
following properties:
\begin{itemize}
\item\target{neutral-var} $\cX\sle\cN$
\hfill\ind{neutral-var}
\item\target{neutral-beta} for all $x,t,u$, $(\l xt)u\in\cN$
\hfill\ind{neutral-beta}
\item\target{neutral-app} if $t\in\cN$ then, for all $u$, $tu\in\cN$
\hfill\ind{neutral-app}
\item\target{neutral-not-lam} for all $x$ and $t$, $\l xt\notin\cN$
\hfill\ind{neutral-not-lam}
\end{itemize}
\end{defi}

\begin{defi}[Computability properties]
\label{def-comp-props}
Given a base type interpretation $I$ and a set $\cN$ of neutral terms,
a set $S$ of terms of type $T$ is an {\em $I$-computability predicate}
if it satisfies the following properties:
\begin{itemize}
\item $S\sle\SN(\gtpt)$  \hfill{\bf(comp-sn)}
\item If $t\in S$, then every $\gtpt$-reduct of
  $t$ is $I$-computable \hfill{\bf(comp-red)}
\item $t\in S$ if $t:T$, $t\in\cN$ and every $\gtpt$-reduct of $t$ is
  $I$-computable \hfill{\bf(comp-neutral)}
\item $\l xt\in S$ if $T=U\a V$, $\l xt:T$ and, for all $u\in\I{U}_I$,
  $t_x^u$ is $I$-computable \hfill{\bf(comp-lam)}
\end{itemize}
\end{defi}

We can then prove that every term is strongly normalizing if the sets
$\I{T}_I$ satisfy some of these properties and function symbols are
$I$-computable, whatever the base type interpretation $I$ and the set
$\cN$ of neutral terms are:

\begin{thm}
\label{thm-wf}
Given a base type interpretation $I$ and a set $\cN$ of neutral terms,
$\gtpt$ is well-founded if:
\begin{itemize}
\item for every type $T$, $\I{T}_I$ satisfies {\em(comp-sn)},
  {\em(comp-neutral)} and {\em(comp-lam)};
\item every function symbol $\sf\in\cF$ is $I$-computable.
\end{itemize}
\end{thm}

\begin{prf}
Because, for every $T$, $\I{T}_I$ satisfies {\em(comp-sn)}, it
suffices to prove that every term is $I$-computable. By
(neutral-var), variables belongs to $\cN$. Because, for every
$T$, $\I{T}_I$ satisfies {\em(comp-neutral)}, variables are
computable. Hence, the identity substitution is computable. We then
prove that, for all $t:T$ and computable $\s$, $t\s\in\I{T}_I$, by
induction on $t$.
\begin{itemize}
\item $t=x$. Then, $t\s=x\s$ is computable since $\s$ is computable.
\item $t=uv$. By induction hypothesis, $u\s$ and $v\s$ are
  computable. Therefore, $t\s=(u\s)(v\s)$ is computable.
\item $t=\l xu$ with $x:V$. By renaming $x$, we can assume that $\s$
  is away from $\{x\}$. Hence, $t\s=\l x(u\s)$. By assumption, $\I{T}_I$
  satisfies (comp-lam). Therefore, $t\s$ is computable if, for all
  computable $v:V$, $(u\s)_x^v$ is computable. Since $\s$ is away from
  $\{x\}$, $(u\s)_x^v=u(\s\cup\{(x,v)\})$ which is computable by
  induction hypothesis.
\item $t=\sf(\vt)$ with $\sf^{|\vT|}:\vT\a U$. By induction
  hypothesis, $\vt\s$ are computable. Thus, $(\sf,\vt\s)\in\S$ and, by
  assumption, $\sf(\vt)\s$ is computable.\qed
\end{itemize}
\end{prf}

\noindent We are therefore left to find a set of neutral terms $\cN$ and a base
type interpretation $I$ so that type interpretations are computability
predicates and function symbols are computable.

First, we are going to study under which conditions the interpretation
of an arrow type $U\a V$, $\I{U\a V}_I$, satisfies the above
computability properties, whatever the base type interpretation $I$
and the set of neutral terms $\cN$ are.

Second, we will define a set of neutral terms $\cN$ and a base type
interpretation $I$ so that, for every type $T$, $\I{T}_I$ satisfies
all the computability properties. Finally, we will prove that function
symbols are computable by induction on $(\gtf,(\gtpt)_\stat)_\lex$,
which is well-founded when the following conditions are satisfied:

\begin{lem}
\label{lem-sig-wf}
Given a base type interpretation $I$ and a well-founded relation on
$I$-computable terms $R$, $(\gtf,R_\stat)_\lex$ is well-founded if,
for all $\sf^{|\vT|}:\vT\a U$, $\I\vT_I\sle\SN(R)$.
\end{lem}

\begin{prf}
If $(\sf,\vt)\in\S_I$, then $\vt\in\I\vT_I$. By assumption,
$\I\vT_I\sle\SN(R)$. Hence, by Lemma \ref{lem-stat-wf},
$\vt\in\SN(R_{\stat(\sg)})$ whatever $\sg$ is. Assume now that there
is an infinite $(\gtf,R_\stat)_\lex$-decreasing sequence
$(\sf_i,\vt_i)_{i\ge 0}$. Then, $(\sf_i)_{i\ge 0}$ is an infinite
$\ge_\cF$-decreasing sequence. Since $\gtf$ is well-founded by
assumption, there must be some
$j$ such that, for all $i\ge j$, $\sf_i\eqf\sf_j$. Since symbols
equivalent in $\eqf$ have the same status by assumption,
$(\vt_i)_{i\ge j}$ is an infinite $R_{\stat(\sf_j)}$-decreasing
sequence, a contradiction.\qed
\end{prf}

\subsection{Computability properties of arrow types}
\label{sec-arrow}

In this sub-section, the results hold for any base type interpretation
$I$ and any set of neutral terms $\cN$. For the sake of simplicity, we
drop the index $I$ in $\I{T}_I$ and simply write $\I{T}$.

\begin{lem}
\label{lem-comp-sn-arrow-core}
$\I{U\a V}$ satisfies {\em(comp-sn)} if:
\begin{itemize}
\item $\I{U}\neq\vide$, which is the case if $\I{U}$ satisfies
  {\em(comp-neutral)};
\item $\I{V}$ satisfies {\em(comp-sn)}.
\end{itemize}
\end{lem}

\begin{prf}
Assume that there is an infinite reduction sequence $t_0\gtpt
t_1\gtpt\ldots$ with $t_0\in\I{U\a V}$ and $t_i:T_i$. By
definition of $\I{U\a V}$, $T_0=U\a V$. By definition of
$\gtpt$, $T_0\ge T_1\ge\ldots$ By assumption,
$\I{U}\neq\vide$. So, let $u\in\I{U}$. By definition of $\I{U\a V}$,
we have $t_0u\in\I{V}$. We now prove that there is an infinite
reduction sequence starting from $t_0u$. Since $\I{V}$ is assumed to
satisfy (comp-sn), this is not possible. Therefore, $\I{U\a V}$
satisfies (comp-sn) too. By \link{typ-arrow}\index{(typ-arrow)|indlem[lem-comp-sn-arrow-core]} there are only two cases:
\begin{itemize}
\item
For all $i$, $T_i=U\a B_i$ for some $B_i$. By monotony (Lemma
\ref{lem-mon-core}), $t_0u\gtpt t_1u\gtpt \ldots$
\item
There is $i$ such that $T_{i+1}$ is a sort or $T_{i+1}=A_{i+1}\a
B_{i+1}$ with $A_{i+1}\neq U$. Let $k$ be the smallest $i$ satisfying
this condition. Hence, for all $i\le k$, there is $B_i$ such that
$T_i=U\a B_i$. By monotony (Lemma \ref{lem-mon-core}),
$t_0u\gtpt\ldots t_ku$. By
\link{typ-arrow}\index{(typ-arrow)|indlem[lem-comp-sn-arrow-core]}, we have
$B_k\ge T_{k+1}$. Hence, by \appsub, $t_ku\gtpt
t_{k+1}$.\qed
\end{itemize}
\end{prf}

\begin{lem}
\label{lem-comp-red-arrow-core}
${\I{U\a V}}$ satisfies {\em(comp-red)} if:
\begin{itemize}
\item $\I{U}\neq\vide$, which is the case if $\I{U}$ satisfies
  {\em(comp-neutral)};
\item $\I{V}$ satisfies {\em(comp-red)}.
\end{itemize}
\end{lem}

\begin{prf}
Let $t\in\I{U\a V}$ and $t':T'$ such that $t\gtpt t'$. By
definition of $\I{U\a V}$, $t:U\a V$. By definition of $\gtpt$,
$U\a V\ge T'$. By
\link{typ-arrow}\index{(typ-arrow)|indlem[lem-comp-red-arrow-core]}, there are
two cases:
\begin{enumerate}
\item
$V\ge T'$. By assumption, $\I{U}\neq\vide$. So, let $u\in\I{U}$. By
  definition of $\I{U\a V}$, $tu\in\I{V}$. By \appsub, we have
  $tu\gtpt t'$. Therefore $t'\in\I{T'}$ since $\I{V}$
  satisfies (comp-red).
\item
$T'=U\a V'$ and $V\ge V'$. Let $u\in\I{U}$. By monotony (Lemma
  \ref{lem-mon-core}), $tu\gtpt t'u$. By definition of
  $\I{U\a V}$, $tu\in\I{V}$. Since $\I{V}$ satisfies (comp-red),
  $t'u\in\I{V'}$. Therefore, $t'\in\I{T'}$.\qed
\end{enumerate}
\end{prf}

\begin{lem}
\label{lem-comp-red-app-core}
Let $t:U\a V$ and $u:U$. Then, every $\gtpt$-reduct of $tu$ is computable if:
\begin{itemize}
\item every $\gtpt$-reduct of $t$ is computable;
\item $u$ is computable;
\item if $t=\l xv$, then $v_x^u$ is computable;
\item for all $u'$ such that $u\gtpt u'$, $tu'$ is computable;
\item $\I{U}$ satisfies {\em(comp-red)};
\item $\I{V}$ satisfies {\em(comp-red)};
\item $\I{V'}$ satisfies {\em(comp-lam)} whenever $V'\le V$.
\end{itemize}
\end{lem}

\begin{prf}
  We prove that every $w:W$ such that $tu\gtpt w$ is computable, by
  induction on $|w|$. By definition of $\gtpt$, we have $V\ge W$.
\begin{itemize}
\item\appsub
\begin{itemize}
\item $t\gep w$. By
  \link{typ-right-subterm}\index{(typ-right-subterm)|indlem[lem-comp-app-no-fun-core]},
  $U\a V>V$. Hence, by transitivity, $U\a V>W$ and $t\gtpt
  w$. Therefore, $w$ is computable by assumption.
\item $u\gept w$. Then, $w$ is computable since, by
  assumption, $u$ is computable and $\I{U}$ satisfies (comp-red).
\end{itemize}

\item\appeq\ $w=t'u'$ and either:
\begin{itemize}
\item $t=t'$ and $u\gtp u'$, in which case $t'u$ is computable by
  assumption since $u\gtpt u'$;
\item or $tu\gtapp t'$ and $tu\gtapp u'$. We prove that, for $v\in\{t',u'\}$, if $tu\gtapp v$ then $v$ is computable. There are three cases:
\begin{itemize}
\item $t\gtpt v$. Then, $v$ is computable by assumption.
\item $u\gept v$. Then, either $u=v$ and $v$ is computable by
  assumption, or $u\gtpt v$ and $v$ is computable since $u$ is
  computable and $\I{U}$ satisfies (comp-red).
\item $tu\gtpt v$. Then, since $v\in\{t',u'\}$, $v$ is computable by
  induction hypothesis.
\end{itemize}
\end{itemize}

\item\applam\ $w=\l yv$, $tu\gtp v$ and $y\notin\FV(v)$ by Lemma
  \ref{lem-fv-core}. Then, there are $A$ and $B$ such that $W=A\a
  B$. By
  \link{typ-right-subterm}\index{(typ-right-subterm)|indlem[lem-comp-app-no-fun-core]},
  $W>B$. Hence, by transitivity, $V>B$ and $tu\gtpt v$. Thus, by
  induction hypothesis, $v$ is computable. Since $\I{W}$ satisfies
  (comp-lam), $w$ is computable if, for all computable term $a:A$,
  $v_y^a$ is computable. Since $y\notin\FV(v)$, $v_y^a=v$. Therefore,
  $w$ is computable.

\item\appvar\ $w\in\vide$. Impossible.

\item\appbeta\ $t=\l xv$ and $v_x^u\ge^\vide w$. Since $\tau((\l
  xv)u)=\tau(v_x^u)$, we have $v_x^u\gept w$. Thus $w$ is computable
  since $v_x^u$ is computable and $\I{V}$ satisfies (comp-red).\qed
\end{itemize}
\end{prf}

\begin{lem}
\label{lem-comp-app-no-fun-core}
Let $t:U\a V$ and $u:U$. Then, $tu$ is computable if:
\begin{itemize}
\item every $\gtpt$-reduct of $t$ is computable;
\item $u$ is computable;
\item if $t=\l xv$, then $v_x^u$ is computable;
\item either $t$ is neutral or $t=\l xv$;
\item $\I{U}$ satisfies {\em(comp-red)} and {\em(comp-sn)};
\item $\I{V}$ satisfies {\em(comp-red)} and {\em(comp-neutral)};
\item $\I{V'}$ satisfies {\em(comp-lam)} whenever $V'\le V$.
\end{itemize}
\end{lem}

\begin{prf}
  We prove that $tu$ is computable by induction on $u$ with $\gtpt$ as
  well-founded relation ($\I{U}$ satisfies (comp-sn) by
  assumption). So, by induction hypothesis, for all $u'$ such that
  $u\gtpt u'$, $tu'$ is computable. Hence, by Lemma
  \ref{lem-comp-red-app-core}, every $\gtpt$-reduct of $tu$ is
  computable. Now, $tu$ is neutral because, either $t$ is neutral and
  $tu$ is neutral by
  \link{neutral-app}\index{(neutral-app|indlem[lem-comp-app-no-fun-core]},
  or $t=\l xv$ and $tu$ is neutral by
  \link{neutral-beta}\index{(neutral-beta)|indlem[lem-comp-app-no-fun-core]}. Therefore,
  $tu$ is computable since $\I{V}$ satisfies (comp-neutral).\qed
\end{prf}

\begin{cor}
\label{cor-comp-neutral-arrow-core}
$\I{U\a V}$ satisfies {\em(comp-neutral)} if:
\begin{itemize}
\item $\I{U}$ satisfies {\em(comp-red)} and {\em(comp-sn)};
\item $\I{V}$ satisfies {\em(comp-red)} and {\em(comp-neutral)};
\item $\I{V'}$ satisfies {\em(comp-lam)} whenever $V'\le V$.
\end{itemize}
\end{cor}

\begin{prf}
Let $t$ be a neutral term of type $U\a V$ such that every
$\gtpt$-reduct of $t$ is computable. By definition, $t\in\I{U\a V}$
if, for all computable $u:U$, $tu$ is computable. Since $t$ is
neutral, by
\link{neutral-not-lam}\index{(neutral-not-lam)|indcor[cor-comp-neutral-arrow-core]},
$t$ is not of the form $\l xv$. Therefore, by Lemma
\ref{lem-comp-app-no-fun-core}, $tu$ is computable since all the
required properties are satisfied.\qed
\end{prf}

\begin{lem}
\label{lem-comp-lam-arrow-core}
Let $x:U$ and $v:V$. Then, $\l xv$ is computable if:
\begin{itemize}
\item for all computable $u:U$, $v_x^u$ is computable;
\item $\I{U}$ satisfies {\em(comp-sn)} and {\em(comp-red)} and
  contains a variable, which is the case if it satisfies
  {\em(comp-neutral)} too;
\item $\I{V}$ satisfies {\em(comp-sn)}, {\em(comp-red)} and
  {\em(comp-neutral)};
\item $\I{V'}$ satisfies {\em(comp-lam)} whenever $V'\le V$.
\end{itemize}
\end{lem}

\begin{prf}
By definition, $\l xv$ is computable if, for all computable $u:U$,
$(\l xv)u$ is computable. By Lemma \ref{lem-comp-app-no-fun-core},
$(\l xv)u$ is computable if every $\gtpt$-reduct of $\l xv$ is
computable, the other conditions being satisfied. Since $\I{U}$
contains a variable, we can wlog assume that this is $x$. So,
$v_x^x=v$ is computable. Since $\I{V}$ satisfies (comp-sn),
$v\in\SN(\gtpt)$. We now prove that every $\gtpt$-reduct $w:W$ of $\l
xv$ is computable, by induction on $(v,|w|)$ with $(\gtpt,>_\bN)_\lex$
as well-founded relation. By definition of $\gtpt$, we have $U\a V\ge
W$.
\begin{itemize}
\item\lamsub\ $v\gept w$. Since $v$ is computable and
  $\I{V}$ satisfies (comp-red), we have $w$ computable.

\item\lameq\ $w=\l xb$ and $v\gtp b$. Then, there is $B$ such
  that $W=U\a B$. Since $U\a V\ge W$, by Lemma \ref{lem-typ-arrow}, we
  have $V\ge B$. Hence, $v\gtpt b$. Thus, by induction hypothesis,
  every $\gtpt$-reduct of $\l xb$ is computable. By Lemma
  \ref{lem-comp-app-no-fun-core}, to prove that $\l xb$ is computable,
  it suffices to check that, for all $u:U$ computable, $b_x^u$ is
  computable. By assumption, $v_x^u$ is computable. By stability by
  substitution (Lemma \ref{lem-subs-core}), $v_x^u\gtpt
  b_x^u$. Therefore, $b_x^u$ is computable since $\I{V}$ satisfies
  (comp-red) by assumption.

\item\lamneq\ $w=\l yb$, $\tx\neq\ty$, $\l xv\gtp b$ and
  $y\notin\FV(b)$ by Lemma \ref{lem-fv-core}. Then, there are $A$ and
  $B$ such that $W=A\a B$. Since $U\neq A$, by
  \link{typ-arrow}\index{(typ-arrow)|indlem[lem-comp-lam-arrow-core]},
  $V\ge W$. Since, by assumption, $\I{W}$ satisfies (comp-lam), it
  suffices to prove that, for all computable $a:A$, $b_y^a$ is
  computable. Since $y\notin\FV(b)$, $b_y^a=b$. By
  \link{typ-right-subterm}\index{(typ-right-subterm)|indlem[lem-comp-lam-arrow-core]},
  $U\a V>V$ and $W>B$. Hence, by transitivity, $U\a V>B$ and $\l
  xv\gtpt b$. Therefore, since $|w|>_\bN|b|$, by induction hypothesis,
  $b$ is computable.

\item\lamvar\ $w\in\vide$. Impossible.

\item\lameta\ $v=ax$, $a\gep w$ and $x\notin\FV(a)$. Since
  $\tau(\l xv)=\tau(a)$, we have $a\gept w$. Let $u:U$
  computable. We have $au=v_x^u$ computable by assumption. Therefore,
  $a$ is computable. By Lemma \ref{lem-comp-red-arrow-core}, $\I{U\a
    V}$ satisfies (comp-red) since $\I{U}\neq\vide$ and $\I{V}$
  satisfies (comp-red). Therefore, $w$ is computable too.\qed
\end{itemize}
\end{prf}

\begin{cor}
\label{cor-comp-lam-arrow-core}
$\I{U\a V}$ satisfies {\em(comp-lam)} if:
\begin{itemize}
\item $\I{U}$ satisfies {\em(comp-sn)}, {\em(comp-red)} and
  {\em(comp-neutral)}\comment{\footnote{Requiring that $\I{U}$ contains a
    variable is enoug.h}};
\item $\I{V}$ satisfies {\em(comp-sn)}, {\em(comp-red)} and
  {\em(comp-neutral)};
\item $\I{V'}$ satisfies {\em(comp-lam)} whenever $V'\le V$.
\end{itemize}
\end{cor}

In conclusion, we can see that $\I{U\a V}_I$ is a computability
predicate if so are $\I{U}_I$ and $\I{V'}_I$ for all $V'\le
V$. Therefore, if we can define a base type interpretation $I$ so
that, for every sort $\sA$, $I(\sA)$ is a computability predicate
then, for all type $T$, $\I{T}_I$ will be a computability predicate.

\subsection{Well-foundedness of core CPO}

We now define a set of neutral terms $\cN$ and a base type
interpretation $I$ for proving the well-foundedness of core CPO.

\begin{defi}[Neutral terms for core CPO]
\label{def-neutral-core}
Let $\cN$ be the smallest set of terms containing the terms of the
form $\sf(\vt)$ and closed by (neutral-var), (neutral-beta) and
(neutral-app).
\end{defi}

One can easily check that $\cN$ satisfies all the properties of
Definition \ref{def-comp-props}.

In contrast with the usual practice, but as in~\cite{jouannaud07jacm},
our interpretation of sorts is not the set of strongly normalizing
terms of that sort. To define the base type interpretation $I$, we
proceed by induction on $\gttwf$ which is well-founded by
\link{typ-sn}. So, let $\sA$ be a sort and assume that $I$ is defined
for all sorts $\sB\lttwf\sA$. Then, let $I(\sA)$ be the least fixpoint of
the monotone function $F_\sA$ defined as follows:\\

\begin{center}
$F_\sA(S)=\{t\in\cL\mid t:\sA
\et(\all u)(\all U)~t\gtpt u\et u:U\A u\in\I{U}_{I\cup\{(\sA,S)\}}\}$\\[5mm]
\end{center}

We now prove that $F_\sA$ is indeed well-defined and monotone. Then,
by Knaster and Tarski's fixpoint theorem \cite{tarski55pjm}, $F_\sA$
admits a (least) fixpoint.

\begin{lem}
\label{lem-sort-int-wd-core}
$F_\sA$ is well-defined.
\end{lem}

\begin{prf}
The recursive call to $\I{U}_{I\cup\{(\sA,S)\}}$ is well-defined
because, by definition of $\gtpt$, we have $\sA\ge U$. Hence, by Lemma
\ref{lem-typ-occ-sort}, $\Sort_{\letwf\sA}(U)$.\qed
\end{prf}

\begin{lem}
\label{lem-sort-int-mon-core}
$F_\sA$ is monotone.
\end{lem}

\begin{prf}
Let $S\sle S'$, $J=I\cup\{(\sA,S)\}$, $J'=I\cup\{(\sA,S')\}$ and $t\in
F_\sA(S)$. Then, (1) $t:\sA$ and (2) $(\all u)(\all U)~t\gtpt u\et
u:U\A u\in\I{U}_J$. Now, we have $t\in F_\sA(S')$ because $t$
satisfies (1) and (2) with $S$ replaced by $S'$. Indeed, assume that
$t\gtpt u$ and $u:U$. By (2), $u\in\I{U}_J$. By definition of $\gtpt$,
$\sA\ge U$. If $\sA=U$, then $u\in\I{U}_{J'}$ since $u\in\I{U}_J=S\sle
S'=\I{U}_{J'}$. Otherwise, by Lemma \ref{lem-typ-occ-sort},
$\Sort_{\lttwf\sA}(U)$. Therefore, by Lemma \ref{lem-bints-agree},
$\I{U}_J=\I{U}_{J'}$ and $u\in\I{U}_{J'}$.\qed
\end{prf}

Note that the least fixpoint of $F_\sA$ can be reached by transfinite
iteration of $F_\sA$ from $\vide$ \cite{kuratowski22fm,cousot79pjm},
that is, there is an ordinal $\ka$, such that
$I(\sA)=F_\sA^\ka(\vide)$ where:
\begin{itemize}
\item $F_\sA^0(S)=S$
\item $F_\sA^{\ka+1}(S)=F_\sA(F_\sA^\ka(S))$
\item $F_\sA^\ka(S)=\bigcup_{\kb<\ka}F_\sA^\kb(S)$ if $\ka$ is a limit ordinal
\end{itemize}

We now check that type interpretations are computability predicates.

\begin{lem}
\label{lem-comp-sort-core}
Given a sort $\sA$, $\I\sA$ satisfies {\em(comp-red)},
{\em(comp-neutral)} and {\em(comp-lam)}.
\end{lem}

\begin{prf} We show each property in turn.
\begin{itemize}
\item (comp-red) Let $t\in\I\sA$ and assume that $t\gtpt u$ and
  $u:U$. Since $\I\sA=F_\sA(\I\sA)$, we have $u\in\I{U}$ by
  definition of $F_\sA$.
\item (comp-neutral) Let $t:\sA$ be a neutral term whose
  $\gtpt$-reducts are all computable. Since $\I\sA=F_\sA(\I\sA)$, we
  have $t\in\I\sA$ by definition of $F_\sA$.
\item (comp-lam) Trivial for typing reasons.\qed
\end{itemize}
\end{prf}

\begin{lem}
\label{lem-comp-sn-sort-core}
Given a sort $\sA$, $\I\sA$ satisfies {\em(comp-sn)} if, for all type
$U<\sA$, $\I{U}$ satisfies {\em(comp-sn)}.
\end{lem}

\begin{prf}
As already mentioned, $\I\sA=F_\sA^\ka(\vide)$ for some ordinal
$\ka$. Since $\vide$ satisfies (comp-sn), it therefore suffices to
check that $F_\sA$ preserves termination: if $S\sle\SN(\gtpt)$, then
$F_\sA(S)\sle\SN(\gtpt)$. So, let $S\sle\SN(\gtpt)$ and let $t\in
F_\sA(S)$. By definition of $F_\sA$, we have $t:\sA$ and, if $t\gtpt
u$ and $u:U$, then $u\in\I{U}_J$ where $J=I\cup\{(\sA,S)\}$. By
definition of $\gtpt$, $\sA\ge U$. If $\sA=U$, then $\I{U}_J=S$ and
$u\in\SN(\gtpt)$ since $S\sle\SN(\gtpt)$. Otherwise, $u\in\SN(\gtpt)$
since $\I{U}\sle\SN(\gtpt)$ by assumption.\qed
\end{prf}

\begin{thm}
\label{thm-comp-typ-core}
For all type $T$, $\I{T}$ is a computability predicates, \ie satisfies
{\em(comp-sn)}, {\em(comp-red)}, {\em(comp-neutral)} and
{\em(comp-lam)}.
\end{thm}

\begin{prf}
We proceed by induction on $\gttwf$ which is well-founded by
assumption \link{typ-sn}\index{(typ-sn)|indthm[thm-comp-typ-core]}. If $T$
is a sort, then we can conclude by Lemma \ref{lem-comp-sort-core},
Lemma \ref{lem-comp-sn-sort-core} and induction hypothesis. Otherwise,
$T=U\a V$. Since $T\gts_l U$, by induction hypothesis, $\I{U}$ is a
computability predicate. Let now $V'$ be a type such that $V\ge
V'$. By
\link{typ-right-subterm}\index{(typ-right-subterm)|indthm[thm-comp-typ-core]}
and transitivity, $T>V'$. Hence, by induction hypothesis, $\I{V'}$ is
a computability predicate. Therefore, $\I{U\a V}$ satisfies (comp-sn)
by Lemma \ref{lem-comp-sn-arrow-core}, (comp-red) by Lemma
\ref{lem-comp-red-arrow-core}, (comp-neutral) by Corollary
\ref{cor-comp-neutral-arrow-core} and (comp-lam) by Corollary
\ref{cor-comp-lam-arrow-core}.\qed
\end{prf}

Now, we are left to prove that every function symbol is computable.

\begin{lem}
\label{lem-comp-big-core}
Let $\sf^{|\vT|}:\vT\a U$ and $\vt\in\I\vT$. Then, $\sf(\vt)\in\I{U}$.
\end{lem}

\begin{prf}
By Theorem \ref{thm-comp-typ-core}, $\I\vT$ satisfies
(comp-sn). Hence, by Lemma \ref{lem-sig-wf},
$(\gtf,{(\gtpt)_\stat})_\lex$ is well-founded. We now prove that, for
all $(\sf,\vt)\in\S$, $\sf(\vt)$ is computable, by induction on
$(\gtf,(\gtpt)_\stat)_\lex$ (1).

First, we check that $t=\sf(\vt)$ is computable if all its
$\gtpt$-reducts so are. This follows from the facts that, by
definition of $\cN$, $t$ is neutral and, by Theorem
\ref{thm-comp-typ-core}, $\I{U}$ satisfies (comp-neutral).

We now prove that, for all finite sets of variables $X$, for all
substitutions $\s$ such that $\dom(\s)\cap\FV(\vt)=\vide$ and $\s$
is computable on $X$, and for all terms $w$ such that $\sf(\vt)\gtp[X]
w$, we have $w\s$ computable, by induction on the size of $w$
(2). Note that $\vt\s=\vt$ since $\dom(\s)\cap\FV(\vt)=\vide$.

\begin{itemize}
\item\bigsub\ $(\ex i)~t_i\gept w$. By stability by substitution
  of $\gept$ (Lemma \ref{lem-subs-core}), we have
  $t_i\s\gept w\s$. Therefore, $w\s$ is computable since, by
  Theorem \ref{thm-comp-typ-core}, $\I{V}$ satisfies (comp-red).

\item\bigeq\ There are $\sg$ and $\vu$ such that $w=\sg(\vu)$,
  $\sf\eqf\sg$, $(\all i)~\sf(\vt)\gtp[X]u_i$ and
  $\vt\,{(\gtpt)_{\stat(\sf)}}\,\vu$. Since $(\all i)~\sf(\vt)\gtp[X]u_i$,
  by induction hypothesis (2), $\vu\s$ are computable. If $t_i\gtpt
  u_j$ then, by stability by substitution (Lemma \ref{lem-subs-core}),
  $t_i=t_i\s\gtpt u_j\s$. Therefore,
  $\vt\,{(\gtpt)_{\stat(\sf)}}\,\vu\s$ and, by induction hypothesis (1),
  $\sg(\vu)\s$ is computable.

\item\biggt\ Since $\sf(\vt)\gtp[X]\vu$, by induction hypothesis (2),
  $\vu\s$ are computable. Hence, $\sg(\vu)\s$ is computable by
  induction hypothesis (1).

\item\bigapp\ Since $\sf(\vt)\gtp[X] u$ and $\sf(\vt)\gtp[X] v$, by
  induction hypothesis (2), $u\s$ and $v\s$ are computable. Therefore,
  $(uv)\s=(u\s)(v\s)$ is computable.

\item\biglam\ Wlog we can assume that $\s$ is away from $\{y\}$ and
  $y\notin\FV(\sf(\vt))$. Hence, $(\l yv)\s=\l y(v\s)$. By Theorem
  \ref{thm-comp-typ-core}, $\l y(v\s)$ is computable if, for all
  computable $u:\ty$, $(v\s)_y^u$ is computable. Since $\s$ is away
  from $\{y\}$, $(v\s)_y^u=(v_y^z)\t$ where $\t=\s\cup\{(z,u)\}$. Let
  $Z=X\cup\{z\}$. Since $\sf(\vt)\gtp[Z] v_y^z$,
  $\dom(\t)\cap\FV(\sf(\vt))=\vide$ and $\t$ is computable on $Z$, we
  have $v_y^z\t$ computable by induction hypothesis (2).

\item\bigvar\ $w\in X$. Then, $w\s$ is computable since $\s$ is
  computable on $X$.\qed
\end{itemize}
\end{prf}

\begin{thm}
The relation $\gtpt$ of Definition \ref{def-cpr-core} is well-founded.
\end{thm}

\begin{prf}
After Theorem \ref{thm-wf}, Theorem \ref{thm-comp-typ-core} and Lemma
\ref{lem-comp-big-core}.\qed
\end{prf}

We can therefore conclude that $\gtpt[+]$ is a monotone, stable,
well-founded order.

The well-foundedness proof of core CPO is actually similar to that of
HORPO, although the proof here is presented in a quite different style
from HORPO's monolithic proof~\cite{jouannaud07jacm}. This similarity
fades away with the two coming extensions, to inductive types and to
small symbols. The reason why we have split the proof into small
pieces is indeed to factor out its structure and those parts
which are common to core CPO and its extensions.

\section{Accessibility}
\label{sec-ind}

In this section, we introduce an extension of the core definition that
will allow us to handle rewrite rules like the ones defining recursors
for strictly positive inductive types as used in the Coq proof
assistant \cite{coquand88colog,coq}.

\newcommand\zero{\ms{zero}}
\newcommand\suc{\ms{suc}}
\renewcommand\lim{\ms{lim}}

\begin{exa}
\label{ex-ord}
Consider the inductive type $\sO$ of ``Brouwer ordinals'' whose
constructors are $\zero:\sO$ for zero, $\suc^1:\sO\a\sO$ for successor,
and $\lim^1:(\sN\a\sO)\a\sO$ for limit, where $\sN$ is the
inductive type of Peano (unary) natural numbers with constructors
$\ms{0}:\sN$ and $\ss^1:\sN\a\sN$. Given a
type $A$, the recursor (of arity $4$) at type $A$
\[\ms{rec}_\sO^A:\sO\a A\a(\sO\a A\a A)\a((\sN\a\sO)\a(\sN\a A)\a A)\a A\]
can be defined by the following rewrite rules:

\begin{rewc}
\ms{rec}_\sO^A(\zero,u,v,w) & u\\
\ms{rec}_\sO^A(\suc(x),u,v,w) & v\,x\,(\ms{rec}_\sO^A(x,u,v,w))\\
\ms{rec}_\sO^A(\lim(y),u,v,w) & w\,y\,(\l n\,\ms{rec}_\sO^A(y\,n,u,v,w))\\
\end{rewc}\medskip
\end{exa}

To capture such a relation, we need the following two comparisons to
succeed:
\[\ms{rec}_\sO^A(\lim(y),u,v,w)\gtp y \quad\mbox{and}\quad \lim(y)\gtpt y\,n.\]

The second comparison cannot succeed unless we allow non empty sets
$X$ of variables in \bigeq\ in order to have $\lim(y)\gtpt[\{n\}]y\,n$,
but we have seen in Section \ref{sec-limit} that this may lead to
non-termination. Instead, we will use a specific ordering: the
{\em structural ordering} introduced by Coquand for dealing with such kind
of definitions in the calculus of constructions \cite{coquand92types}.

For the first comparison to succeed, since the type of $y$ is bigger
than the type of $\lim(y)$, we must {\em not} check types in
\bigsub, but we have seen in Section \ref{sec-limit} that this may
lead to non-termination. To solve this problem, we will compare in
\bigsub\ the right-hand side with possibly deep subterms of the
left-hand side.

We cannot take any deep subterm however, as shown by the following
example: assuming the signature $\sc^1:(\sA\a\sB)\a\sA$ and
$\sf^1:\sA\a(\sA\a\sB)$, the deep subterm comparison $\sf(\sc(x))\gtpt
x$ leads to non-termination, since, taking $t=\l x\,\sf(x)\,x$, we have
$\sf(\sc(t))\,\sc(t)\gtpt t\,\sc(t)\gtpt \sf(\sc(t))\,\sc(t)$ by
monotony and \appbeta. There are two cases where deep subterms can be
used: as for the first, deep subterms whose type is a (sufficiently
small) sort~\cite{jouannaud07jacm}; as for the second, Mendler showed
that pattern-matching on constructors of a sort $\sA$ having an
argument whose type has a negative occurrence of $\sA$ wrt the arrow
type constructor (see Definition \ref{def-pos} just after), leads to
non-termination, while, on the contrary, $\b$-reduction combined with
recursion combinators for positive inductive types
terminates~\cite{mendler87phd,mendler91apal}.

\subsection{Accessible subterms}

In this sub-section, we first define the sets of positive and negative
positions of a type, and the notions of accessible and structurally
smaller term, before we prove some properties of these notions.

\begin{defi}[Positive and negative positions in a type]
\label{def-pos}
The sets $\Pos(T)$, $\Pos(\sA,T)$, $\Pos^+(T)$ and $\Pos^-(T)$ of
positions, positions of $\sA$, positive positions and negative
positions in a type $T$ are inductively defined as follows:
\begin{itemize}
\item $\Pos(\sA)=\Pos^+(\sA)=\Pos(\sA,\sA)=\{\vep\}$
\item $\Pos^-(\sA)=\vide$
\item $\Pos(\sA,\sB)=\vide$ if $\sA\neq\sB$
\item $\Pos(T\a U)=\{1p\mid p\in\Pos(T)\}\cup\{2p\mid p\in\Pos(U)\}$
\item $\Pos(\sA,T\a U)=\{1p\mid p\in\Pos(\sA,T)\}\cup\{2p\mid p\in\Pos(\sA,U)\}$
\item $\Pos^+(T\a U)=\{1p\mid p\in\Pos^-(T)\}\cup\{2p\mid p\in\Pos^+(U)\}$
\item $\Pos^-(T\a U)=\{1p\mid p\in\Pos^+(T)\}\cup\{2p\mid p\in\Pos^-(U)\}$
\end{itemize}
A sort $\sA$ {\em occurs only positively} (resp. {\em negatively}) in
$T$ if $\Pos(\sA,T)\sle\Pos^+(T)$ (resp. $\Pos(\sA,T)\sle\Pos^-(T)$).
\end{defi}

For instance, for $T=(\sA\a\sB)\a\sB$ with $\sA\neq\sB$, we have
$\Pos(T)=\{\vep,1,2,11,12\}$, $\Pos^+(T)=\{11,2\}$,
$\Pos^-(T)=\{12\}$, $\Pos(\sA,T)=\{11\}$ and
$\Pos(\sB,T)=\{12,2\}$. Hence, $\sA$ occurs only positively in $T$,
but $\sB$ has both positive and negative occurrences in $T$.

\begin{defi}[Accessible arguments]
\label{def-acc}
For every $\sf^\af:\vT\a\sA$, we assume given a set $\Acc(\sf)$ of
{\em accessible} arguments of $\sf$ such that $i\in\Acc(\sf)$ implies
$\Sort_{\letwf\sA}(T_i)$ and $\Pos(\sA,T_i)\sle\Pos^+(T_i)$.
\end{defi}

Note that, if $\af<|\vT|$, the output type of $\sf$ is functional.

Let us consider Example \ref{ex-ord} and assume that
$\sO\gtt\sN$. Then, $\sO$ occurs only positively in the type of the
first argument of $\suc$, and we can take $\Acc(\suc)=\{1\}$. Similarly,
we can take $\Acc(\lim)=\{1\}$.

We can now introduce those sorts $\sA$ which are not bigger than any
arrow type and will be interpreted by $\SN_\sA(\gtpt)$ later:

\begin{defi}[Basic sorts]
A sort $\sA$ is {\em basic} if, for all type $T<\sA$, $T$ is a basic
sort and, for all $\sf^\af:\vT\a\sA$ and $i\in\Acc(\sf)$, $T_i=\sA$ or
$T_i$ is a basic sort.
\end{defi}

In particular, are basic all first-order data types like unary natural
numbers, lists, trees, etc. whose constructors do not take a function as
argument.

Accessibility blends accessible arguments and subterms of basic sort:

\begin{defi}[Accessibility]
A term $u$ is said to be {\em accessible} in a term $t$ if:
\begin{itemize}
\item $u$ is a subterm of basic sort of $t$ such that
  $\FV(u)\subseteq\FV(t)$, written $t\gtb u$, or
\item there are $\sf^\af:\vT\a\sA$, $\vt:\vT$ and $i\in\Acc(\sf)$
  such that $t=\sf(t_1,\ldots,t_\af)t_{\af+1}\ldots t_{|\vT|}$ and $t_i\gea
  u$, written $t\gta u$,
\end{itemize}
where $\ges^s_b$ and $\gea$ are the reflexive closures of $\gts^s_b$
and $\gta$ respectively.
\end{defi}

Coming back to Example \ref{ex-ord}, we have $x$ accessible in
$\suc(x)$ since $\suc(x)\gta x$, and $y$ accessible in $\lim(y)$ since
$\lim(y)\gta y$.

\begin{lem}
\label{lem-acc-pos}
If $t\gta u:U$, then there are two sorts $\sA$ and $\sB$ such
that $t:\sA$, $\sB\letwf\sA$, $\Sort_{\letwf\sB}(U)$ and
$\Pos(\sB,U)\sle\Pos^+(U)$.
\end{lem}

\begin{prf}
By induction on $\gta$, which is clearly well-founded. Assume that
there are $\sf^\af:\vT\a\sA$, $\vt:\vT$ and $i\in\Acc(\sf)$ such that
$t=\sf(t_1,\ldots,t_\af)t_{\af+1}\ldots t_{|\vT|}$ and $t_i\gea u$.
By definition of $\Acc$,
$\Sort_{\letwf\sA}(T_i)$ and $\Pos(\sA,T_i)\sle\Pos^+(T_i)$. If $t_i=u$
then $U=T_i$ and we are done with $\sB=\sA$. Assume now that
$t_i\gta u$. Then, by induction hypothesis, there are two sorts $\sB$
and $\sC$ such that $t_i:\sB$, $\sC\letwf\sB$, $\Sort_{\letwf\sC}(U)$ and
$\Pos(\sC,U)\sle\Pos^+(U)$. Since $\Sort_{\letwf\sA}(T_i)$ and $T_i=\sB$,
we have $\sB\letwf\sA$. By transitivity, we get $\sC\letwf\sA$.\qed
\end{prf}

\begin{cor}
\label{cor-acc-pos}
If $t:\sA$, $t\gta u:U$ and $\sA$ occurs in $U$, then
$\Sort_{\letwf\sA}(U)$ and $\Pos(\sA,U)\sle\Pos^+(U)$.
\end{cor}

\begin{prf}
By Lemma \ref{lem-acc-pos}, there is a sort $\sB\letwf\sA$ such that
$\Sort_{\letwf\sB}(U)$ and $\Pos(\sB,U)\sle\Pos^+(U)$. Since $\sA$ occurs
in $U$, $\sA\letwf\sB$. Therefore, $\sB=\sA$ and we are done.\qed
\end{prf}

\begin{defi}[\cite{coquand92types}]
Given a finite set $X$ of variables, we say that $u$ is {\em
  structurally smaller than} $t$ wrt $X$, written $t\,\gti[X]\,u$,
if there are $\sA$, $v$ and $\vx:\vU$ such that $t:\sA$, $u:\sA$,
$u=v\vx$, $t\gta v$, $\vx\in X$ and $\Pos(\sA,\vU)=\vide$.
\end{defi}

One can easily check:

\begin{lem}
\label{lem-subs-acc}
$\gtb$ and $\gta$ (resp. $\gti[X]$) are stable by substitution
(resp. away from $X$).
\end{lem}

\subsection{CPO with accessible subterms}

\begin{defi}[CPO with accessible subterms]
\label{def-cpr-ind}
The relation $\gtp[X]$ is extended by replacing the rules \bigsub\ and
\bigeq\ of Figure \ref{fig-cpr-core} by the ones of Figure
\ref{fig-cpr-ind}.
\end{defi}

\begin{figure}[ht]
\caption{New CPO rules with accessible subterms\label{fig-cpr-ind}}
\fbox{\begin{tabular}{rl}
\bigsub & $\sf(\vt)\gtp[X] v$ if $\sf\in\cF_b$ and $\vt\,\geb\gea\gept v$\\
\bigeq &
$\sf(\vt)\gtp[X] \sg(\vu)$ if $\sf\in\cF_b$, $\sf\eqf\sg$, $\sf(\vt)\gtp[X]\vu$
and $\vt\,({\gtpt}\cup{\gti[X]\!\gept})_{\stat(\sf)}\,\vu$
  \end{tabular}}
\end{figure}

The rules of Example \ref{ex-ord} are now easily oriented by CPO. Take
for instance the third rule. It is included in CPO since, by \bigapp:
\begin{itemize}
\item $l=\ms{rec}_\sO^A(\lim(y),u,v,w)\gtp w\,x$ by \bigapp since:
\begin{itemize}
\item $l>w$ by \bigsub,
\item $l>y$ by \bigsub\ since $\lim(y)\gta y$,
\end{itemize}
\item $l\gtp \l n\,\ms{rec}_\sO^A(y\,n,u,v,w)$ by \biglam\ since
  $l\gtp[\{n\}] \ms{rec}_\sO^A(y\,n,u,v,w)$ because, by \bigeq:
\begin{itemize}
\item $l\gtp[\{n\}]y\,n$ since by \bigapp:
\begin{itemize}
\item $l>y$ as already seen,
\item $l>n$ by \bigvar,
\end{itemize}
\item $l>u,v,w$ by \bigsub,
\item $\lim(y)\,\gti[\{n\}]\,y\,n$.
\end{itemize}
\end{itemize}

\noindent Following \cite{blanqui06tr}, we could strengthen CPO further by
defining $\gti[X]$ and $\gtp[X]$ simultaneously, by replacing in
\bigeq, $\gti[X]$ by $\gti[\sf(\vt),X]$ and, in the definition of
$\gti[X]$, $\vx\in X$ by $\sf(\vt)\gtp[X]\vx$.

\subsection{Comparison with CHORPO}
\label{sec-chorpo}

CHORPO is a variant of HORPO which was also aiming at ordering
recursors of inductive types like Brouwer's ordinals. In rules (1),
(3), (4) and (7) of the 12 rules of HORPO as recalled in
Section~\ref{sec-comp}, one has to show recursively that every direct
subterm of the left-hand side $\sf(\vt)$ is bigger than (or equal to) the
right-hand side. In CHORPO, one can also use in addition to the direct
subterms, any term of the \emph{computability closure}
$\CC(\sf(\vt),\vide)$ of the left-hand side, a set inductively defined by
6 rules ($\CC 1$) to ($\CC 6$) that, for most of them, correspond to
CPO rules as follows.

$\CC(\sf(\vt),X)$ must contain $\{\vt\}$, which corresponds to
\bigsub, and $X$, which corresponds to \bigvar; $(\CC
1)$ says that $\CC(\sf(\vt),\vide)$ contains any term $u$ of minimal
type such that $\vt\gts^s u$, where $t\gts^s u$ if $t\gts u$ and
$\FV(u)\sle\FV(t)$, which corresponds to \bigsub, \appsub\
and \lamsub; $(\CC 2)$ corresponds to \biggt; $(\CC 3)$
corresponds to \bigeq\ with $\gtpt$ replaced by
${\gtpt}\cup{\gts^s_\tau}$; $(\CC 4)$ corresponds to \bigapp\ and
$(\CC 5)$ to \biglam.

On the other hand, $(\CC 6)$ says that $\CC(\sf(\vt),\vide)$ is
closed by $>_\horpo$. Capturing such a rule in CPO requires to
consider the transitive closure of $\gtpt$ in \bigeq\ which would
most presumably turn CPO into an undecidable relation, as it is
probably already the case of CHORPO for the same reason.

In conclusion, while CHORPO and CPO look incomparable, the restriction
of CHORPO to $(\CC 1)$, $(\CC 2)$, $(\CC 3)$, $(\CC 4)$, $(\CC 5)$ is
included in CPO. In fact, CPO can be seen as a decidable reformulation
of CHORPO integrating in a simple, uniform and more powerful way both
HORPO and the notion of computability closure. Note finally that HORPO
and the computability closure are themselves already related, as shown
in~\cite{blanqui06tr}. More precisely, the first version of HORPO
\cite{jouannaud99lics} is included in the fixpoint of the monotone
function mapping $>$ to the relation $>_\CC^\vide$ such that $t>_\CC^X
u$ if $u\in\CC(t,X)$, RPO being equal to this fixpoint when restricted
to first-order terms.

\subsection{Computability with accessible subterms}

\begin{lem}[Basic properties]\hfill
\begin{itemize}
\item\label{lem-well-def-ind}
$\gtp[X]$ is well-defined.
\item\label{lem-mon-ind}
$\gtpt$ is monotone.
\item\label{lem-fv-ind}
If $a\gtp[X] b$, then $\FV(b)\sle\FV(a)\cup X$.
\item\label{lem-alpha-ind}
$\gtp[X]$ is stable by $\al$-equivalence. 
\item\label{lem-subs-ind}
$\gtp[X]$ is stable by substitution away from $X$.
\item\label{lem-alpha-X-ind}
If $e,e'\in\cX$, $\tau(e)=\tau(e')$, $t\gtp[X]u$ and $e'\notin\FV(\l
eu)$, then $t\gtp[X-\{e\}\cup\{e'\}]u_e^{e'}$.
\end{itemize}
\end{lem}

\begin{prf}
As for the core definition using Lemma \ref{lem-subs-acc} and the fact that,
if $a\gta b$, then $\FV(b)\sle\FV(a)$.\qed
\end{prf}

In order to extend the well-founded proof of core CPO to accessible
subterms, we need to define a set of neutral terms and a base type
interpretation so that accessible arguments of a computable term
$\sf(\vt)$ are computable. Hence, the following definitions:

\begin{defi}[Neutral terms for CPO with accessible subterms]
\label{def-neutral-ind}
Let $\cN$ be the smallest set of terms containing the terms of the
form $\sf(\vt)$ with $\Acc(\sf)=\vide$, and closed by (neutral-var),
(neutral-beta) and (neutral-app).
\end{defi}

One can easily check that $\cN$ satisfies all the properties of
Definition \ref{def-comp-props}. Note that, now, a term is neutral if
and only if it is of the form $x\,\vv$, $(\l xa)\,b\,\vv$, or
$\sf(\vt)\,\vv$ with $\Acc(\sf)=\vide$.

To define the base type interpretation $I$, we proceed as for core CPO
by well-founded induction on $\gttwf$. So, let $\sA$ be a sort and
assume that $I$ is defined for all sorts $\sB\lttwf\sA$. Then, let
$I(\sA)$ be the least fixpoint of the monotone function $F_\sA$
defined as follows:\\

$F_\sA(S)=\{t\in\cL\mid t:\sA
\et(\all u)(\all U)~t\gtpt u\et u:U\A u\in\I{U}_{I\cup\{(\sA,S)\}}\\\hsp[15mm]
\et(\all\sf)(\all\vT)(\all\vt)(\all i)~
\sf^\af:\vT\a\sA\,\et\,
t=\sf(t_1,\ldots,t_\af)t_{\af+1}\ldots t_{|\vT|}\,\et\,
i\in\Acc(\sf)\\\hsp[115mm]
\A t_i\in\I{T_i}_{I\cup\{(\sA,S)\}}\}$\\

Note that, by this definition, a term
$\sf(t_1,\ldots,t_\af)t_{\af+1}\ldots t_n:\sA$ is computable if all
its $\gtpt$-reducts and all its accessible arguments $t_i$ with
$i\in\Acc(\sf)$ are computable. This makes the terms of this form
behave like neutral terms when $\vt$ are computable.

We now prove that $F_\sA$ is indeed well-defined and monotone.

\begin{lem}
\label{lem-sort-int-wd}
$F_\sA$ is well-defined.
\end{lem}

\begin{prf}
  The calls to $\I{U}_{I\cup\{(\sA,S)\}}$ and
  $\I{T_i}_{I\cup\{(\sA,S)\}}$ with $i\in\Acc(\sf)$ are well-defined
  because every sort occurring in $U$ or $T_i$ is $\letwf$ to
  $\sA$. Indeed, by definition of $\gtpt$, we have $\sA\ge U$. Hence,
  by Lemma \ref{lem-typ-occ-sort}, $\Sort_{\letwf\sA}(U)$. As for $T_i$,
  it follows by definition of $\Acc(\sf)$.\qed
\end{prf}

\begin{lem}
\label{lem-typ-int-mon}
Let $T$ be a type such that $\Sort_{\letwf\sA}(T)$. Then, the function
$S\to\I{T}_{I\cup\{(\sA,S)\}}$ is monotone (resp. anti-monotone) wrt
set inclusion if $\Pos(\sA,T)\sle\Pos^+(T)$
(resp. $\Pos(\sA,T)\sle\Pos^-(T)$).
\end{lem}

\begin{prf}
Let $S\sle S'$, $J=I\cup\{(\sA,S)\}$ and $J'=I\cup\{(\sA,S')\}$. We
proceed by induction on $T$.
\begin{itemize}
\item $T=\sA$. Then, $\I{T}_J=S\sle S'=\I{T}_{J'}$.
\item $T=\sB\lttwf\sA$. Then, $\I{T}_J=I(\sB)=\I{T}_{J'}$.
\item $T=U\a V$ and $\Pos(\sA,T)\sle\Pos^+(T)$. Let $t\in\I{T}_J$. By
  definition of $\I{T}$, $t\in\I{T}_{J'}$ if, for all
  $u\in\I{U}_{J'}$, $tu\in\I{V}_{J'}$. By definition,
  $\Pos(\sA,T)=\{1p\mid p\in\Pos(\sA,U)\}\cup\{2p\mid
  p\in\Pos(\sA,V)\}$ and $\Pos^+(T)=\{1p\mid
  p\in\Pos^-(U)\}\cup\{2p\mid p\in\Pos^+(V)\}$. Hence,
  $\Pos(\sA,U)\sle\Pos^-(U)$ and
  $\Pos(\sA,V)\sle\Pos^+(V)$. Therefore, by induction hypothesis,
  $\I{U}_{J'}\sle\I{U}_J$ and $\I{V}_J\sle\I{V}_{J'}$. So,
  $u\in\I{U}_J$ and, since $t\in\I{T}_J$, we have
  $tu\in\I{V}_J\sle\I{V}_{J'}$.
\item $T=U\a V$ and $\Pos(\sA,T)\sle\Pos^-(T)$. Let
  $t\in\I{T}_{J'}$. By definition of $\I{T}$, $t\in\I{T}_J$ if, for
  all $u\in\I{U}_J$, $tu\in\I{V}_J$. By definition,
  $\Pos(\sA,T)=\{1p\mid p\in\Pos(\sA,U)\}\cup\{2p\mid
  p\in\Pos(\sA,V)\}$ and $\Pos^-(T)=\{1p\mid
  p\in\Pos^+(U)\}\cup\{2p\mid p\in\Pos^-(V)\}$. Hence,
  $\Pos(\sA,U)\sle\Pos^+(U)$ and
  $\Pos(\sA,V)\sle\Pos^-(V)$. Therefore, by induction hypothesis,
  $\I{U}_J\sle\I{U}_{J'}$ and $\I{V}_{J'}\sle\I{V}_J$. So,
  $u\in\I{U}_{J'}$ and, since $t\in\I{T}_{J'}$, we have
  $tu\in\I{V}_{J'}\sle\I{V}_J$.\qed
\end{itemize}
\end{prf}

\begin{lem}
\label{lem-sort-int-mon}
$F_\sA$ is monotone.
\end{lem}

\begin{prf}
Let $S\sle S'$, $J=I\cup\{(\sA,S)\}$, $J'=I\cup\{(\sA,S')\}$ and $t\in
F_\sA(S)$. Then, (1) $t:\sA$, (2) $(\all u)(\all U)~t\gtpt u\et u:U\A
u\in\I{U}_J$, and (3) $(\all\sf)(\all\vT)(\all\vt)(\all{}i)
~\sf^\af:\vT\a\sA\,\et\,t=\sf(t_1,\ldots,t_\af)t_{\af+1}\ldots t_{|\vT|}
\,\et\,i\in\Acc(\sf)\A t_i\in\I{T_i}_J$. We have $t\in F_\sA(S')$
because $t$ satisfies (1), (2) and (3) with $S$ replaced by $S'$:
\begin{enumerate}
\item $t:\sA$ by (1).
\item Assume that $t\gtpt u$ and $u:U$. By (2), $u\in\I{U}_J$. By
  definition of $\gtpt$, $\sA\ge U$. If $\sA=U$, then $u\in\I{U}_{J'}$
  since $u\in\I{U}_J=S\sle S'=\I{U}_{J'}$. Otherwise, by Lemma
  \ref{lem-typ-occ-sort}, $\Sort_{\lttwf\sA}(U)$. Therefore, by Lemma
  \ref{lem-bints-agree}, $\I{U}_J=\I{U}_{J'}$ and $u\in\I{U}_{J'}$.
\item Assume that $\sf^\af:\vT\a\sA$,
  $t=\sf(t_1,\ldots,t_\af)t_{\af+1}\ldots t_{|\vT|}$ and
  $i\in\Acc(\sf)$. By definition of $\Acc$, $\Sort_{\letwf\sA}(T_i)$,
  $\Pos(\sA,T_i)\sle\Pos^+(T_i)$ and, by Lemma \ref{lem-typ-int-mon},
  $\I{T_i}_J\sle\I{T_i}_{J'}$. Therefore, $t_i\in\I{T_i}_{J'}$ since
  $t_i\in\I{T_i}_J$ by (3).\qed
\end{enumerate}
\end{prf}

\subsection{Well-foundedness of the structural term ordering}

\begin{lem}
\label{lem-sort-int-rk-mon}
The function $\ka\to F_\sA^\ka(\vide)$ is monotone.
\end{lem}

\begin{prf}
Let $J^\ka=F_\sA^\ka(\vide)$. We prove by induction on $\kb$ that, for
all $\ka<\kb$, $J^\ka\sle J^\kb$.
\begin{itemize}
\item $\kb=\kc+1$. Then, $J^\kb=F_\sA(J^\kc)$. If $\ka=\kc$, then
  $J^\ka\sle J^\kb$ by definition of $F_\sA$. Otherwise, $\ka<\kc$
  and, by induction hypothesis, $J^\ka\sle J^\kc$. By Lemma
  \ref{lem-sort-int-mon}, $J^{\ka+1}\sle J^{\kc+1}$. Since $J^\ka\sle
  J^{\ka+1}$ by definition of $F_\sA$, we have $J^\ka\sle J^\kb$.
\item $\kb$ is a limit ordinal. Then, $J^\ka\sle J^\kb$ by definition
  of $J^\kb$.\qed
\end{itemize}
\end{prf}

The functions $F_\sA^\ka$ provide us with a well-founded relation that
is the basis of the well-foundedness of the structural term
ordering when it is instantiated by computable terms:

\begin{defi}[Rank ordering]
\label{def-comp-ord}
Let the rank of a term $t\in\I\sA$, $\rk_\sA(t)$, be the
smallest ordinal $\ka$ such that $t\in F_\sA^\ka(\vide)$. Then, let
$t\gto u$ if there is a sort $\sA$ such that $t\in \I\sA$,
$u\in\I\sA$ and $\rk_\sA(t)>\rk_\sA(u)$.
\end{defi}

We now prove that $\gtpt$ is included in $\gto$ and that their
union is strongly normalizing on computable terms.

\begin{lem}
\label{lem-comp-ord-red}
If $t\in\I\sA$, $u\in\I\sA$ and $t\gtpt u$, then $t\gto u$.
\end{lem}

\begin{prf}
By definition, we have $t\in F_\sA^\ka(\vide)$ where
$\ka=\rk_\sA(t)$. We can neither have $\ka=0$ nor $\ka$ be a limit
ordinal. So, there is $\kb$ such that $\ka=\kb+1$. Hence,
$F_\sA^\ka(\vide)=F_\sA(F_\sA^\kb(\vide))$ and $u\in F_\sA^\kb(\vide)$
by definition of $F_\sA$. Therefore, $t\gto u$.\qed
\end{prf}

\begin{lem}
\label{lem-sn-qgt}
$\I{T}\sle\SN(\gtpt\cup\gto)$ if, for all $T'\le T$, $\I{T'}$
satisfies {\em(comp-sn)}.
\end{lem}

\begin{prf}
Assume that there is an infinite $(\gtpt\cup\gto)$-decreasing sequence
$(t_i)_{i\ge 0}$ such that $t_0\in\I{T}$ and $t_i:T_i$. Then,
$(T_i)_{i\ge 0}$ is an infinite $\ge$-decreasing sequence. Since $>$
is well-founded by \link{typ-sn}\index{(typ-sn)|indlem[lem-sn-qgt]}, there
must be some $j$ such that, for all $i\ge j$, $T_i=T_j$. If $T_j$ is a
sort then, by Lemma \ref{lem-comp-ord-red}, $(t_i)_{i\ge j}$ is an
infinite $\gto$-decreasing sequence, which is not possible since
$\gto$ is well-founded. If $T_j$ is not a sort, then $(t_i)_{i\ge j}$
is an infinite $\gtpt$-decreasing sequence since $\gto$ only
compares terms of base type. But this is not possible since $T\ge T_j$
and, by assumption, $\I{T_j}$ satisfies (comp-sn).\qed
\end{prf}

We now show that $\gta$ preserves computability, and that the
structural term ordering $\gti[X]$ is stable by computable substitutions
of domain $X$.

\begin{lem}
\label{lem-comp-acc}
If $t$ is computable and $t\gea u$, then $u$ is computable.
\end{lem}

\begin{prf}
  By induction on the definition of $\gea$. If $t=u$, this is
  immediate. Otherwise, there are $\sf^\af:\vT\a\sA$, $\vt:\vT$ and
  $i\in\Acc(\sf)$ such that $t=\sf(t_1,\ldots,t_\af)t_{\af+1}\ldots
  t_{|\vT|}$ and $t_i\gea u$. Since $t$ is computable, by definition
  of $\I\sA$, $t_i$ is computable. Therefore, by induction hypothesis,
  $u$ is computable.\qed
\end{prf}

\begin{lem}
\label{lem-comp-acc-rk}
If $t:\sA$ is computable, $t\gta u:U$ and $\sA$ occurs in $U$,
then there is $\kb$ such that $\rk_\sA(t)=\kb+1$ and $u\in\I{U}_J$,
where $J(\sA)=F_\sA^\kb(\vide)$ and $J(\sB)=I(\sB)$ if $\sB\neq\sA$.
\end{lem}

\begin{prf}
  First note that, by Corollary \ref{cor-acc-pos},
  $\Sort_{\letwf\sA}(U)$ and $\Pos(\sA,U)\sle\Pos^+(U)$. We now
  proceed by induction on the definition of $\gta$. Assume that there
  are $\sf^\af:\vT\a\sA$, $\vt:\vT$ and $i\in\Acc(\sf)$ such that
  $t=\sf(t_1,\ldots,t_\af)t_{\af+1}\ldots t_{|\vT|}$ and $t_i\gea
  u$. By definition, $\rk_\sA(t)$ can be neither $0$ nor a limit
  ordinal. Therefore, there must be $\kb$ such that $\rk_\sA(t)=\kb+1$
  and $t_i\in\I{T_i}_J$. If $t_i=u$, then we are done. Assume now that
  $t_i\gta u$. Then, there is $\sB$ such that $t_i:\sB$. By definition
  of $\Acc$, $\sB\letwf\sA$. By Corollary \ref{cor-acc-pos},
  $\Sort_{\letwf\sB}$.  Since $\sA$ occurs in $U$, we have
  $\sA\letwf\sB$ and thus $\sB=\sA$ because $\gttwf$ is well-founded
  by \link{typ-sn}\index{(typ-sn)|indlem[lem-comp-acc-rk]}. Hence, by
  induction hypothesis, there is $\kc$ such that $\rk_\sA(t_i)=\kc+1$
  and $u\in\I{U}_K$, where $K(\sA)=F_\sA^\kc(\vide)$ and
  $K(\sB)=I(\sA)$ if $\sB\neq\sA$. Therefore, $u\in\I{U}_J$ by Lemma
  \ref{lem-typ-int-mon} since $\kc\le\kb$, $\Sort_{\letwf\sA}(U)$ and
  $\Pos(\sA,U)\sle\Pos^+(U)$.\qed
\end{prf}

\begin{lem}
\label{lem-comp-acc-rk-succ}
If $t\,\gti[X]\,u$, $\s$ is computable on $X$ and $t\s$ is computable,
then $u\s$ is computable and $t\s\gto u\s$.
\end{lem}

\begin{prf}
Since $t\,\gti[X]\,u$, there are $\sA$, $v$ and $\vx:\vW$ such that
$t:\sA$, $u:\sA$, $u=v\vx$, $t\gta v$, $\vx\in X$ and
$\Pos(\sA,\vW)=\vide$. Therefore, $u\s=(v\s)(\vx\s)$ and, since
$\gta$ is stable by substitution, $t\s\gta v\s$. By Lemma
\ref{lem-comp-acc-rk}, there is $\kb$ such that $\rk_\sA(t\s)=\kb+1$
and $v\s\in\I{\vW\a A}_J$, where $J(\sA)=F_\sA^\kb(\vide)$ and
$J(\sB)=I(\sA)$ if $\sB\neq\sA$. Since $\s$ is computable on $X$, we
have $\vx\s$ computable. Since $\Pos(\sA,\vW)=\vide$, by Lemma
\ref{lem-bints-agree}, we have $\vx\s\in\I\vW_J$. Therefore,
$u\s\in\I\sA_J$ and $t\s\gto u\s$.\qed
\end{prf}

\subsection{Well-foundedness of CPO with accessible subterms}

We now check that type interpretations are computability predicates,
and that function symbols are computable.

One can easily check that all the lemmas of Section \ref{sec-arrow}
are still valid, as well as the lemmas \ref{lem-comp-sort-core} and
\ref{lem-comp-sn-sort-core} (since they do not depend on $(\cF_b\_)$
rules). Therefore, following the proof of Theorem
\ref{thm-comp-typ-core}, we get:

\begin{thm}
\label{thm-comp-typ-ind}
For all type $T$, $\I{T}$ is a computability predicate, \ie satisfies
{\em(comp-sn)}, {\em(comp-red)}, {\em(comp-neutral)} and
{\em(comp-lam)}.
\end{thm}

\begin{lem}
\label{lem-basic-ind}
If $\sA$ is a basic sort, then $\I\sA=\SN_\sA(\gtpt)$.
\end{lem}

\begin{prf}
By Lemma \ref{lem-comp-sn-sort-core}, it suffices to prove that, for
all $t\in\SN_\sA(\gtpt)$, we have $t\in\I\sA$. By
\link{typ-sn}\index{(typ-sn)|indlem[lem-basic-ind]}, $\gttwf$ is
well-founded. By Lemma \ref{lem-mon-ind}, $\gtpt$ is monotone and thus
${{\gtpt}\cup{\gts}}$ is well-founded on $\SN(\gtpt)$. We can
therefore proceed by induction on $(\sA,t)$ with
$(\gttwf,{\gtpt}\cup{\gts})$ as well-founded relation.

We first prove that every $\gtpt$-reduct $u$ of $t$ is
computable. Since $t\in\SN_\sA(\gtpt)$, we have $u\in\SN_U(\gtpt)$. By
definition of $\gtpt$, $\sA\ge U$. Therefore, $U$ is a basic sort and,
by induction hypothesis, $u\in\I{U}$ since $\sA\gttwf U$ or else $\sA=U$ and
$t\gtpt u$.

Hence, if $t$ is neutral, then $t\in\I\sA$ since $\I\sA$ satisfies
(comp-neutral). Otherwise, $t=\sf(t_1,\ldots,t_\af)t_{\af+1}\ldots
t_{|\vT|}$ with $\sf^\af:\vT\a\sA$ and $\Acc(\sf)\neq\vide$. Let
$i\in\Acc(\sf)$. Then, $\Sort_{\letwf\sA}(T_i)$. Since $\sA$ is basic,
$T_i=\sA$ or $T_i$ is a basic sort. In both cases, $T_i\letwf\sA$ and
$T_i$ is a basic sort. Therefore, $t_i\in\I{T_i}$ since
$t_i\in\SN_{T_i}(\gtpt)$ and $\sA\gttwf T_i$ or else $\sA=T_i$ and
$t\gts t_i$.\qed
\end{prf}

\begin{lem}
\label{lem-comp-big-ind}
Let $\sf^{|\vT|}:\vT\a U$ and $\vt\in\I\vT$. Then, $\sf(\vt)$ is computable.
\end{lem}

\begin{prf}
  There are $\vU$ and $\sA$ such that $U=\vU\a\sA$. By definition,
  $\sf(\vt)$ is computable if, for every $\vu\in\I\vU$, $\sf(\vt)\vu$
  is computable. By Theorem \ref{thm-comp-typ-ind}, $\I\vT$ and
  $\I\vU$ satisfy (comp-sn) and, by Lemma \ref{lem-sn-qgt},
  $\I\vT\sle\SN(\gtpt\cup\gto)$. Therefore, by Lemma \ref{lem-sig-wf},
  $(\gtf,(\gtpt\cup\gto)_\stat)_\lex$ is well-founded. We can
  therefore prove that, for all $((\sf,\vt),\vu)$ such that
  $\sf(\vt)\vu$ is of base type, $\sf(\vt)\vu$ is computable, by
  induction on $((\gtf,(\gtpt\cup\gto)_\stat)_\lex,(\gtpt)_\lex)_\lex$
  (0).

  Since $\sf(\vt)\vu$ is of base type and all its accessible arguments
  are computable by assumption, it suffices to prove that all its
  $\gtpt$-reducts are computable. To this end, we prove that, for all
  $k\le n=|\vu|$, every $\gtpt$-reduct of $\sf(\vt)u_1\ldots u_k$ is
  computable, by induction on $k$ (1).
\begin{itemize}
\item $k=0$. The proof is the same as for Lemma
  \ref{lem-comp-big-core} except for the new cases:
\begin{itemize}
\item\bigsub\ There are $i$, $u:U$ and $v:V$ such that
  $t_i\,\geb u\,\gea\,v\gept w$. By stability by substitution of
  $\geb$, $\gea$ and $\gept$ (Lemma \ref{lem-subs-acc} and
  \ref{lem-subs-ind}), we have $t_i=t_i\s\,\geb u\s\,\gea\,v\s\gept
  w\s$. Since $\vt$ are computable and $\I\vT$ satisfies (comp-sn), we
  have $\vt\in\SN(\gtpt)$. Since $\gtpt$ is monotone (Lemma
  \ref{lem-mon-ind}), we have $u\s\in\SN(\gtpt)$. Hence, by Lemma
  \ref{lem-basic-ind}, $u\s$ is computable and, by Lemma
  \ref{lem-comp-acc}, $v\s$ is computable. Therefore, $w\s$ is
  computable since, by Theorem \ref{thm-comp-typ-ind}, $\I{V}$
  satisfies (comp-red).

\item\bigeq\ There are $\sg$ and $\vu$ such that $w=\sg(\vu)$,
  $\sf\eqf\sg$, $\vt\,({\gtpt}\cup{\gti[X]\gept})_{\stat(\sf)}\,\vu$
  and $\sf(\vt)\gtp[X]\vu$. Since $\sf(\vt)\gtp[X]\vu$, by induction
  hypothesis (2), $\vu\s$ are computable. If $t_i\gtpt u_j$ then, by
  stability by substitution (Lemma \ref{lem-subs-ind}),
  $t_i=t_i\s\gtpt u_j\s$. If $t_i\,\gti[X]v\gept u_j$ then, by
  Lemma \ref{lem-comp-acc-rk-succ}, $t_i\s\gto v\s$ and, by stability
  by substitution again, $v\s\gept u_j\s$. Therefore, by Lemma
  \ref{lem-comp-ord-red} and transitivity, $t_i\s\gto u_j\s$. Thus, in
  both cases, $\vt\,(\gtpt\cup\gto)_{\stat(\sf)}\,\vu\s$ and, by
  induction hypothesis (0), $\sg(\vu)\s$ is computable.
\end{itemize}

\item $k>0$. Then, $\sf(\vt)\vu=tu_k$ where $t=\sf(\vt)u_1\ldots
  u_{k-1}$. By induction hypothesis (1), every $\gtpt$-reduct of $t$
  is computable. Now, if $u_k\gtpt u_k'$, then $tu_k'$ is computable
  by induction hypothesis (0). Therefore, by Lemma
  \ref{lem-comp-red-app-core}, every $\gtpt$-reduct of $tu_k$ is
  computable.\qed
\end{itemize}
\end{prf}

\begin{thm}
The relation $\gtpt$ of Definition \ref{def-cpr-ind} is well-founded.
\end{thm}

\begin{prf}
After Theorem \ref{thm-wf}, Theorem \ref{thm-comp-typ-ind} and Lemma
\ref{lem-comp-big-ind}.\qed
\end{prf}

\subsection{Using semantic comparisons}

The extension of CPO described here is still not able to orient the
terminating rules defining the recursor of the type $\sC$ in Example
\ref{ex-cont}:

\begin{exa}
Given an arbitrary type $A$, the recursor (of arity $3$) at type $A$
of the type $\sC$ of continuations of Example~\ref{ex-cont} has type
$\ms{rec}_\sC^A:\sC\a A\a(\neg\neg C\a\neg\neg A\a A)\a A$. Its
rewrite rules are the following:

\begin{rewc}
\ms{rec}_\sC^A(\sd,u,v) & u\\
\ms{rec}_\sC^A(\sc(x),u,v) &
v\,x\,(\l y^{\neg A}x\,(\l z^\sC y\,\ms{rec}_\sC^A(z,u,v)))
\end{rewc}
\end{exa}

The problem is that we do not have $\sc(x)\gti[\{y,z\}]z$. Indeed,
$\sC$ is {\em non-strictly} positive and the structural term ordering
can only handle {\em strictly} positive types.

To handle such rules, we know two solutions. The first one is to
define the interpretation of $\sC$ so that $\ms{rec}_\sC^A$ is
computable by definition \cite{werner94phd,matthes98phd,blanqui05fi},
which is possible since positivity conditions are satisfied. However,
this solution lacks flexibility for the user who is forced to define
all other functions on $\sC$ via the recursor.

The second, flexible solution consists in considering types with size
annotations (to be interpreted by ordinals) and, in \bigeq, 
compare terms by their size annotations, an approach
initiated independently in \cite{hughes96popl,gimenez96phd} and later
developed in various works,
\eg~\cite{abel04ita,barthe04mscs,blanqui04rta,blanqui06lpar-sbt}. Indeed,
assuming that $\sc(x)$ has type $\sC^{\al+1}$, then $x$ has type
$\neg\neg C^\al$ and, thus, the bound variable $z$ gets the type
$\sC^\al$ which size annotation is smaller than the one of $\sc(x)$.

Including semantics in RPO was pioneered by Kamin and
L\'evy~\cite{kamin80note}, and extended to HORPO
in~\cite{borralleras01lpar}. In both cases, semantics was added by
replacing the precedence by a semantic order on terms. The use of size
annotations is a different way to include semantics in these orders.
These two different ways of including semantics in recursive path
orders are however related: both can be seen as an instance of the
more general semantic labeling
schema~\cite{zantema95fi,hamana07ppdp,blanqui09csl}.

\section{Small symbols}
\label{sec-small}

In this section, we consider a further extension of CPO that
originated from some draft version of \cite{jouannaud15tlca} and try
to answer the following general question: can we relax the constraints
on the precedence? More precisely, to which extent can a function
symbol be smaller than an application or an abstraction? We are going
to show that this is indeed possible if the rules governing these {\em
  small} symbols are more restrictive than the ones for {\em big}
symbols.

We first define the extension of CPO to small symbols, and then show
the computability properties including a specific one for small
symbols. Unlike before, this will reveal a circularity among the
dependencies between the different computability properties, hence
strong normalization does not follow. Breaking this circularity will
require assumptions on the types of small symbols that are then
investigated for practical purposes. It will appear that, for
instance, any constructor of a strictly-positive inductive type can be
considered a small symbol.

\subsection{CPO with small symbols}

\begin{defi}[CPO with small symbols]
\label{def-cpr-small}
\renewcommand\ind[1]{{\bf(#1)}\index{(#1)|inddef[def-cpr-small]}}
We assume that the set of function symbols is partitioned into a set
$\cF_b$ of {\em big} symbols and a set $\cF_s$ of {\em small} symbols 
so that:
\begin{itemize}
\item\target{small-lt-big} no small symbol is greater or
  equivalent to a big symbol\hfill\ind{small-lt-big}
\item\target{small-acc} small symbols with arrow output type have no
  accessible argument\hfill\ind{small-acc}
\end{itemize}
We then extend $\gtp[X]$ by adding the rules of Figure \ref{fig-cpr-small}.
\end{defi}

We will add conditions on the types of small symbols after Definition
\ref{def-comp-pos} (see Figure \ref{fig-small-conds}).

\newcommand\appsmall{(@$\cF_s$)}
\newcommand\lamsmall{($\l$$\cF_s$)}

\newcommand\smallsub{($\cF_s$$\gts$)}
\newcommand\smalleq{($\cF_s$=)}
\newcommand\smallgt{($\cF_s$>)}
\newcommand\smallapp{($\cF_s$@)}
\newcommand\smalllam{($\cF_s$$\l$)}
\newcommand\smallvar{($\cF_s$$\cX$)}

\begin{figure}[ht]
\caption{Additional CPO rules for small symbols\label{fig-cpr-small}}
\fbox{\begin{tabular}{rl}
\appsmall & $tu\gtp[X] \sf(\vv)$
if $\sf\in\cF_s$ and $(\all i)~tu\gtpt[X] v_i$
\\\lamsmall & $\l xt\gtp[X] \sf(\vv)$
if $\sf\in\cF_s$ and $(\all i)~\l xt\gtpt[X] v_i$
\medskip

\\\smallsub & $\sf(\vt)\gtp[X] v$ if $\sf\in\cF_s$ and $(\ex i)~t_i\gept v$
\\\smalleq & $\sf(\vt)\gtp[X] \sg(\vu)$
if $\sf\in\cF_s$, $\sf\eqf\sg$, $(\all i)~\sf(\vt)\gtpt[X] u_i$
and $\vt~({\gtpt}\cup{\gti[X]~\gept})_{\stat(\sf)}~\vu$
\\\smallgt & $\sf(\vt)\gtp[X] \sg(\vu)$
if $\sf\in\cF_s$, $\sf\gtf\sg$ and $(\all i)~\sf(\vt)\gtpt[X] u_i$
\\\smallapp & $\sf(\vt)\gtp[X] u\,v$
if $\sf\in\cF_s$, $\sf(\vt)\gtpt[X] u$ and $\sf(\vt)\gtpt[X] v$
\\\smallvar & $\sf(\vt)\gtp[X] x$ if $\sf\in\cF_s$ and $x\in X$
  \end{tabular}}
\end{figure}

Because of the rules \appsmall\ and \smallapp, one may think that the
relation is not terminating anymore, but this is not the case for
typing reasons. Indeed, in contrast with rules for big symbols, rules
for small symbols require type checking the recursive calls
systematically.

For instance, assume that $\sf:o\a o$ and $\sg^2:o\a o\a o$. Then,
although we have $\sf\,\sa\gtpt\sg(\sa,\sa)$ by \appsmall\ since
$\sf\,\sa\gtpt\sa$ by \appsub, we do not hopefully have
$\sg(\sa,\sa)\gtpt \sf\,\sa$ by \smallapp\ because we do not have
$\sg(\sa,\sa)\gtpt \sf$ for typing reasons.

On the other hand, there is no rule \smalllam\ such that
$\sf(\vt)\gtp[X]\l yv$ if $\sf(\vt)\gtp[X]v$ and $y\notin\FV(v)$
because, together with the rule \lamsmall, it would lead to
non-termination as shown by the following example: given small symbols
$\sa:o\a o\gtf\sb:o$, $\l x\sb\gtpt\sa$ by \lamsmall, and $\sa\gtpt\l
x\sb$ by \smalllam\ since $\sa\gtpt\sb$ by \smallgt. It is however
possible to have \smalllam\ if one removes \lamsmall. We choose to
present the case of \smalllam\ because it seems more useful, but the
termination proof can be easily adapted if \smalllam\ is replaced by
\lamsmall. Note however that this does not lead to the same definition
for the sets $\SPos$, $\LPos$, \ldots (Definition \ref{def-comp-pos})
studied in Section \ref{sec-check-pos}.

Two potential improvements are left. First, take a rule \smallsub\
similar to the rule \bigsub\ of Figure \ref{fig-cpr-ind}. Second, get
rid of the assumption \link{small-acc}, if possible.

\subsection{Computability properties}

\begin{lem}[Basic properties]\hfill
\begin{itemize}
\item\label{lem-well-def-small}
$\gtp[X]$ is well-defined.
\item\label{lem-mon-small}
$\gtpt$ is monotone.
\item\label{lem-fv-small}
If $a\gtp[X] b$, then $\FV(b)\sle\FV(a)\cup X$.
\item\label{lem-alpha-small}
$\gtp[X]$ is stable by $\al$-equivalence.
\item\label{lem-subs-small}
$\gtp[X]$ is stable by substitution away from $X$.
\item\label{lem-alpha-X-small}
If $e,e'\in\cX$, $\tau(e)=\tau(e')$, $t\gtp[X]u$ and $e'\notin\FV(\l
eu)$, then $t\gtp[X-\{e\}\cup\{e'\}]u_e^{e'}$.
\end{itemize}
\end{lem}

Keeping the same definitions for neutral terms and the base type
interpretation as in Section~\ref{sec-prf}, it is easy to check that
Lemma \ref{lem-comp-sn-arrow-core} and Lemma \ref{lem-comp-red-arrow-core}
still hold. However, because of the new rules \appsmall\ and
\lamsmall, Corollary \ref{cor-comp-neutral-arrow-core} and
Corollary \ref{cor-comp-lam-arrow-core}, hence
Lemma~\ref{lem-comp-app-no-fun-core}
and Lemma~\ref{lem-comp-lam-arrow-core} reveal new dependencies that require
introducing the following new computability property for a set $S$ of
terms of type $T$:\\

\begin{itemize}
\item[]{\bf(comp-small)} $\sf(\vt)\in S$ if $\sf(\vt):T$, $\sf\in\cF_s$ and
  $\vt$ are computable.\\
\end{itemize}

Note that big symbols do not need any computability property because
they are bigger than everybody else, and therefore other computability
properties do not depend upon the computability of big symbols. It
follows that they cannot be implied in any circularity.

\begin{lem}
\label{lem-comp-red-app-small}
Let $t:U\a V$ and $u:U$. Then, every $\gtpt$-reduct of $tu$ is computable if:
\begin{itemize}
\item every $\gtpt$-reduct of $t$ is computable;
\item $u$ is computable;
\item if $t=\l xv$, then $v_x^u$ is computable;
\item for all $u'$ such that $u\gtpt u'$, $tu'$ is computable;
\item $\I{U}$ satisfies {\em(comp-red)};
\item $\I{V}$ satisfies {\em(comp-red)};
\item $\I{V'}$ satisfies {\em(comp-lam)} and {\em(comp-small)}
  whenever $V'\le V$.
\end{itemize}
\end{lem}

\begin{prf}
The proof is the same as for Lemma \ref{lem-comp-red-app-core}
except for the new case:
\begin{itemize}
\item\appsmall\ $w=\sf(\vv)$, $\sf\in\cF_s$ and $(\all i)~tu\gtpt
  v_i$. By induction hypothesis, $\vv$ are computable. Since $\I{W}$
  satisfies (comp-small) by assumption, $w$ is computable.\qed
\end{itemize}
\end{prf}

\begin{lem}
\label{lem-comp-app-no-fun-small}
Let $t:U\a V$ and $u:U$. Then, $tu$ is computable if:
\begin{itemize}
\item $u$ is computable;
\item every $\gtpt$-reduct of $t$ is computable;
\item if $t=\l xv$, then $v_x^u$ is computable;
\item either $t$ is neutral or $t=\l xv$;
\item $\I{U}$ satisfies {\em(comp-red)} and {\em(comp-sn)};
\item $\I{V}$ satisfies {\em(comp-red)} and {\em(comp-neutral)};
\item $\I{V'}$ satisfies {\em(comp-lam)} and {\em(comp-small)}
  whenever $V'\le V$.
\end{itemize}
\end{lem}

\begin{prf}
  As for Lemma \ref{lem-comp-app-no-fun-core} but using Lemma
  \ref{lem-comp-red-app-small} instead.\qed
\end{prf}

\begin{cor}
\label{cor-comp-neutral-arrow-small}
$\I{U\a V}$ satisfies {\em(comp-neutral)} if:
\begin{itemize}
\item $\I{U}$ satisfies {\em(comp-sn)} and {\em(comp-red)};
\item $\I{V}$ satisfies {\em(comp-red)} and {\em(comp-neutral)};
\item $\I{V'}$ satisfies {\em(comp-lam)} and {\em(comp-small)}
  whenever $V'\le V$.
\end{itemize}
\end{cor}

\begin{prf}
As for Corollary \ref{cor-comp-neutral-arrow-core} but using Lemma
\ref{lem-comp-app-no-fun-small} instead.\qed
\end{prf}

\begin{lem}
\label{lem-comp-lam-arrow-small}
Let $x:U$ and $v:V$. Then, $\l xv$ is computable if:
\begin{itemize}
\item for all computable $u:U$, $v_x^u$ is computable;
\item $\I{U}$ satisfies {\em(comp-sn)} and {\em(comp-red)} and
  contains a variable, which is the case if it satisfies
  {\em(comp-neutral)} too;
\item $\I{V}$ satisfies {\em(comp-sn)}, {\em(comp-red)} and
  {\em(comp-neutral)};
\item $\I{V'}$ satisfies {\em(comp-lam)} whenever $V'\le V$;
\item $\I{W}$ satisfies {\em(comp-small)} whenever $W\le U\a V$.
\end{itemize}
\end{lem}

\begin{prf}
The proof is the same as for Lemma \ref{lem-comp-lam-arrow-core}
except for the new case:
\begin{itemize}
\item\lamsmall\ $w=\sf(\vv)$, $\sf\in\cF_s$ and $(\all i)~\l xv\gtpt v_i$.
By induction hypothesis, $\vv$ are
  computable. Thus, $w$ is computable since, by assumption, $\I{W}$
  satisfies (comp-small).\qed
\end{itemize}
\end{prf}

\begin{cor}
\label{cor-comp-lam-arrow-small}
$\I{U\a V}$ satisfies {\em(comp-lam)} if:
\begin{itemize}
\item $\I{U}$ satisfies {\em(comp-sn)}, {\em(comp-red)} and
  {\em(comp-neutral)};
\item $\I{V}$ satisfies {\em(comp-sn)}, {\em(comp-red)} and
  {\em(comp-neutral)};
\item $\I{V'}$ satisfies {\em(comp-lam)} whenever $V'\le V$;
\item $\I{W}$ satisfies {\em(comp-small)} whenever $W\le U\a V$.
\end{itemize}
\end{cor}

We are left with the new computability property for small symbols:

\begin{lem}
\label{lem-comp-small-hyp}
$\I{U}$ satisfies {\em(comp-small)} if:
\begin{itemize}
\item $\I{U}$ satisfies {\em(comp-neutral)};
\item $\I{U'}$ satisfies {\em(comp-small)} whenever $U'<U$;
\item for every small $\sf^{|\vT|}:\vT\a U$, $\I\vT$
  satisfies {\em(comp-sn)} and {\em(comp-red)}.
\end{itemize}
\end{lem}

\begin{prf}
  Assume that $\sf^\af:\vT\a U$ is small. By assumption, $\I\vT$
  satisfies (comp-sn) and, by Lemma \ref{lem-sn-qgt},
  $\I\vT\sle\SN(\gtpt\cup\gto)$. Therefore, by Lemma \ref{lem-sig-wf},
  $({\gtf,}{(\gtpt\cup\gto)_\stat})_\lex$ is well-founded when
  restricted to small symbols. We can therefore prove that, for all
  $(\sf,\vt)\in\S$ with $\sf\in\cF_s$, $\sf(\vt)$ is computable, by
  induction on $(\gtf,{(\gtpt\cup\gto)_\stat})_\lex$ (1).

  We first prove that $\sf(\vt)$ is computable if all its
  $\gtpt$-reducts so are. If $U$ is a sort, then the result holds
  since $\vt$ are computable. Otherwise, by
  \link{small-acc}\index{(small-acc)|indlem[lem-comp-small-hyp]},
  $\Acc(\sf)=\vide$ and $\sf(\vt)$ is neutral. Therefore, the result
  holds since $\I{U}$ satisfies (comp-neutral) by assumption.

  We now prove that every $\gtpt$-reduct $w:W$ of $\sf(\vt)$
  is computable by induction on $w$ (2). By definition of $\gtpt$, we
  have $U\ge W$.
\begin{itemize}
\item\smallsub\ $(\ex i)~t_i\gept w$. By assumption, $\I\vT$ satisfies
  (comp-red). Thus, $w$ is computable.

\item\smalleq\ There are $\sg:\vU\a W$ and $\vu:\vU$ such that
  $w=\sg(\vu)$, $(\all i)~\sf(\vt)\gtpt u_i$, $\sf\eqf\sg$ and
  $\vt\,({\gtpt}\cup{\gti[\emptyset]\gept})_{\stat(\sf)}\,\vu$. By
  \link{small-lt-big}\index{(small-lt-big)|indlem[lem-comp-small-hyp]},
  $\sg$ is small. Since $\sf(\vt)\gtpt\vu$, by induction hypothesis
  (2), $\vu$ are computable. We distinguish two cases:
\begin{itemize}
\item $U>W$. Then, $\sg(\vu)$ is computable since $\I{W}$ satisfies
  (comp-small).
\item $U=W$. If $t_i\gti[\emptyset]v\gept u_j$ then, by Lemma
  \ref{lem-comp-acc-rk-succ}, $t_i\gto v$ and, by Lemma
  \ref{lem-comp-ord-red}, $v\geo u_j$. Therefore, by transitivity,
  $t_i\gto u_j$. Hence, $\vt\,(\gtpt\cup\gto)_{\stat(\sf)}\,\vu$ and,
  by induction hypothesis (1), $\sg(\vu)$ is computable.
\end{itemize}

\item\smallgt\ There are $\sg:\vU\a W$ and $\vu:\vU$ such that
  $w=\sg(\vu)$, $(\all i)~\sf(\vt)\gtpt u_i$ and $\sf\gtf\sg$. By
  \link{small-lt-big}\index{(small-lt-big)|indlem[lem-comp-small-hyp]},
  $\sg$ is small. Since $\sf(\vt)\gtpt\vu$, by induction hypothesis
  (2), $\vu$ are computable. We distinguish two cases:
\begin{itemize}
\item $U>W$. Then, $\sg(\vu)$ is computable since $\I{W}$ satisfies
  (comp-small).
\item $U=W$. Then, $\sg(\vu)$ is computable by induction hypothesis (1).
\end{itemize}

\item\smallapp\ There are $u$ and $v$ such that $w=uv$ and
  $\sf(\vt)\gtpt uv$. By induction hypothesis (2), $u$ and $v$
  are computable. Therefore, $uv$ is computable.

\item\smallvar\ Not possible.\qed
\end{itemize}
\end{prf}

\subsection{Well-foundedness of CPO with small symbols}

In contrast with the previous cases, we cannot conclude from the above
lemmas that, for every type $T$, $\I{T}$ is a computability predicate,
because of circularities.

Indeed, for $\I\sA$ to satisfy (comp-small), we need, for every small
symbol $\sf:T\a\sA$, $\I{T}$ to satisfy (comp-sn); but for $\I{T}$ to
satisfy (comp-sn) when $T=U\a V$, we need $\I{U}$ to satisfy
(comp-neutral); but for $\I{U}$ to satisfy (comp-neutral) when
$U=W\a\sA$, we need $\I\sA$ to satisfy (comp-small).
To break this circularity, we will make these dependencies more precise by
introducing sets of positions in types that reflect how these
computability properties depend from each other. The idea here is
that if there is no problematic occurrence of $\sA$ in $T$, then
$\I{T}$ satisfies (comp-sn), and similarly for the other properties.

Instead of sets of positions, we could have simply considered boolean
functions returning true if $T$ contains a problematic occurrence of
$\sA$. Considering positions allows us to pinpoint precisely which
occurrences are problematic, and therefore to obtain sharper
conditions on $\cF_s$ ensuring the absence of cycle in the dependency
graph of the computability properties. Of course, one may think that
there are different ways to carry out these proofs, resulting in
different dependency graphs. We believe that these relationship are
intrinsic to the computability properties, although we have not been
able to substantiate this claim so far.

\begin{defi}[Computability-property positions]
\label{def-comp-pos}
For each computability property ($S$ standing for (comp-sn), $R$ for
(comp-red), $N$ for (comp-neutral), $L$ for (comp-lam) and $C$ for
(comp-small)), we inductively define a set of
positions in a type $T$ wrt a sort $\sA$ as follows:
\begin{itemize}
\item $\CPos_\sA(\sA)=\{\vep\}$ and $\CPos_\sA(\sB)=\vide$ if $\sB\neq\sA$
\item $\SPos_\sA(\sB)=\RPos_\sA(\sB)=\NPos_\sA(\sB)=\LPos_\sA(\sB)=\vide$
whatever $\sA$ and $\sB$ are
\item $\CPos_\sA(U\a V)=\NPos_\sA(U\a V)$
\item $\SPos_\sA(U\a V)= \RPos_\sA(U\a V)=
\{1p\mid p\in \NPos_\sA(U)\}\cup\{2p\mid p\in \SPos_\sA(V)\}$
\item $\NPos_\sA(U\a V)=
\{1p\mid p\in \SPos_\sA(U)\cup \RPos_\sA(U)\}\\
\cup\{2p\mid p\in \RPos_\sA(V)\cup \NPos_\sA(V)\cup \LPos_\sA(V)\cup \CPos_\sA(V)\}$
\item $\LPos_\sA(U\a V)=\CPos_\sA(U\a V)\\
\cup\{1p\mid p\in \SPos_\sA(U)\cup \RPos_\sA(U)\cup \NPos_\sA(U)\}\\
\cup\{2p\mid p\in \SPos_\sA(V)\cup \RPos_\sA(V)\cup \NPos_\sA(V)\cup \LPos_\sA(V) \cup \CPos_\sA(V)\}$
\end{itemize}
\end{defi}

\noindent Note that $\RPos_\sA(T)= \SPos_\sA(T) \sle \LPos_\sA(T)$ and
$\NPos_\sA(T)\sle \CPos_\sA(T)$.
Straightforward simplifications then yield:
\begin{itemize}
\item $\NPos_\sA(U\a V)=\{1p\mid p\in \SPos_\sA(U)\} 
\cup\{2p\mid p\in \LPos_\sA(V)\cup CPos_\sA(V)\}$
\item $\LPos_\sA(U\a V)\\=\CPos_\sA(U\a V)\cup
\{1p\mid p\in \SPos_\sA(U)\cup \NPos_\sA(U)\}
\cup\{2p\mid p\in \LPos_\sA(V)\cup CPos_\sA(V)\}$
\end{itemize}\medskip


\noindent We can now express in Figure \ref{fig-small-conds} conditions on the
types of the small symbols ensuring, as we shall show next, the
absence of cycles in the dependency graph.

\begin{figure}[ht]
\caption{Conditions on types of small symbols\label{fig-small-conds}}
\fbox{\begin{minipage}{15cm}
    \target{small-sort} $\all\sf^{|\vT|}:\vT\a\sA,
    (\all{}i)~\Sort_{\letwf\sA}(T_i)\et\SPos_\sA(T_i)=\vide$
    \hfill{\bf(small-sort)}\\
    \hsp[-1mm]\target{small-arrow} $\all\sf^{|\vT|}:\vT\a\vU\a\sA$
    with $|\vU|>0$, $\Acc(\sf)=\vide\et(\all{}i)~\Sort_{\letwf\sA}(T_i)\et
    T_i\letwf\vU\a\sA$\hfill{\bf(small-arrow)}
\end{minipage}}
\end{figure}

Consider the (small-sort) case and assume that $T_i\letwf\sA$.
Then, either $T_i=\sA$ and $\SPos_\sA(T_i)=\vide$ by definition, or
$T_i\lttwf\sA$ and $\SPos_\sA(T_i)=\vide$ by
Lemma~\ref{lem-typ-ord-comp-pos}. The condition for base types is
therefore (strictly) weaker than the one for arrow types. This weaker
form will indeed be important later for deciding if a function symbol
of base output type can be declared small.


\begin{lem}
\label{lem-comp-pos-lt}
If $\Sort_{\lttwf\sA}(T)$, then
$\SPos_\sA(T)=\NPos_\sA(T)=\LPos_\sA(T)=\CPos_\sA(T)=\vide$.
\end{lem}

\begin{prf}
By induction on $T$.
\begin{itemize}
\item $T=\sB$. Then, $\SPos_\sA(T)=\NPos_\sA(T)=\LPos_\sA(T)=\vide$ by
  definition. Since $\Sort_{\lttwf\sA}(T)$, we have $\sB\neq\sA$ and thus
  $\CPos_\sA(T)=\vide$ too.
\item $T=U\a V$. Since $\Sort_{<\sA}(U)$ and $\Sort_{<\sA}(V)$,
  $\SPos_\sA(U)=\NPos_\sA(U)=\LPos_\sA(U)=\CPos_\sA(U)=\vide$ and
  $\SPos_\sA(V)=\NPos_\sA(V)=\LPos_\sA(V)=\CPos_\sA(V)=\vide$ by
  induction hypothesis. Thus,
  $\SPos_\sA(T)=\NPos_\sA(T)=\LPos_\sA(T)=\CPos_\sA(T)=\vide$.\qed
\end{itemize}
\end{prf}

\begin{lem}
\label{lem-typ-ord-comp-pos}
If $T>T'$ and $\Sort_{\letwf\sA}(T)$ then:
\begin{itemize}
\item $\SPos_\sA(T')=\vide$ whenever $\SPos_\sA(T)=\vide$,
\item $\NPos_\sA(T')=\vide$ whenever $\NPos_\sA(T)=\vide$,
\item $\LPos_\sA(T')=\vide$ whenever $\LPos_\sA(T)=\vide$,
\item $\CPos_\sA(T')=\vide$ whenever $\CPos_\sA(T)=\vide$.
\end{itemize}
\end{lem}

\begin{prf}
We proceed by induction on $T$. Note that $\Sort_{\letwf\sA}(T')$ by
Lemma \ref{lem-typ-le-sort}.
\begin{itemize}
\item $T=\sB$. Since $\Sort_{\letwf\sA}(T)$, we have $\sB\le\sA$. By
  transitivity, $T'<\sA$. Hence, by Lemma \ref{lem-typ-occ-sort},
  $\Sort_{\lttwf\sA}(T')$. Therefore,
  $\SPos_\sA(T')=\NPos_\sA(T')=\LPos_\sA(T')=\CPos_\sA(T')=\vide$ by
  Lemma \ref{lem-comp-pos-lt}.

\item $T=U\a V$. Then, $\Sort_{\letwf\sA}(U)$ and $\Sort_{\letwf\sA}(V)$.
\begin{itemize}
\item $\SPos_\sA(T)=\vide$. Then, $\NPos_\sA(U)=\vide$ and
  $\SPos_\sA(V)=\vide$. By
  \link{typ-arrow}\index{(typ-arrow)|indlem[lem-typ-ord-comp-pos]}, there
  are two cases:
\begin{itemize}
\item $V\ge T'$. Then, $\SPos_\sA(T')=\vide$ by induction hypothesis.
\item $T'=U\a V'$ and $V>V'$. By induction hypothesis,
  $\SPos_\sA(V')=\vide$. Therefore, $\SPos_\sA(T')=\vide$.
\end{itemize}

\item $\CPos_\sA(T)=\NPos_\sA(T)=\vide$. Then, $\SPos_\sA(U)=\vide$
  and $\SPos_\sA(V)=\NPos_\sA(V)=\LPos_\sA(V)=\CPos_\sA(V)=\vide$. By
  \link{typ-arrow}\index{(typ-arrow)|indlem[lem-typ-ord-comp-pos]}, there
  are two cases:
\begin{itemize}
\item $V\ge T'$. Then, $\CPos_\sA(T')=\NPos_\sA(T')=\vide$ by
  induction hypothesis.
\item $T'=U\a V'$ and $V>V'$, hence
  $\SPos_\sA(V')=\NPos_\sA(V')=\LPos_\sA(V')=\CPos_\sA(V')=\vide$
by induction hypothesis.
  $\CPos_\sA(T')=\NPos_\sA(T')=\vide$ follows.
\end{itemize}

\item $\LPos_\sA(T)=\vide$. Then, $\SPos_\sA(U)=\NPos_\sA(U)=\vide$ and
  $\SPos_\sA(V)=\NPos_\sA(V)=\LPos_\sA(V)=\CPos_\sA(V)=\vide$. By
  \link{typ-arrow}\index{(typ-arrow)|indlem[lem-typ-ord-comp-pos]}, there
  are two cases:
\begin{itemize}
\item $V\ge T'$. Then, $\LPos_\sA(T')=\vide$ by induction hypothesis.
\item $T'=U\a V'$ and $V>V'$. By induction hypothesis,
  $\SPos_\sA(V)=\NPos_\sA(V)=\LPos_\sA(V)=\CPos_\sA(V)=\vide$.
  Therefore, $\LPos_\sA(T')=\vide$.
\end{itemize}
\end{itemize}
\end{itemize}
\end{prf}

\begin{lem}
\label{lem-comp-pos}
Assume that the condition \link{small-arrow} of Figure \ref{fig-small-conds} holds.  Let $\sA$
be a sort such that, for all sort $\sB\lttwf\sA$, $\I\sB$ satisfies
{\em(comp-small)}, and let $T$ be a type such that
$\Sort_{\letwf\sA}(T)$. Then:
\begin{itemize}
\item $\I{T}$ satisfies {\em(comp-sn)} and {\em(comp-red)} if
  $\SPos_\sA(T)=\vide$,
\item $\I{T}$ satisfies {\em(comp-neutral)} if $\NPos_\sA(T)=\vide$,
\item $\I{T}$ satisfies {\em(comp-lam)} if $\LPos_\sA(T)=\vide$,
\item $\I{T}$ satisfies {\em(comp-small)} if $\CPos_\sA(T)=\vide$.
\end{itemize}
\end{lem}

\begin{prf}
We proceed by induction on $\gttwf$ which is well-founded by
\link{typ-sn}\index{(typ-sn)|indlem[lem-comp-pos]}.
\begin{itemize}
\item $T=\sB$. Since $\Sort_{\letwf\sA}(T)$, we have $\sB\letwf\sA$.
\begin{itemize}
\item $\SPos_\sA(T)=\vide$. $\I{T}$ satisfies (comp-red) by Lemma
  \ref{lem-comp-sort-core}. By Lemma \ref{lem-comp-sn-sort-core},
  $\I{T}$ satisfies (comp-sn) if, for all $U<T$, $\I{U}$ satisfies
  (comp-sn). So, let $U<T$. By transitivity, $U\lttwf\sA$. Hence, by Lemma
  \ref{lem-typ-occ-sort}, $\Sort_{\lttwf\sA}(U)$ and, by Lemma
  \ref{lem-comp-pos-lt}, $\SPos_\sA(U)=\vide$. Therefore, by induction
  hypothesis, $\I{U}$ satisfies (comp-sn).
\item $\NPos_\sA(T)=\vide$. $\I{T}$ satisfies (comp-neutral) by Lemma
  \ref{lem-comp-sort-core}.
\item $\LPos_\sA(T)=\vide$. $\I{T}$ satisfies (comp-lam) by Lemma
  \ref{lem-comp-sort-core}.
\item $\CPos_\sA(T)=\vide$. Then, $\sB\lttwf\sA$ and, by assumption, $\I{T}$
  satisfies (comp-small).
\end{itemize}

\item $T=U\a V$. Then, $\Sort_{\letwf\sA}(U)$ and $\Sort_{\letwf\sA}(V)$.
\begin{itemize}
\item $\SPos_\sA(T)=\vide$. Then, $\NPos_\sA(U)=\vide$ and
  $\SPos_\sA(V)=\vide$. By induction hypothesis, $\I{U}$ satisfies
  (comp-neutral) and $\I{V}$ satisfies (comp-sn) and
  (comp-red). Hence, $\I{T}$ satisfies (comp-sn) and (comp-red) by
  Lemmas \ref{lem-comp-sn-arrow-core} and
  \ref{lem-comp-red-arrow-core}.

\item $\NPos_\sA(T)=\vide$. Then, $\SPos_\sA(U)=\vide$ and
  $\SPos_\sA(V)=\NPos_\sA(V)=\LPos_\sA(V)=\CPos_\sA(V)=\vide$. By
  induction hypothesis, $\I{U}$ satisfies (comp-sn) and
  (comp-red). Let now $V'\le V$. By Lemma \ref{lem-typ-le-sort},
  $\Sort_{\letwf\sA}(V')$. By Lemma \ref{lem-typ-ord-comp-pos},
  $\SPos_\sA(V')=\NPos_\sA(V')=\LPos_\sA(V')=\CPos_\sA(V')=\vide$. By
  \link{typ-right-subterm}\index{(typ-right-subterm)|indlem[lem-comp-pos-lt]}
  and transitivity, $T>V'$. Hence, by induction hypothesis, $\I{V'}$
  satisfies (comp-red), (comp-neutral), (comp-lam) and
  (comp-small). Therefore, by Corollary
  \ref{cor-comp-neutral-arrow-small}, $\I{T}$ satisfies (comp-neutral).

\item $\CPos_\sA(T)=\vide$. Then, $\NPos_\sA(T)=\vide$, hence
  $\SPos_\sA(U)=\vide$ and
  $\SPos_\sA(V)=\NPos_\sA(V)=\LPos_\sA(V)=\CPos_\sA(V)=\vide$. We
  now check the conditions of Lemma \ref{lem-comp-small-hyp}:

\begin{itemize}
\item $\I{T}$ satisfies (comp-neutral) since $\NPos_\sA(T)=\vide$.

\item Let $W<T$. We prove that $\CPos_\sA(W)=\vide$. By
  \link{typ-arrow}\index{(typ-arrow)|indlem[lem-typ-ord-comp-pos]}, there
  are two cases:
\begin{itemize}
\item $V\ge W$. Then, by Lemma \ref{lem-typ-ord-comp-pos},
  $\CPos_\sA(W)=\vide$.
\item $W=U\a V'$ and $V>V'$. Then, by Lemma
  \ref{lem-typ-ord-comp-pos},
  $\SPos_\sA(V')=\NPos_\sA(V')=\LPos_\sA(V')=\CPos_\sA(V')=\vide$. Thus,
  $\CPos_\sA(W)=\vide$.
\end{itemize}
\noindent
Hence, by induction hypothesis, $\I{W}$ satisfies (comp-small).

\item Let now $\sf^{|\vT|}:\vT\a T$ be small. There are $\vB$ and
  $\sB$ such that $V=\vV\a\sB$. So, by
  \link{small-arrow}\index{(small-arrow)|indlem[lem-typ-ord-comp-pos]},
  $(\all i)~\Sort_{\letwf\sB}(T_i)$ and $T_i\letwf T$.

  We first prove that, if $\Sort_{\letwf\sA}(\vS\a\sB)$ and
  $\CPos_\sA(\vS\a\sB)=\vide$, then $\sB\lttwf\sA$, by induction on
  $\vS$. If $\vS$ is empty, then $\Sort_{\letwf\sA}(\sB)$ and
  $\CPos_\sA(\sB)=\vide$. Thus, $\sB\lttwf\sA$. If $\vS=U\vV$, then
  $\CPos_\sA(\vS\a\sB)=\vide$ implies that
  ${\CPos_\sA(\vV\a\sB)}=\vide$. Hence, by induction hypothesis,
  $\sB\lttwf\sA$.

We therefore have $\sB\lttwf\sA$ for $T=U\vV\a\sB$, $\Sort_{\letwf\sA}(T)$
and $\CPos_\sA(T)=\vide$.

Hence, $\Sort_{\lttwf\sA}(\vT)$ and, by Lemma \ref{lem-comp-pos-lt},
$\SPos_\sA(\vT)=\vide$. If $T_i\lttwf T$, then $\I{T_i}$ satisfies
(comp-red) and (comp-sn) by induction hypothesis. Otherwise, $T_i=T$
and $\I{T_i}$ satisfies (comp-red) and (comp-sn) as shown previously.
\end{itemize}

\item $\LPos_\sA(T)=\vide$. Then, $\SPos_\sA(U)=\NPos_\sA(U)=\vide$ and
  $\SPos_\sA(V)=\NPos_\sA(V)=\LPos_\sA(V)=\CPos_\sA(V)=\vide$. By
  induction hypothesis, $\I{U}$ satisfies (comp-sn), (comp-red) and
  (comp-neutral).

Let now $V'\le V$. By Lemma \ref{lem-typ-le-sort},
$\Sort_{\letwf\sA}(V')$. By Lemma \ref{lem-typ-ord-comp-pos},
$\SPos_\sA(V')=\NPos_\sA(V')=\LPos_\sA(V')=\CPos_\sA(V')=\vide$. By
\link{typ-right-subterm}\index{(typ-right-subterm)|indlem[lem-comp-pos-lt]}
and transitivity, $T>V'$. Hence, by induction hypothesis, $\I{V'}$
satisfies (comp-sn), (comp-red), (comp-neutral), (comp-lam) and
(comp-small).

Let now $W\le T$. By Lemma \ref{lem-typ-le-sort},
$\Sort_{\letwf\sA}(W)$. Since $\LPos_\sA(T)=\vide$, we have
$\CPos_\sA(T)=\vide$. Hence $\CPos_\sA(W)=\vide$ by Lemma
\ref{lem-typ-ord-comp-pos}. If $W=T$, we have already seen that
$\I{T}$ satisfies (comp-small). Otherwise, $W>T$ and, by induction
hypothesis, $\I{W}$ satisfies (comp-small).

Therefore, by Corollary \ref{cor-comp-lam-arrow-small}, $\I{T}$
satisfies (comp-lam).\qed
\end{itemize}
\end{itemize}
\end{prf}

\begin{thm}
\label{thm-comp-typ-small}
Assume that the conditions of Figure \ref{fig-small-conds} hold.  For all types $T$, $\I{T}$ is a computability predicate, \ie satisfies
{\em(comp-sn)}, {\em(comp-red)}, {\em(comp-neutral)}, {\em(comp-lam)}
and {\em(comp-small)}.
\end{thm}

\begin{prf}
We proceed by induction on $\gttwf$ which is well-founded by
assumption \link{typ-sn}\index{(typ-sn)|indthm[thm-comp-typ-small]}. We
distinguish two cases:
\begin{itemize}
\item $T$ is a sort $\sA$. By Lemma \ref{lem-comp-sort-core}, $\I\sA$
  satisfies (comp-red), (comp-neutral), (comp-lam). By Lemma
  \ref{lem-comp-sn-sort-core} and induction hypothesis, $\I\sA$
  satisfies (comp-sn).

Let $W<\sA$. By Lemma \ref{lem-typ-occ-sort}, $\Sort_{\lttwf\sA}(W)$. By
Lemma \ref{lem-comp-pos-lt}, $\CPos_\sA(W)=\vide$. Therefore,  $\I{W}$ satisfies (comp-small) by Lemma
\ref{lem-comp-pos}.

Let now $\sf^\af:\vT\a\sA$ be small. By
\link{small-sort}\index{(small-sort)|indthm[thm-comp-typ-small]},
we have $(\all i)~\Sort_{\letwf\sA}(T_i)$ and $\SPos_\sA(T_i)=\vide$. Therefore,
 $\I\vT$ satisfies (comp-sn) and
(comp-red) by Lemma \ref{lem-comp-pos}. Hence, $\I\sA$
satisfies (comp-small)  by Lemma \ref{lem-comp-small-hyp}.

\item Otherwise, $T=U\a V$. Since $T\gts_l U$, by induction
  hypothesis, $\I{U}$ is a computability predicate. Let now $V'$ be a
  type such that $V\ge V'$. By
  \link{typ-right-subterm}\index{(typ-right-subterm)|indthm[thm-comp-typ-small]}
  and transitivity, $T>V'$.  By induction hypothesis, $\I{V'}$ is a
  computability predicate. Therefore, $\I{U\a V}$ satisfies (comp-sn)
  by Lemma \ref{lem-comp-sn-arrow-core}, (comp-red) by Lemma
  \ref{lem-comp-red-arrow-core}, (comp-neutral) by Corollary
  \ref{cor-comp-neutral-arrow-small}, (comp-small) by Lemma
  \ref{lem-comp-small-hyp} and (comp-lam) by Corollary
  \ref{cor-comp-lam-arrow-small}.\qed
\end{itemize}
\end{prf}

\begin{thm}
  If the conditions of Figure \ref{fig-small-conds} hold, then the relation $\gtpt$ of
  Definition \ref{def-cpr-small} is well-founded.
\end{thm}

\begin{prf}
After Theorem \ref{thm-wf}, Theorem \ref{thm-comp-typ-small} and Lemma
\ref{lem-comp-big-ind}.\qed
\end{prf}

\subsection{Checking computability assumptions for small symbols}
\label{sec-check-pos}

We explore here simple sufficient conditions under which the set
$\SPos_\sA(T)$ is empty, and therefore, which symbols whose output
type is a sort $\sA$ can be declared as small. The order of a type
plays an important role here. In case these conditions are not met, it
is of course always possible to check \link{small-sort} and
\link{small-arrow}, which are both decidable.

\begin{lem}
\label{lem-pos-sr-fst}
$\SPos_\sA(T)=\vide$ if $o(T)\le 1$.
\end{lem}

\begin{prf}
We proceed by induction on $T$.
\begin{itemize}
\item $T=\sB$. Then, $\SPos_\sA(T)=\vide$ by definition.
\item $T=U\a V$. Since $o(T)\le 1$, $o(U)\le 0$ and $o(V)\le 1$. $U$
  being a sort, $\NPos_\sA(U)=\vide$. Since $o(V)\le 1$,
  $\SPos_\sA(V)=\vide$ by induction hypothesis. Hence
  $\SPos_\sA(T)=\vide$.\qed
\end{itemize}\medskip
\end{prf}

\noindent Can therefore be declared as small, any symbol whose type is of order
less than or equal to $2$ since its arguments have then a type of
order less than or equal to $1$. This is in particular the case of the
constructors of first-order data types.

More generally, can be declared as small every constructor of a
strictly-positive inductive type, whatever its order is, which is the
class of inductive types allowed in the Coq proof assistant
\cite{coq}:

\begin{lem}
\label{lem-pos-sr-strict}
Given types $\vT$ and a sort $\sA$, 
$\SPos_\sA(\vT\a\sA)=\vide$ if 
$\Sort_{\lttwf\sA}(\vT)$.
\end{lem}

\begin{prf}
By induction on $T$.
\begin{itemize}
\item $T=\sA$. Immediate.
\item $T=U\a V$. Then, $\Sort_{\lttwf\sA}(U)$ and $V$ is of the form
  $\vT\a\sA$ with $\Sort_{\lttwf\sA}(\vT)$. By Lemma \ref{lem-comp-pos-lt},
  $\NPos_\sA(U)=\vide$. By induction hypothesis,
  $\SPos_\sA(V)=\vide$. Therefore, $\SPos_\sA(T)=\vide$.\qed
\end{itemize}\medskip
\end{prf}

\noindent Non-strictly positive types are not available in Coq because strong
elimination rules may cause non-terminating computations in Coq's
richer type system \cite{coquand88colog}. Nothing such that can happen in
our simple type system in which constructors of non-strictly positive
inductive types of order $\le 2$ can be declared as small:

\begin{lem}
\label{lem-pos-nlc-fst}
$\NPos_\sA(T)\!=\!\LPos_\sA(T)\!=\!\CPos_\sA(T)\!=\!\vide$
if $o(T)\le 1$, $\Sort_{\letwf\sA}(T)$ and $\Pos(\sA,T)\sle\Pos^-(T)$.
\end{lem}

\begin{prf}
We proceed by induction on $T$.
\begin{itemize}
\item $T=\sB$. Then, $\NPos_\sA(T)=\LPos_\sA(T)=\vide$ by
  definition. Since $\Sort_{\letwf\sA}(T)$, we have $\sB\le\sA$. Since
  $\Pos(\sA,T)\sle\Pos^-(T)$ and $\Pos^-(T)=\vide$, we have
  $\sB\neq\sA$. Therefore, $\CPos_\sA(T)=\vide$.
\item $T=U\a V$. Since $o(T)\le 1$, we have $o(U)\le 0$ and $o(V)\le
  1$. Thus, $U$ is a sort and $\SPos_\sA(U)=\NPos_\sA(U)=\vide$. By
  Lemma \ref{lem-pos-sr-fst}, $\SPos_\sA(V)=\vide$. Since
  $\Sort_{\letwf\sA}(T)$, we have $\Sort_{\letwf\sA}(V)$. Since
  $\Pos(\sA,T)\sle\Pos^-(T)$, we have
  $\Pos(\sA,V)\sle\Pos^-(V)$. Hence, by induction hypothesis,
  $\NPos_\sA(V)=\LPos_\sA(V)=\CPos_\sA(V)=\vide$. Therefore,
  $\NPos_\sA(T)=\LPos_\sA(T)=\CPos_\sA(T)=\vide$.\qed
\end{itemize}
\end{prf}

\begin{lem}
\label{lem-pos-sr-snd}
$\SPos_\sA(T)=\vide$ if $o(T)\le 2$, $\Sort_{\letwf\sA}(T)$ and
$\Pos(\sA,T)\sle\Pos^+(T)$.
\end{lem}

\begin{prf}
We proceed by induction on $T$.
\begin{itemize}
\item $T=\sB$. Then, $\SPos_\sA(T)=\vide$ by definition.
\item $T=U\a V$. Since $o(T)\le 2$, we have $o(U)\le 1$ and $o(V)\le
  2$. Since $\Sort_{\letwf\sA}(T)$, we have $\Sort_{\letwf\sA}(U)$ and
  $\Sort_{\letwf\sA}(V)$. Since $\Pos(\sA,T)\sle\Pos^+(T)$, we have
  $\Pos(\sA,U)\sle\Pos^-(U)$ and $\Pos(\sA,V)\sle\Pos^+(V)$. Hence, by
  Lemma \ref{lem-pos-nlc-fst}, $\NPos_\sA(U)=\vide$ and, by induction
  hypothesis, $\SPos_\sA(V)=\vide$. Therefore,
  $\SPos_\sA(T)=\vide$.\qed
\end{itemize}\medskip
\end{prf}

\noindent But positivity is not always sufficient as shown by the following
example. Assume that $\sf:T\a\sA$ with $T=(\sB\a N)\a\sA$,
$N=(\sB\a\sA)\a\sB$ and $\sB<\sA$. The sort $\sA$ occurs negatively in
$N$ and positively in $T$, which is a 3rd order type. We cannot
declare $\sf$ as small since we do not know how to prove that $\I{T}$
satisfies (comp-sn) by using our lemmas. Indeed, to prove that $\I{T}$
satisfies (comp-sn), we need to prove that $\I{\sB\a N}$ satisfies
(comp-neutral) (Lemma \ref{lem-comp-sn-arrow-core}). To prove that
$\I{\sB\a N}$ satisfies (comp-neutral), we need to prove that $\I{N}$
satisfies (comp-lam) (Corollary
\ref{cor-comp-neutral-arrow-small}). To prove that $\I{N}$ satisfies
(comp-lam), we need to prove that $\I{\sB\a\sA}$ satisfies
(comp-neutral) (Corollary \ref{cor-comp-lam-arrow-small}). To prove
that $\I{\sB\a\sA}$ satisfies (comp-neutral), we need to prove that
$\I\sA$ satisfies (comp-small) (Corollary
\ref{cor-comp-neutral-arrow-small}). But, to prove that $\I\sA$
satisfies (comp-small), we need to prove that $\I{T}$ satisfies
(comp-sn) (Lemma \ref{lem-comp-small-hyp}). The circularity has not
been broken here, but we can of course declare $\sf$ as being big
instead of small.

\subsection{Examples}
\label{sec-small-ex}

In this section, we analyze two examples that show the need for small
symbols and their use. We will see that CPO with small symbols
contains not only core CPO, but also a subset of its transitive
closure. But CPO with small symbols is not transitive either, as shown
by the second example which needs the use of both small symbols and
the transitive closure.

\begin{exa}
\label{ex-small-one}
Taken from the Termination Problems Data Base (TPDB) \cite{tpdb} under
the name \verb|Applicative_05__TreeFlatten|.
Let $\sa$ be a sort. Consider the function symbols $\ms{nil}:\sa$,
$\ms{flatten}:\sa\a\sa$, $\ms{concat}^1:\sa\a\sa$,
$\ms{cons}^2:\sa\a\sa\a\sa$, $\ms{append}^2:\sa\a\sa\a\sa$,
$\ms{node}^2:\sa\a\sa\a\sa$ and $\ms{map}^2:(\sa\a\sa)\a\sa\a\sa$.

The higher-order rewrite system
\begin{rewc}
\ms{map}(F,\ms{nil}) & \ms{nil}\\
\ms{map}(F,\ms{cons}(x,v)) & \ms{cons}(F~x,\ms{map}(F,v))\\
\ms{flatten}~\ms{node}(x,v)
& \ms{cons}(x,\ms{concat}(\ms{map}(\ms{flatten},v)))\\
\ms{concat}(\ms{nil}) & \ms{nil}\\
\ms{concat}(\ms{cons}(x,v)) & \ms{append}(x,\ms{concat}(v))\\
\ms{append}(\ms{nil},v) & v\\      
\ms{append}(\ms{cons}(x,u),v) & \ms{cons}(x,\ms{append}(u,v))\\
\end{rewc}
can be proved terminating with CPO by considering $\ms{concat}$,
$\ms{append}$, $\ms{map}$, $\ms{cons}$ and $\ms{nil}$ small, while
$\ms{node}$ and $\ms{flatten}$ can be either small or big (we consider
them as big in the following). All symbols can have multiset
status. Let the precedence be
$\ms{concat}\gtf\ms{append}\gtf\ms{cons}$,
$\ms{node}\gtf\ms{map}\gtf\ms{nil}$, $\ms{node}\gtf\ms{flatten}$ and
$\ms{map}\gtf\ms{cons}$.

Let us show the proof of the third rule, which is the most interesting
one. Since $\ms{cons}$ is small, we apply first \appsmall\ and then we
recursively need $\ms{flatten}~\ms{node}(x,v) \gtpt x$, which holds by
\appsub\ and then \bigsub, and $\ms{flatten}~\ms{node}(x,v)\gtpt
\ms{concat}(\ms{map}(\ms{flatten},v))$, which needs \appsmall\
again. We then recursively need $\ms{flatten}~\ms{node}(x,v)\gtpt
\ms{map}(\ms{flatten},v)$, which generates the subgoal
$\ms{node}(x,v)\gtpt \ms{map}(\ms{flatten},v)$ by \appsub, and then
the new subgoals $\ms{node}(x,v)\gtpt\ms{flatten}$ and
$\ms{node}(x,v)\gtpt v$ by \biggt. We conclude by \biggt\ and \bigsub.

The above example cannot be shown by core CPO because $\ms{flatten}$
is curried and the head symbol of the third rule is then an
application. There is however a way out with the transitive closure of
core CPO if one allows the introduction of new symbols. Let
$\ms{flattenunc}^1:\sa\a\sa$ be a new symbol. Assuming for example
$\ms{flatten}\gtf\ms{flattenunc}$ and
$\ms{node}\gtf\{\ms{cons},\ms{concat},\ms{map},\ms{flatten}\}$, we can
then show the successive ordering comparisons:

\begin{rewc}[~~\gtpt~~] 
\ms{flatten}~\ms{node}(x,v) & (\l x\,\ms{flattenunc}(x))\,\ms{node}(x,v)\\
(\l x\,\ms{flattenunc}(x))\,\ms{node}(x,v) & \ms{flattenunc}(\ms{node}(x,v)) \\
\ms{flattenunc}(\ms{node}(x,v))
& \ms{cons}(x,\ms{concat}(\ms{map}(\ms{flatten},v)))\\
\end{rewc}

\noindent
The first reduces to $\ms{flatten}\gtpt \l x\,\ms{flattenunc}(x)$, the
second is a $\beta$-reduction, and the third is a classical RPO-like
computation. Details are left to the reader.
\end{exa}

The use of small symbols can therefore help showing termination of
examples that would otherwise require the use of the transitive
closure of core CPO (as well as a signature extension in the above
case).  Small symbols, however, do not make CPO transitive. Our second
example requires indeed using both small symbols \emph{and} the transitive
closure:

\begin{exa}
\label{ex-small-two}
Taken from TPDB under the name \verb|AotoYamada_05__014|.
Let $\sa$ and $\sb$ be sorts. Consider the function symbols
$\ms{0}:\sb$, $\ms{nil}:\sa$, $\ms{inc}:\sa\a\sa$,
$\ms{double}:\sa\a\sa$, $\ss^1:\sb\a\sb$, $\ms{plus}^1:\sb\a\sb\a\sb$,
$\ms{times}^1:\sb\a\sb\a\sb$, $\ms{map}^1:{(\sb\a\sb)}\a\sa\a\sa$, and
$\ms{cons}^2:\sb\a\sa\a\sa$.

The higher-order rewrite system 
\begin{rewc}
\ms{plus}(\ms{0})~x & x\\
\ms{plus}(\ss(y))~x & \ss(\ms{plus}(y)~x)\\
\ms{times}(\ms{0})~x & \ms{0}\\
\ms{times}(\ss(y))~x & \ms{plus}( \ms{times}(y)~x )~x \\
\ms{map}(F)~\ms{nil} & \ms{nil} \\
\ms{map}(F)~\ms{cons}(x,v) & \ms{cons}(F~x,\ms{map}(F)~v))\\
\ms{inc} & \ms{map}(\ms{plus}(\ss(\ms{0})))\\
\ms{double} & \ms{map}(\ms{times}(\ss(\ss(\ms{0}))))\\
\end{rewc}
can be proved terminating with CPO by taking $\sa=\sb$ in the type
ordering, $\ms{cons}$ and $\ss$ as small symbols, the precedence
$\ms{times}\gtf\ms{plus}$,
$\ms{inc}\gtf\{\ms{map},\ms{plus},\ms{0}\}$,
$\ms{double}\gtf\{\ms{map},\ms{times},\ms{0}\}$, and status multiset
for all symbols.

We consider the 4th rule, for which we shall use the transitive
closure of CPO, and the 6th rule, for which small symbols are needed
(for the second rule too).

For the 4th rule, we exhibit the middle term $(\l
z\,\ms{plus}(\ms{times}(y)\,z)\,z)\,x$ which is smaller than the
lefthand side and $\b$-reduces to the righthand side of the rule.

To prove that $\ms{times}(\ss(y))\,x$ is greater than this middle term,
we apply \appeq, and since the second arguments are equal, we have to
show that $\ms{times}(\ss(y))\gtpt(\l
z\,\ms{plus}(\ms{times}(y)\,z)\,z)$. Since, both terms have the same
type, by \biglam\ and then \bigapp, we are left to show
$\ms{times}(\ss(y))\gtp[\{z\}]\ms{plus}(\ms{times}(y)\,z)$, since
$\ms{times}(\ss(y))\gtp[\{z\}] z$ holds by \bigvar. For this last
check, we apply first \biggt\ and then \bigapp, since
$\ms{times}(\ss(y))\gtp[\{z\}]\ms{times}(y)$ holds by \bigeq\ and then
\smallsub, and $\ms{times}(\ss(y))\gtp[\{z\}] z$ holds by \bigvar.

For the 6th rule, we apply first \appsmall, which requires to
check $\ms{map}(F)\,\ms{cons}(x,v) \gtpt F\,x$ and
$\ms{map}(F)\,\ms{cons}(x,v) \gtpt \ms{map}(F)\,v$. Since
the types of both sides are equivalent, the first one holds by applying
\appeq\ and then \bigsub\ to the first argument and
\smallsub\ to the second one.  Finally, for
$\ms{map}(F)\,\ms{cons}(x,v) \gtpt \ms{map}(F)\,v$, we apply
\appeq\ and then \smallsub\ to the second argument.
\end{exa}

\section{Conclusion}
\label{sec-conclu}

We have defined in this paper a well-founded relation on algebraic
lambda-terms following a type discipline accepting simple types in the
sense of Church, and inductive types in the sense of
Martin-L\"of. Further, we could easily cope with (implicitly)
universally quantified type variables as in~\cite{jouannaud07jacm}, a
type discipline called weak polymorphism.

We want to stress that core CPO has reached a point where we cannot
expect any major improvement, as indicated by the counter-examples
found to our own attempts to improve it. We are in great debt with
Cynthia Kop and Femke van Raamsdonk for igniting this quest, by
providing us with an example that removing the type check in the rule
\bigeq\ results in losing the well-foundedness property
\cite{kop08pc}. The very existence of these counter-examples supports
our conviction that CPO defines an extremely sharp decidable
approximation of sets of rules for which there exists a computability
predicate.

Of course, all these counter-examples still hold when adding inductive
types and small symbols. We did our best to exploit the idea of small
symbols as much as possible within our proof frame, but cannot argue
that the conditions on the signature of small symbols are all
necessary and that the corresponding recursive calls cannot be
improved: we did not extend our quest for counter-examples to this
question. We finally believe that there is also some room left for
improving the accessibility relationship, which is restricted so far
to terms headed by a function symbol, possibly applied to extra
arguments.

A more challenging problem to be investigated now is the
generalization of this new definition to the calculus of constructions
along the lines of \cite{walukiewicz03jfp} and the suggestions made in
\cite{jouannaud07jacm}, where an RPO-like ordering on types was
proposed which allowed to give a single definition for terms and
types. Generalizing CPO to dependent types appears to follow the
classical route initiated in~\cite{harper93jacm}, albeit
non-trivial~\cite{jouannaud15tlca}. We therefore believe that this
work should be applicable to
Dedukti~\cite{boespflug12pxtp-dedukti,dedukti} with limited effort. On
the other hand, we have failed so far to generalize CPO to truly
polymorphic types: its use in the proof assistant Coq \cite{coq} will
require much more effort.

Finally, note that HORPO \cite{koprowski09aaecc} on the one hand, and
the notion of computability closure on the other hand
\cite{blanqui13coq-cc}, have already been formalized in the proof
assistant Coq \cite{coq}. These works could serve as a basis for
formalizing the results presented in this paper and develop a
termination certificate verifier for CPO.

\medskip
{\bf Acknowledgements.} The authors thank the reviewers for their
suggestions.

\bibliographystyle{plain}

\end{document}